\documentclass[aps,prb,twocolumn,superscriptaddress,floatfix]{revtex4-2}

\usepackage{lmodern}
\usepackage{amsmath}
\usepackage{amssymb}
\usepackage{graphicx}
\usepackage{xcolor}
\usepackage{hyperref}
\usepackage{caption}
\usepackage{subcaption}
\usepackage{amsthm}
\usepackage{mathtools}
\usepackage{enumitem}

\newif\iffigplaceholders
\figplaceholderstrue      
\makeatletter
\iffigplaceholders
  \let\orig@includegraphics\includegraphics
  \renewcommand{\includegraphics}[2][]{%
    \IfFileExists{#2}{\orig@includegraphics[#1]{#2}}%
      {\PackageWarning{figcheck}{MISSING FIGURE: #2}%
       \fbox{\parbox[c][2.6cm][c]{0.72\linewidth}{\centering\scriptsize\ttfamily
         [figure placeholder]\\ \detokenize{#2}}}}}
\fi
\makeatother

\newcommand{\C}{\mathbb{C}}

\newcommand{\abs}[1]{\left|#1\right|}
\newcommand{\dH}{d_{\mathrm{H}}}

\DeclareMathOperator{\Res}{Res}
\DeclareMathOperator{\dist}{dist}
\newcommand{\nn}{\mathbf{n}}
\newcommand{\adj}{\operatorname{adj}}
\makeatletter
\renewcommand*\env@matrix[1][*\c@MaxMatrixCols c]{%
  \hskip -\arraycolsep
  \let\@ifnextchar\new@ifnextchar
  \array{#1}}
\makeatother

\begin{document}

\title{Spectral preservation under momentum-dependent similarity transformations in non-Hermitian lattice systems}

\author{Ma Ye}
\email{e0376923@u.nus.edu}

\date{\today}

\begin{abstract}
We investigate the conditions under which momentum-dependent similarity transformations preserve spectral properties of lattice Hamiltonians with open boundary conditions (OBC). While such transformations exactly preserve spectra in infinite systems, their application to finite systems introduces subtleties due to the long-range nature of the inverse transformation in real space. For general traceless $2\times 2$ Hamiltonians, we derive necessary and sufficient conditions for reduction to skew-diagonal form via constant similarity transforms, providing explicit transformation matrices for all cases. We then establish rigorous conditions for bulk spectral preservation under momentum-dependent transformations: the generalized Brillouin zone of $H$ must lie inside the smallest zero of $\det S(z)$ (the two-radius condition $r_{\mathrm{GBZ}}^{\max}<z_{\min}$), together with a spectral-stability (no critical non-Hermitian skin effect) condition on $H$. Two-sidedness of $S(z)$ governs only the modification of a finite number of boundary eigenvalues, not the bulk. Our results establish when bulk topological invariants computed in transformed coordinates reliably predict boundary physics, with implications for non-Hermitian systems, photonic crystals, and other platforms where chiral or hidden symmetries emerge only after appropriate basis changes.
\end{abstract}

\maketitle

\section{Introduction}

Similarity transformations provide a powerful tool for simplifying Hamiltonians while preserving spectral properties. In momentum space, where translationally invariant Hamiltonians $H(k)$ depend on crystal momentum $k$, momentum-dependent similarity transformations $S(k)$ have proven particularly valuable for revealing hidden symmetries and topological structures. Examples include non-Hermitian systems where chiral symmetry becomes apparent only after appropriate basis changes \cite{ref1,ref2}, photonic crystals with emergent sublattice structure \cite{ref3}, and dissipative quantum systems where unitary transformations expose topological invariants \cite{ref4}.

The theoretical framework for momentum-dependent transformations is well-established for periodic boundary conditions (PBC) or infinite systems, where $H'(k) = S(k)H(k)S^{-1}(k)$ preserves all eigenvalues exactly. However, finite systems with open boundary conditions (OBC) present fundamental challenges. This issue is particularly acute in non-Hermitian systems exhibiting the skin effect, where OBC and PBC spectra differ dramatically \cite{ref5,ref6}.

The central difficulty lies in the mathematical structure of the inverse transformation. While $S(z)$ (where $z = e^{ik}$) typically contains only finitely many powers of $z$ as a Laurent polynomial, its inverse $S^{-1}(z)$ generically involves rational functions with poles. When transformed to real space for OBC calculations, this manifests as long-range couplings extending throughout the system, creating boundary effects that can fundamentally alter the spectrum.

Recent progress has been made in understanding specific classes of transformations. Zhong, Wang and Fan~\cite{ref7} proved the equivalence of the generalized-Brillouin-zone and pole--zero edge-state invariants for sublattice-symmetric two-band models, and extended the pole--zero construction to Hamiltonians that are not already in off-diagonal form. Wu, Xie, Zhou and An~\cite{Wu2022} showed that the periodic-boundary spectrum, the open-boundary spectrum and the GBZ are connected by a family of similarity transformations, and used this to compute them numerically. Both use similarity transformations as a tool at the level of the symbol; what is missing is an account of what survives when such a transformation is realised as a finite-range operator on a finite lattice, which is the question we address. However, a systematic treatment of when arbitrary traceless Hamiltonians can be reduced to forms revealing hidden symmetries, and what constraints this imposes on the transformation's momentum dependence, has remained incomplete.

This work addresses two fundamental questions. First, for traceless $2\times 2$ Hamiltonians—a ubiquitous class including many topological and non-Hermitian models—when can constant (momentum-independent) similarity transformations reveal hidden chiral symmetry by bringing the Hamiltonian to skew-diagonal form? Second, when momentum dependence is unavoidable, under what precise conditions does a transformation preserve physically relevant spectral properties under OBC?

We establish that constant transformations exist if and only if the Hamiltonian's matrix elements satisfy a linear dependence relation with non-isotropic normal. Momentum dependence does not by itself force a loss of locality: because $S^{-1}=\adj S/\det S$, the transformed system remains a finite-range lattice model precisely when $\det S(z)$ is a monomial, and this \emph{unimodular} class --- strictly larger than the constant one --- is exact on any finite chain. Everything that follows concerns the complementary case, where $\det S$ has genuine zeros and no exact finite-range realisation exists. We provide explicit formulas for the transformation matrices covering all cases, including degenerate scenarios. For momentum-dependent transformations, we derive rigorous necessary and sufficient conditions for bulk preservation: the Hamiltonian must not be critically sensitive to perturbations (spectral stability), and the generalized Brillouin zone of $H$ must lie inside the smallest zero of $\det S(z)$, i.e.\ $r_{\mathrm{GBZ}}^{\max}<z_{\min}$. These conditions ensure that the truncated inverse transformation converges on the GBZ contour, permitting bulk spectral preservation despite boundary truncation. A separate two-sidedness property of $S(z)$ controls whether boundary eigenvalues are modified.

Importantly, we demonstrate that even optimal transformations cannot preserve all eigenvalues in finite systems—boundary states are necessarily modified. This reflects a fundamental incompatibility between momentum-space methods (which assume translational invariance) and real-space boundaries. Our results clarify when bulk topological invariants computed after transformation reliably predict boundary phenomena, and identify the error scaling.

The remainder of this paper proceeds as follows. Section II establishes the framework for similarity transformations and provides explicit transformation matrices for bringing traceless Hamiltonians to skew-diagonal form. Section III states our main theorems on spectral preservation, with detailed proofs relegated to the Appendix. Section IV presents numerical verification demonstrating both the utility of constant transformations for revealing topological zero modes and the limitations of momentum-dependent transformations near boundaries. Section V discusses implications for topological analysis and experimental platforms. Section VI concludes.

\section{Framework and explicit transformations}

\subsection{General two-band Hamiltonian}

We consider a general single-particle two-band Hamiltonian with translational invariance, expressed in momentum space as
\begin{equation}
H(z) = \begin{pmatrix}
h_{11}(z) & h_{12}(z)\\
h_{21}(z) & h_{22}(z)
\end{pmatrix}, \quad z = e^{ik}
\end{equation}
where $k$ is the crystal momentum and each matrix element is a Laurent polynomial
\begin{equation}
h_{ij}(z) = \sum_{n=-N}^{M} h_{ij,n}z^{n}.
\end{equation}
In real space, this corresponds to a tight-binding Hamiltonian with hopping range determined by the powers of $z$.

\subsection{Similarity transformations}

A similarity transformation is defined as
\begin{equation}
H'(z) = S^{-1}(z)H(z)S(z)
\end{equation}
where $S(z)$ is an invertible matrix. If $S(z)$ depends on momentum, its matrix elements are Laurent polynomials, but $S^{-1}(z)$ will generically have matrix elements that are rational functions (ratios of polynomials). In real space, this corresponds to long-range hopping that may decay as a power law rather than exponentially.

For OBC calculations, such long-range terms pose significant complications since boundaries truncate the system. We therefore distinguish two cases: constant transformations where $S$ is independent of $z$, and momentum-dependent transformations where $S(z)$ contains powers of $z$.

\subsection{Reduction to skew-diagonal form}

For traceless Hamiltonians—a natural class since trace shifts do not affect topological properties—we write
\begin{equation}
H(z) = \begin{pmatrix}
d(z) & q(z)\\
r(z) & -d(z)
\end{pmatrix}.
\end{equation}

A Hamiltonian in skew-diagonal form
\begin{equation}
H_{\text{skew}}(z) = \begin{pmatrix}
0 & a(z)\\
b(z) & 0
\end{pmatrix}
\end{equation}
exhibits manifest chiral symmetry and admits well-defined topological invariants. The question is: which Hamiltonians $H(z)$ can be brought to this form via constant similarity transformation?

\subsection{Necessary and sufficient condition}

We establish the following result (proof in Appendix):

\textbf{Theorem~1} A traceless $2\times 2$ Hamiltonian $H(z)$ with entries $d(z)$, $q(z)$, $r(z)$ can be reduced to skew-diagonal form by a constant similarity transformation if and only if there exist constants $k_1$, $k_2$, $k_3$ (not all zero) such that
\begin{equation}
k_1 d(z) + k_2 q(z) + k_3 r(z) = 0 \quad \forall z,
\label{eq:constraint}
\end{equation}
with the non-degeneracy condition $k_1^2+4k_2k_3\neq0$ (equivalently, at least one dependence relation has a non-isotropic normal).

This linear dependence relation is necessary (following from requiring vanishing diagonal elements in the transformed Hamiltonian) and, together with the non-degeneracy condition, sufficient (allowing explicit construction of an invertible $S$). The non-degeneracy is essential: the explicit transformation has $\det S=\sqrt{k_1^2+4k_2k_3}/k_3$, so when the only available relation satisfies $k_1^2+4k_2k_3=0$ the construction returns a singular $S$ and, as shown in Appendix~A, no invertible $S$ exists. Geometrically, writing $H=\tfrac{q+r}{2}\sigma_x+\tfrac{i(q-r)}{2}\sigma_y+d\,\sigma_z$, the relation's normal $\vec n$ obeys $\vec n\!\cdot\!\vec n=k_1^2+4k_2k_3$; skew-diagonalisation requires rotating $\vec n$ onto $\hat e_3$ by $SO(3,\mathbb C)$, which is impossible for an isotropic ($\vec n\!\cdot\!\vec n=0$) normal.

\subsection{Explicit transformation matrices}

Given the constraint in Eq.~(\ref{eq:constraint}), we can explicitly construct the transformation matrix $S$. Setting $s_{11} = s_{12} = 1$ (columns can be rescaled), the remaining matrix elements $s_{21}$ and $s_{22}$ satisfy
\begin{equation}
s_{21} + s_{22} = -\frac{k_1}{k_3}, \quad s_{21}s_{22} = -\frac{k_2}{k_3}
\end{equation}
which are the Vieta formulas for the quadratic equation
\begin{equation}
t^2 + \frac{k_1}{k_3}t - \frac{k_2}{k_3} = 0.
\end{equation}

The solutions are
\begin{equation}
s_{21}, s_{22} = \frac{-k_1 \pm \sqrt{k_1^2 + 4k_2k_3}}{2k_3}.
\end{equation}

Table \ref{tab:transform} lists the transformation matrices for all cases, including degenerate scenarios where one or more coefficients vanish. The derivation of each case appears in Appendix Section II.

\begin{table*}[htbp]
\caption{Explicit similarity transformation matrices $S$ for reducing traceless $H(z)$ to skew-diagonal form, given the constraint $k_1d(z) + k_2q(z) + k_3r(z) = 0$. Each column can be rescaled arbitrarily. The transformation matrix has elements $S = \begin{pmatrix} s_{11} & s_{12} \\ s_{21} & s_{22} \end{pmatrix}$. The generic row requires the non-degeneracy $k_1^2+4k_2k_3\neq0$; when the only relation has $k_1^2+4k_2k_3=0$ (isotropic normal) the two roots coincide, $S$ is singular, and no reduction exists.}
\label{tab:transform}
\centering
\begin{tabular}{cccccc}
\hline
$k$ values & $s_{11}$ & $s_{12}$ & $s_{21}$ & $s_{22}$ & Comments \\
\hline
$k_{1,2,3}\neq 0$ & $1$ & $1$ & $\frac{-k_1 \pm \sqrt{k_1^2 + 4k_2k_3}}{2k_3}$ & $\frac{2k_2}{k_1 \mp \sqrt{k_1^2 + 4k_2k_3}}$ & Generic case\\[0.3cm]
$k_1=0\neq k_{2,3}$ & $1$ & $1$ & $\pm \sqrt{\frac{k_2}{k_3}}$ & $\mp \sqrt{\frac{k_2}{k_3}}$ & Symmetric constraint\\[0.3cm]
$k_2=0\neq k_{1,3}$ & $1$ & $1$ & $-\frac{k_1}{k_3}$ & $0$ & Special degeneracy\\[0.3cm]
$k_3=0\neq k_{1,2}$ & $0$ & $1$ & $1$ & $\frac{k_2}{k_1}$ & Linear constraint\\[0.3cm]
$k_{1,2}=0\neq k_{3}$ & --- & --- & --- & --- & $d(z),q(z)$ independent\\[0.3cm]
$k_{1,3}=0\neq k_{2}$ & --- & --- & --- & --- & $d(z),r(z)$ independent\\[0.3cm]
$k_{2,3}=0\neq k_{1}$ & $1$ & $0$ & $0$ & $1$ & Already skew-diagonal\\
\hline
\end{tabular}
\end{table*}

\subsection{Physical interpretation}

The transformation matrices in Table \ref{tab:transform} reveal important physical structure. When all matrix elements are constants (independent of $z$), the transformation $S$ represents a purely local change of basis that affects each unit cell identically. Such transformations preserve all locality properties and introduce no complications for OBC calculations.

In contrast, when the constraint parameters $k_i$ themselves depend on momentum through the functions $d(z)$, $q(z)$, $r(z)$, or when mathematical operations like square roots introduce additional $z$-dependence, the resulting transformation becomes momentum-dependent. This corresponds to a non-local change of basis that mixes degrees of freedom from different spatial locations.

Consider the generic case where all three constraint parameters are nonzero. The transformation matrix elements involve terms like $\sqrt{k_1^2 + 4k_2k_3}$. If any of the $k_i$ contain $z$-dependence from the original Hamiltonian functions, this square root operation can introduce additional powers of $z$ beyond what appeared in the original constraint. When expanded in real space, such transformations introduce couplings between unit cells with range determined by the highest and lowest powers of $z$ in the matrix elements.

The critical consequence for finite systems is that $S^{-1}(z)$ will generically involve rational functions of $z$, with poles arising from denominators in the explicit formulas or from the matrix inversion process itself. The location of these poles relative to the \emph{generalized} Brillouin zone of $H$—not the unit circle—determines whether bulk spectral preservation is achievable in finite systems: the nearest pole must lie outside the largest GBZ radius, $z_{\min}>r_{\mathrm{GBZ}}^{\max}$, as we establish rigorously in the next section. The unit-circle criterion is only the Hermitian special case $r_{\mathrm{GBZ}}^{\max}=1$.

\subsection{Momentum-dependent transformations that remain finite range}
\label{sec:unimodular}

The obstruction just described is sharper than it looks. Since $S^{-1}=\adj S/\det S$ and
the adjugate of a Laurent-polynomial matrix is again a Laurent-polynomial matrix, the only
source of unbounded range is a zero of $\det S$ away from the origin. Hence
\begin{equation}
S^{-1}(z)\ \text{is a Laurent polynomial}
\iff \det S(z)=c\,z^{m},
\label{eq:unimodular}
\end{equation}
that is, iff $S$ is a unit of $GL_2(\C[z,z^{-1}])$. For such \emph{unimodular} transformations
the transformed system is again a finite-range lattice model, the real-space transformation
is exactly invertible on a chain of any length, and the entire spectrum --- bulk
\emph{and} boundary --- is preserved exactly. Since $\det S$ then has no zeros away from the
origin, the radial condition of Theorem~4 below is vacuous on this class.

Unimodular transformations are strictly more powerful than the constant ones of
Table~\ref{tab:transform}. Taking $S=\left(\begin{smallmatrix}1&p(z)\\0&1\end{smallmatrix}\right)$,
which has $\det S=1$ for any Laurent polynomial $p$, gives $(S^{-1}HS)_{11}=d-pr$, so $H$ is
brought to skew-diagonal form whenever $r(z)$ divides $d(z)$. For example
$d=(z+1)(z-2)$, $q=z+3+z^{-1}$, $r=z+1$ admits no constant transformation --- the only
solution of Eq.~(\ref{eq:constraint}) is $k_1=k_2=k_3=0$ --- yet $p=z-2$ yields
\begin{equation}
S^{-1}HS=\begin{pmatrix} 0 & z^{3}-3z^{2}+z+7+z^{-1}\\[2pt] z+1 & 0\end{pmatrix}
\end{equation}
exactly, with hoppings of range three. A complete characterisation of which $H$ can be
skew-diagonalised by \emph{some} unimodular $S$ remains open; since $\C[z,z^{-1}]$ is
Euclidean in the width $\deg_{\mathrm{top}}-\deg_{\mathrm{bot}}$, its $GL_2$ is generated by
elementary matrices, so one expects a greatest-common-divisor criterion generalising the
divisibility condition above.

The rest of this paper treats the complementary case, $\det S$ with genuine zeros, where no
exact finite-range transformation exists and truncation is unavoidable.

\section{Main results on spectral preservation}

Having established when and how traceless Hamiltonians can be reduced to skew-diagonal form, we now address spectral preservation under general momentum-dependent transformations. Our main results concern three distinct scenarios: infinite systems, finite systems with bulk preservation, and the inevitable modification of boundary states.

\subsection{Exact preservation in infinite systems}

For bi-infinite lattices (extending to $\pm\infty$ with no boundaries), momentum-dependent transformations preserve all spectral properties exactly. The real-space representation of $S(z) = \sum_k S_k z^k$ takes the block-Toeplitz form
\begin{equation}
S_{\text{bi-inf}} = \begin{pmatrix}[cc|cccc]
\ddots & & & & & \\
& S_0 & S_{-1} & S_{-2} & \cdots & \\
\hline
& S_1 & S_0 & S_{-1} & \cdots &\\
& S_2 & S_1 & S_0 & \cdots &\\
& & & & & \ddots
\end{pmatrix}
\end{equation}
extending indefinitely in both directions. Similarly, the inverse transformation $S^{-1}(z) = \sum_k S^-_k z^k$ has real-space representation with coefficients $S^-_k$.

\textbf{Theorem~2} For a bi-infinite lattice with Hamiltonian $H(z)$ and any transformation $S(z)$ (Laurent polynomial) with $\det S(z)\neq0$ on the unit circle $|z|=1$, the transformed Hamiltonian $H'(z) = S^{-1}(z)H(z)S(z)$ has identical spectrum to $H(z)$ when both are represented in real space.

The real-space identity $S_{\mathrm{real}}S^{-1}_{\mathrm{real}}=I$ (proven in Appendix~C) is a purely algebraic consequence of $S(z)S^{-1}(z)=I$ and holds for any invertible symbol. The \emph{spectral} conclusion, however, requires $S_{\mathrm{real}}$ to be a boundedly invertible operator on $\ell^2(\mathbb Z)$, i.e.\ $\det S(z)\neq0$ for $|z|=1$; otherwise $S^{-1}(z)$ is unbounded on the circle and the similarity is not implemented by a bounded operator.

The key technical point, proven in Appendix Section C, is that the real-space inverse of the matrix $S$ equals the real-space representation of the symbol inverse $S^{-1}(z)$. This follows from the orthogonality relations
\begin{equation}
\sum_m S_m S^-_{-m} = I, \quad \sum_m S_{n+m}S^-_{-m} = 0 \text{ for } n\neq 0
\end{equation}
which ensure that multiplication in real space corresponds to convolution of Laurent coefficients.

\subsection{Conditions for bulk preservation in OBC systems}

For finite systems with OBC, spectral preservation becomes subtle. The real-space transformation matrix must be truncated at boundaries, introducing errors in the relationship $S^{-1}S \approx I$ near edges. Whether these errors remain localized or propagate depends critically on the analytic properties of $S^{-1}(z)$. However before analyzing such properties of $S(z)$, we would need to check the stability of OBC spectra of $H(z)$ w.r.t. perturbances.

\textbf{Theorem~3 (OBC spectral stability, sharp form)}
Let $H(z)$ and $C(z)$ be fixed $2\times2$ Laurent-polynomial symbols with the same $z$-powers, write $H_\varepsilon(z)=H(z)+\varepsilon C(z)$, and let $\sigma_N$ denote the OBC spectrum on $N$ sites. Denote by $\tilde{P}(z,E)$ the characteristic polynomial $\det[H(z)-EI]$ cleared of denominators (a genuine polynomial in $z$ and $E$) and by $\mathcal{G}$ the associated generalized Brillouin zone (GBZ). Then, in the thermodynamic limit,

\emph{(i) Stability.} If $\tilde{P}(z,E)$ is \textbf{irreducible} over $\mathbb{C}[z,E]$, then $\lim_N\sigma_N(H_\varepsilon)\to\lim_N\sigma_N(H)$ as $\varepsilon\to0$, with Hausdorff distance
\begin{equation}
d_{\mathrm H}\!\left[\lim_N\sigma_N(H),\ \lim_N\sigma_N(H_\varepsilon)\right]=O\!\left(|\varepsilon|^{1/m_{\max}}\right)
\end{equation}
where $m_{\max}\geq1$ is the maximal order of vanishing of the modulus-gap function on the GBZ (generically $m_{\max}=2$, giving $O(|\varepsilon|^{1/2})$).

\begin{figure*}[htbp]
\centering
\includegraphics[width=0.92\linewidth]{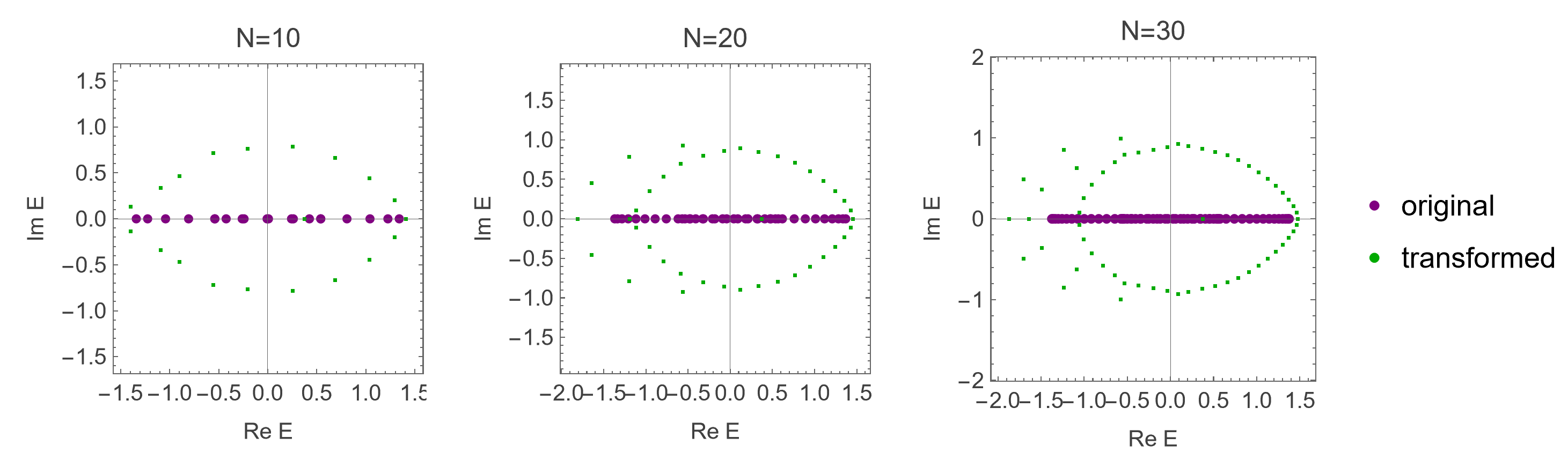}
\caption{Violation of Condition~2 (critical NHSE). When the combined GBZ of $H$ is not the union of its factor bands---here two decoupled chains---the exponentially small truncation defect drives an $O(1)$ discontinuous change of the OBC spectrum that grows with system size. Original spectrum (purple) versus transformed (green) for $N=10,20,30$.}
\label{fig:critnhse}
\end{figure*}

\emph{(ii) Instability (critical NHSE).} Suppose instead that $H_\varepsilon=H_0+\varepsilon C$ couples decoupled subsystems: $H_0=\bigoplus_i H_i$ is a direct sum of blocks (equivalently, $H_0(z)$ is constant--similar to block--diagonal form), with subsystem symbols multiplying to $\tilde{P}=\prod_i f_i$, and $C$ is a generic coupling. Then the coupled OBC spectrum is discontinuous as $\varepsilon\to0$ if and only if the combined GBZ of the product symbol differs from the union of the factor bands,
\begin{equation}
\mathcal{G}_{\mathrm{comb}}\!\Big[\textstyle\prod_i f_i\Big]\;\neq\;\bigcup_i \mathcal{G}[f_i],
\label{eq:sharp_converse}
\end{equation}
Figure~\ref{fig:critnhse} shows the resulting discontinuity for two decoupled chains. Equivalently, the spectrum is discontinuous iff some factor is \emph{removed} --- its OBC band enters another factor's GBZ region. This criterion is \textbf{definitive for the coupled--subsystem case}: it follows from a model--independent backbone plus an $\varepsilon$--sensitivity scoping argument, is closed for two subsystems by an explicit self--energy bound, and is fully explicit for single--band (Hatano--Nelson) factors; the Appendix delimits its scope for general subsystems.

\emph{Refinement.} The naive reading ``unequal skin depth $\Rightarrow$ discontinuity'' is \textbf{necessary but not sufficient}: subsystems of different decay rate whose bands are GBZ-separated stay continuous. The sharp, symbol-decidable criterion is Eq.~(\ref{eq:sharp_converse}) --- the critical non-Hermitian skin effect~\cite{Li2020} (multicomponent version~\cite{QinMa2023}; see also the scaling rule of Ref.~\cite{YokomizoMurakami2021}, the coupled-chain hybridisation analysis of Ref.~\cite{Rafi2022}, and the exactly solvable size-dependent boundary effects of Ref.~\cite{Guo2021}) --- and expresses the non-commutation of the thermodynamic ($N\to\infty$) and zero-coupling ($\varepsilon\to0$) limits. Direction (i) is proved in the Appendix for any $H$; direction (ii) is proved there for two subsystems by an explicit Schur self-energy bound and for any number by the $\varepsilon$-sensitivity scoping argument, the residual multi-block pairwise combinatorics being bypassed. \emph{Remark (general reducible $H$).} If $\tilde{P}$ merely factorises but $H(z)$ admits no constant block structure --- characteristic-polynomial reducibility is strictly weaker than block-diagonalisability --- the same criterion is expected to govern $\sigma_N(H)$, but establishing it requires a companion/transfer-matrix realisation of the factors~\cite{KunstDwivedi2019} and is left open.

\textbf{Theorem~4}
Consider a finite lattice of length $L$ with OBC, Hamiltonian $H(z)$ stable under conditions of Theorem 3, and transformation $S(z)$ whose inverse $S^{-1}(z)$ is holomorphic at $z=0$ (so that the Taylor truncation $P_d[S^{-1}]$ is well defined; for a one-sided $S$ this is the familiar $\det S(0)\neq0$, but for a genuinely two-sided $S$ the two conditions are logically independent --- see Appendix~E). Let $r_{\mathrm{GBZ}}^{\max}$ be the largest generalized-Brillouin-zone radius of $H$ over its bulk spectrum, and $z_{\min}$ the modulus of the zero of $\det S(z)$ nearest the origin. The bulk spectrum of $H(z)$ is preserved under the truncated transformation $\tilde H_d(z) = P_d[S^{-1}](z)H(z)S(z)$ as $d,L\to\infty$ if and only if:
\begin{enumerate}
\item $r_{\mathrm{GBZ}}^{\max} < z_{\min}$, i.e.\ the whole GBZ of $H$ lies inside the smallest zero of $\det S$, AND
\item $H(z)$ satisfies the spectral-stability condition of Theorem 3, i.e.\ its combined GBZ equals the union of its factor bands, $\mathcal{G}_{\mathrm{comb}}=\bigcup_i\mathcal{G}[f_i]$ (in particular, any irreducible $H(z)$ qualifies).
\end{enumerate}

What Conditions~1--2 mean is simplest stated on the transformed model itself. Unless
$\det S$ is a monomial (Sec.~\ref{sec:unimodular}), $S^{-1}(z)H(z)S(z)$ is a \emph{rational}
symbol, so the transformed system is a lattice model whose hoppings have unbounded range.
Truncating at range $d$ replaces $S^{-1}$ by its degree-$d$ Taylor polynomial, whose defect
on a circle $|z|=r$ is controlled by the single factor $(r/z_{\min})^{d}$, with $z_{\min}$ the
modulus of the pole nearest the origin. The bulk spectrum, however, is read not on the unit
circle but on the GBZ contour $|z|=r_{\mathrm{GBZ}}(E)$, which the skin effect moves off it.
Condition~1 therefore says exactly this:
\begin{quote}
\emph{the truncated approximation converges on the GBZ.}
\end{quote}
When it holds, keeping hoppings out to range $d$ is a controlled approximation and the bulk
survives; when it fails the truncation does not converge on the contour where the spectrum
lives, and no finite-range approximation represents it. Note that $z_{\min}$ need only exceed the GBZ radius,
not unity --- the familiar Hermitian rule ``no zero of $\det S$ in the unit disk'' is the
special case $r_{\mathrm{GBZ}}^{\max}=1$, and it misclassifies zeros in both directions once
$r_{\mathrm{GBZ}}^{\max}\neq1$.

Condition~2 is what converts a statement about the \emph{symbol} into one about
\emph{eigenvalues}. Theorem~3 is precisely the response function of the OBC spectrum to a
symbol-level perturbation: H\"older-continuous with exponent $1/m$ in the stable case, and
discontinuous under critical NHSE, where an exponentially small truncation defect produces an
$O(1)$ jump. The two conditions are thus a bound on the input and a certificate for the
transfer function, and Theorem~4 is their composition. The proof --- an exact identity for the
truncated characteristic polynomial, a radial Cauchy bound on the inverse defect, and the
Smith normal form locating the poles of $S^{-1}$ at the zeros of $\det S$ --- is given in
Appendix~E.

\begin{figure*}[htbp]
\centering
\includegraphics[width=0.92\linewidth]{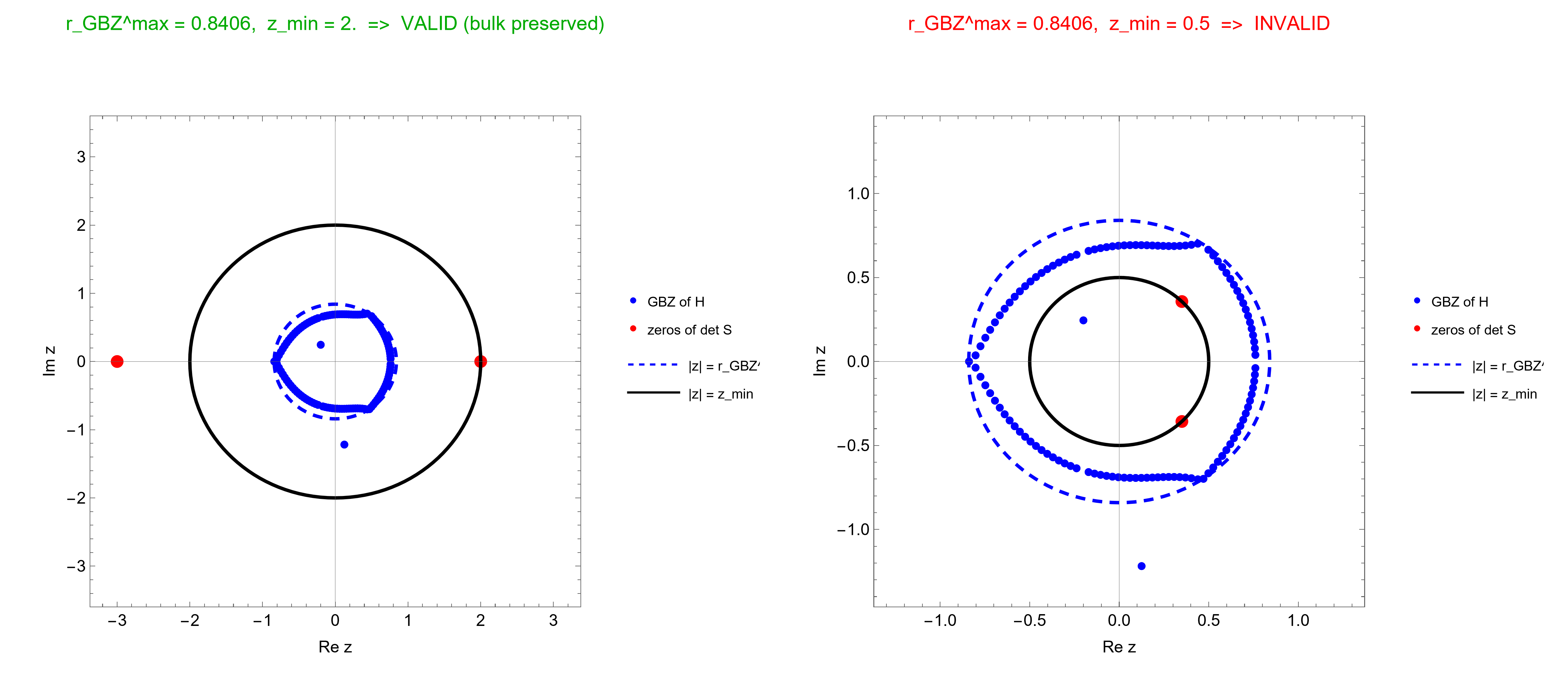}
\caption{The radial condition $r_{\mathrm{GBZ}}^{\max}<z_{\min}$ in the complex-$z$ plane. Blue: the generalized Brillouin zone of $H$; red: the zeros of $\det S(z)$; dashed blue circle $|z|=r_{\mathrm{GBZ}}^{\max}$; solid black circle $|z|=z_{\min}$. (a)~A valid transformation: the whole GBZ lies inside the nearest zero of $\det S$, and the bulk spectrum is preserved. (b)~An invalid transformation: the zeros of $\det S$ (a conjugate pair of modulus $\tfrac12$) lie inside the GBZ, so $z_{\min}<r_{\mathrm{GBZ}}^{\max}$ and preservation fails.}
\label{fig:radial}
\end{figure*}

Notably, bulk preservation does \emph{not} require $S(z)$ to contain both positive and negative powers of $z$. A one-sided $S$ (only non-negative powers) preserves the bulk exactly whenever Conditions~1--2 hold: its real-space matrix $S_L$ is block-lower-triangular, so $(S_L)^{-1}=(S^{-1})_L$ and for $d\geq L$ the truncated transform is an exact similarity, preserving the \emph{entire} spectrum.

\subsection{Modification of boundary states}

Even when Theorem 4's conditions are satisfied, finite systems exhibit persistent boundary effects.

\textbf{Theorem~5} Let $S(z)$ be a genuinely two-sided transformation (containing both positive and negative powers of $z$) and let $H(z)$ satisfy the conditions of Theorem 4. Then for all sufficiently large $L$, at least one eigenvalue of the OBC Hamiltonian is modified by the truncated transformation, provided a boundary trace defect (below) is nonzero. Proof in the appendix.

The mechanism is an exact trace identity rather than any failure of similarity. Because $S_L$ is not block-triangular for a two-sided $S$, the finite truncation breaks the similarity: $(S^{-1})_L S_L\neq I$, leaving a boundary-localized finite-rank defect. Writing $M_L=S_L\,P_d[S^{-1}]_L$, the sum of eigenvalue shifts equals the boundary trace defect
\begin{equation}
\begin{split}
\sum_i\big[\lambda_i(\tilde H_L)-\lambda_i(H_L)\big]&=\operatorname{tr}\tilde H_L-\operatorname{tr}H_L\\
&=\operatorname{tr}\big[(M_L-I)H_L\big]=:T_L .
\end{split}
\label{eq:trace5}
\end{equation}
Since $M_L-I$ is supported within $O(\text{range}+d)$ of the two boundaries, $T_L$ is boundary-localized and becomes exactly $L$-independent once $L\gtrsim2(\text{range}+d)$ --- which is why ``sufficiently large $L$'' is the natural setting. If $T_L\neq0$ the eigenvalue multisets cannot coincide, so at least one eigenvalue is modified; if $T_L$ vanishes by symmetry (e.g.\ a diagonal $S$ with skew $H$), the modification is certified by a higher power-trace defect $T_L^{(k)}=\operatorname{tr}\tilde H_L^{\,k}-\operatorname{tr}H_L^{\,k}$. All $T_L^{(k)}$ vanish for large $L$ only if the boundary defect preserves the characteristic polynomial exactly, which we do not observe for any two-sided $S$. Eigenstates with support near boundaries are most affected, while extended bulk states remain insensitive.

\section{Numerical verification}

We verify our theoretical results through explicit numerical calculations on finite chains with OBC. All computations employ chain length $N=80$ sites unless otherwise noted.

\subsection{Constant transformation: Hidden chiral symmetry}
\label{sec:constant}

We first demonstrate a case where constant similarity transformation reveals hidden chiral symmetry, enabling reliable topological prediction. Consider the Hamiltonian
\begin{equation}
H(z) = \begin{pmatrix}
d(z) & q(z)\\
r(z) & -d(z)
\end{pmatrix}
\end{equation}
where
\begin{align}
d(z) &= z + \frac{4}{3} - \frac{2}{z}\\
q(z) &= z - \frac{1}{2} - \frac{1}{10z}\\
r(z) &= 6z + \frac{27}{2} + \frac{11i}{3} - \frac{177}{10z} - \frac{19i}{5z}.
\end{align}

These satisfy the constraint $(-9-2i)d(z) + (3+2i)q(z) + r(z) = 0$, allowing constant transformation via Table \ref{tab:transform}:
\begin{equation}
S = \begin{pmatrix}
1 & 1\\
9.351+2.134i & -0.3513-0.1337i
\end{pmatrix}.
\end{equation}

The transformed Hamiltonian
\begin{equation}
S^{-1}HS = \begin{pmatrix}
0 & a'(z)\\
b'(z) & 0
\end{pmatrix}
\end{equation}
has off-diagonal elements $a'(z) = (0.65-0.13i)z + (1.51+0.07i) - \frac{1.97-0.013i}{z}$ and $b'(z) = (10.35+2.13i)z - (3.34+1.07i) - \frac{2.94+0.21i}{z}$.

\begin{figure}[htbp]
\centering
\includegraphics[width=0.9\linewidth]{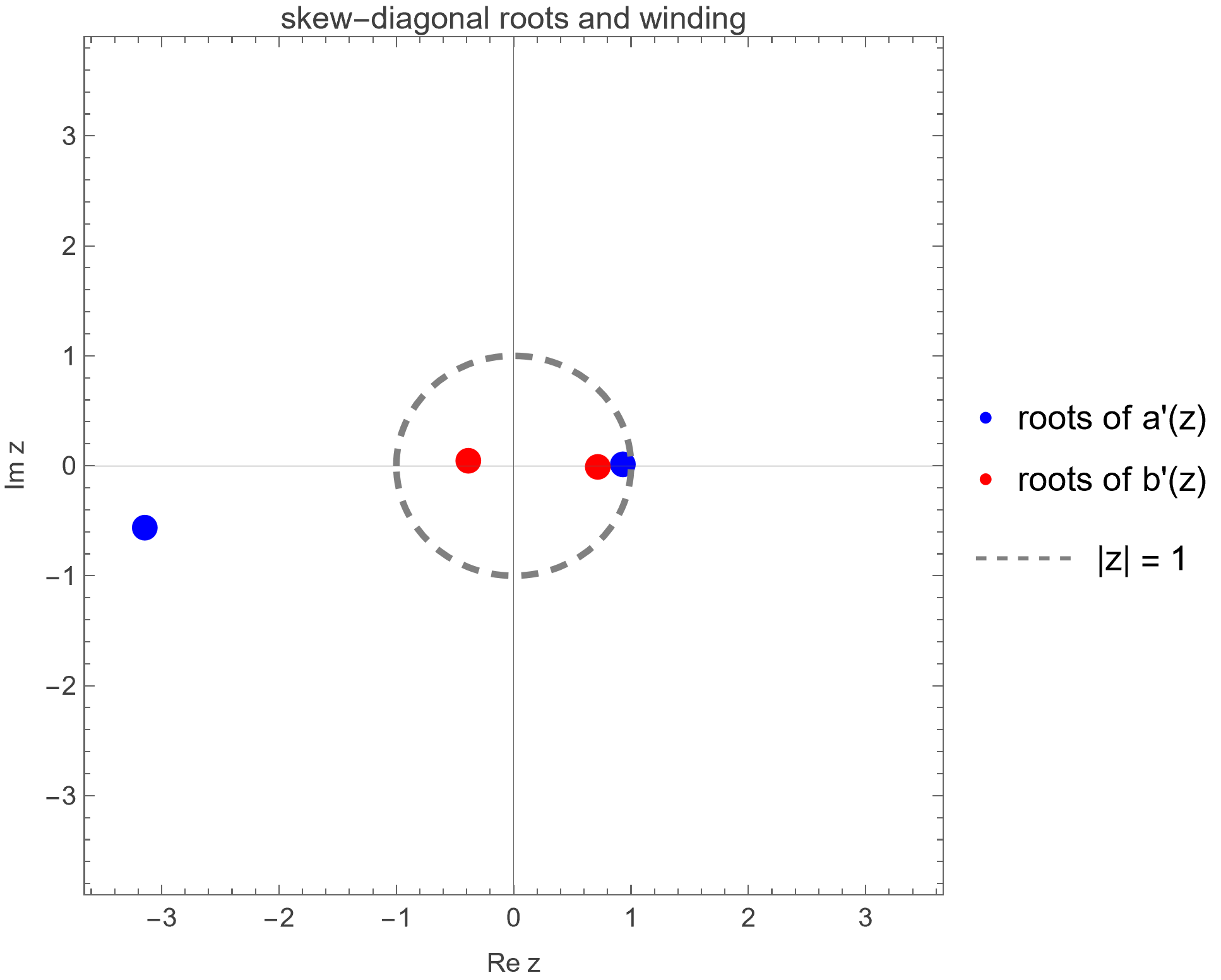}
\caption{Roots of the skew-diagonal components $a'(z)$ (blue) and $b'(z)$ (red) after the constant transformation, with the unit circle (dashed) as the leading-power reference. The generalized Brillouin zone encloses the two largest-modulus roots, both from $a'(z)$, giving winding number $w=-1$ and predicting the single zero-energy mode of Fig.~\ref{fig:constant_zero}.}
\label{fig:winding}
\end{figure}

The roots of $a'(z)$ and $b'(z)$ are plotted in Fig.~\ref{fig:winding}: those of $a'(z)$ have moduli $\{3.19, 0.93\}$, while those of $b'(z)$ have moduli $\{0.72, 0.39\}$. With $r_a = r_b = 1$ (leading powers), the generalized Brillouin zone encloses the two largest-modulus roots. These are both from $a'(z)$, giving winding number $w = r_a - 2 = -1$. This predicts one topological zero-energy edge state.

\begin{figure}[htbp]
\centering
\includegraphics[width=0.9\linewidth]{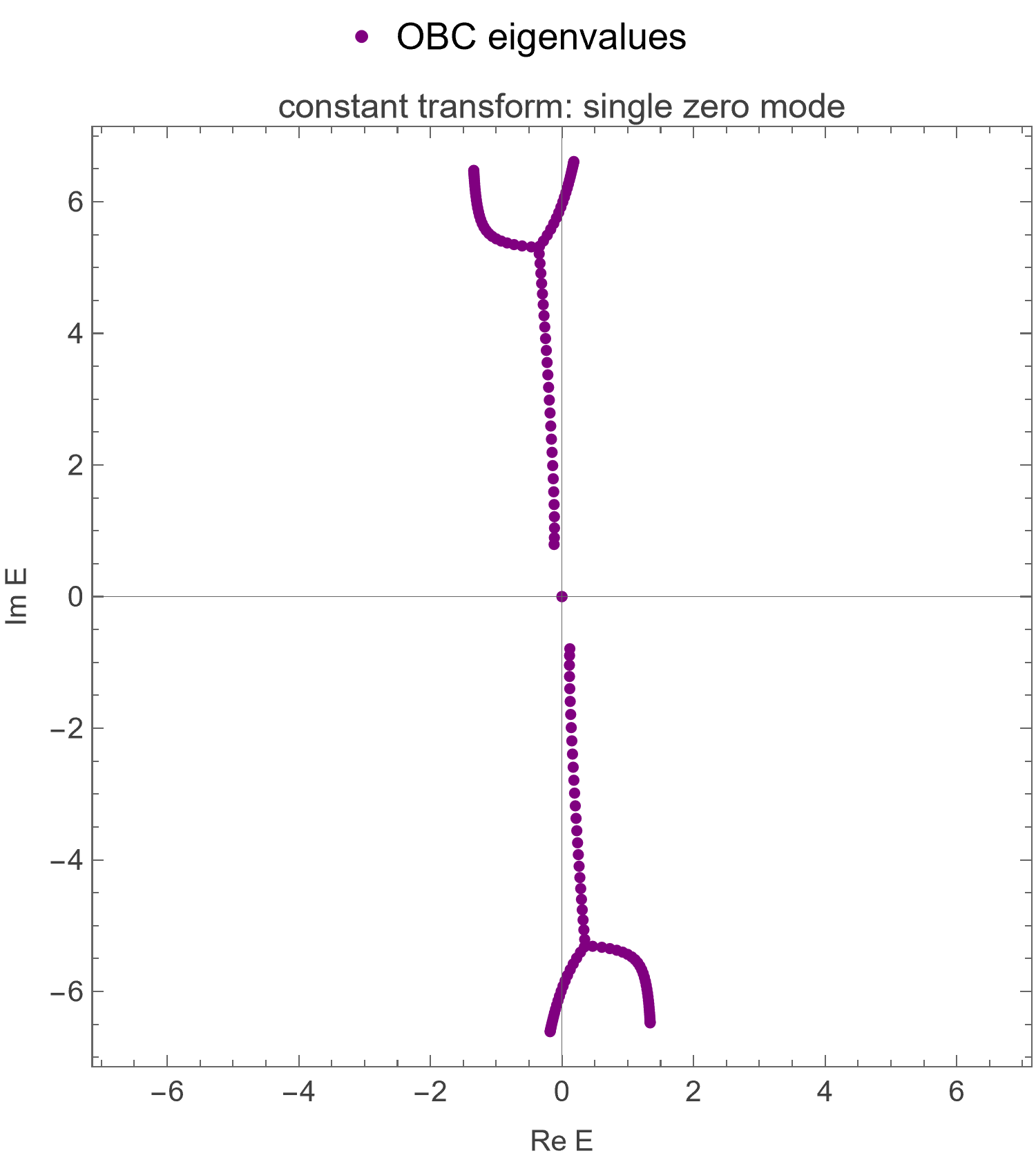}
\caption{Finite OBC spectrum for the Hamiltonian with hidden chiral symmetry revealed by constant transformation. Chain length $N=80$ sites. A single zero-energy state is clearly visible.}
\label{fig:constant_zero}
\end{figure}

Figure~\ref{fig:constant_zero} shows the finite OBC spectrum, confirming a single zero-energy state clearly separated from the bulk continuum. 

\subsection{Momentum-dependent transformation: Non-converging case}
\label{sec:nonconv}

Throughout Secs.~\ref{sec:nonconv}--\ref{sec:conv} we fix the skew-diagonal Hamiltonian
\begin{equation}
H(z)=\begin{pmatrix} 0 & z-\frac14+\frac{3}{2z}\\[2pt] z+\frac25+\frac{1}{10z} & 0\end{pmatrix},
\label{eq:Hexample}
\end{equation}
whose largest GBZ radius is $r_{\mathrm{GBZ}}^{\max}=0.841$, and vary only $S(z)$. This is the same $H$ used in Fig.~\ref{fig:radial} and as System~1 of Sec.~\ref{sec:twosystems}, so a single Hamiltonian runs through all the numerics.

We first examine a transformation violating Theorem 4's radial condition,
\begin{equation}
S(z)=zI-W,\qquad W=\begin{pmatrix} \frac{7}{20} & -\frac12\\[2pt] \frac{51}{200} & \frac{7}{20}\end{pmatrix},
\label{eq:Sbad}
\end{equation}
for which $\det S(z)=z^2-\frac{7}{10}z+\frac14$ has the conjugate pair of zeros $\frac{7}{20}\pm\frac{\sqrt{51}}{20}i$, both of modulus exactly $\frac12$. Thus $z_{\min}=0.5<r_{\mathrm{GBZ}}^{\max}=0.841$: the zeros lie \emph{inside} the GBZ and Condition~1 (the radial condition) is violated.

\begin{figure}[htbp]
\centering
\includegraphics[width=0.9\linewidth]{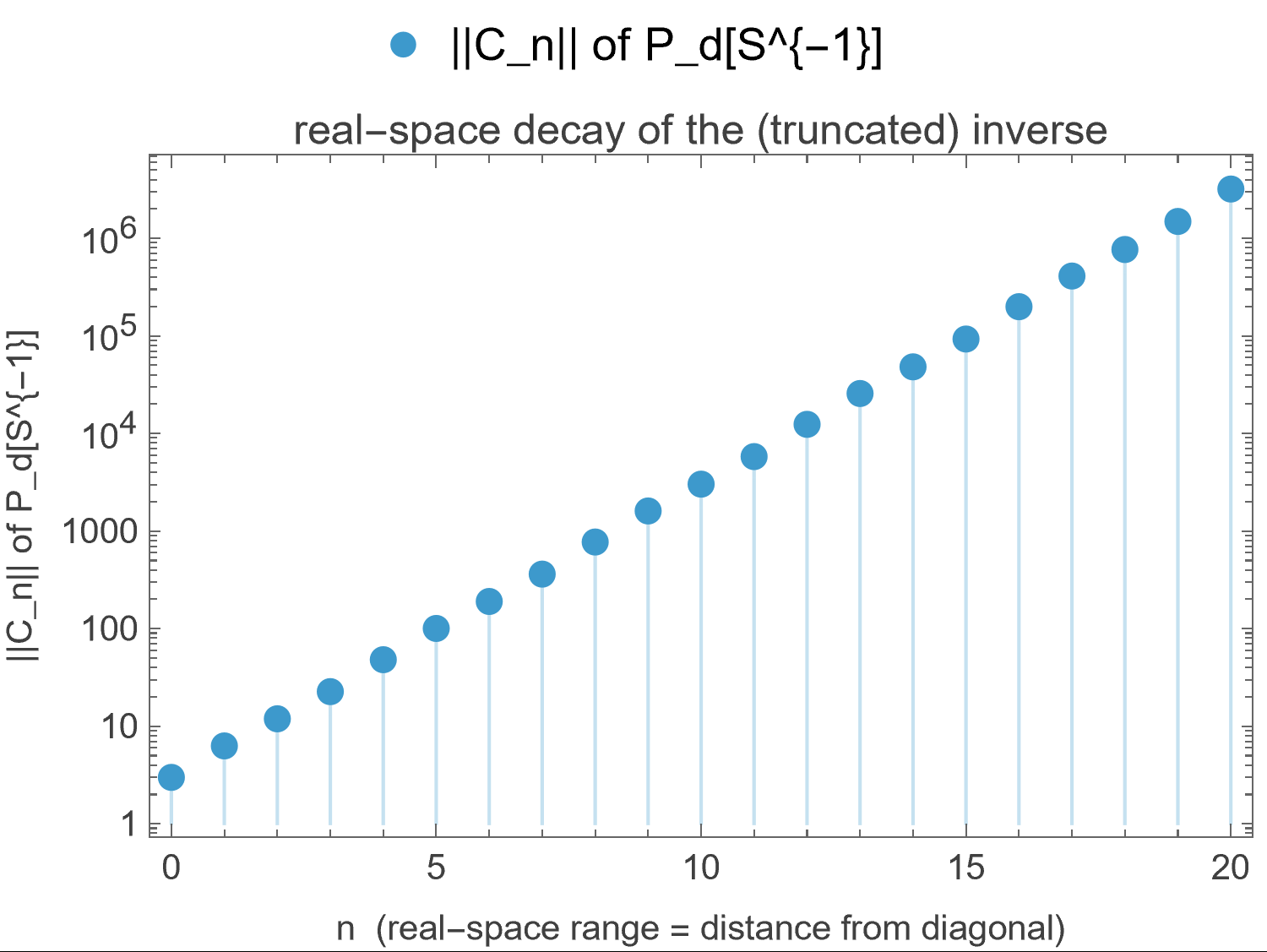}
\caption{Real-space Taylor coefficients $\|C_n\|$ of $P_d[S^{-1}]$. Linear horizontal axis, logarithmic vertical axis. The lack of decay indicates the transformation does not satisfy Theorem 4's conditions.}
\label{fig:nonconv_inv}
\end{figure}

Figure~\ref{fig:nonconv_inv} shows the modulus of real-space entries of $S^{-1}$. Matrix elements fail to decay, indicating non-convergence. The product $S^{-1} \cdot S$ exhibits large errors extending throughout, not just at boundaries.

\begin{figure}[htbp]
\centering
\includegraphics[width=0.9\linewidth]{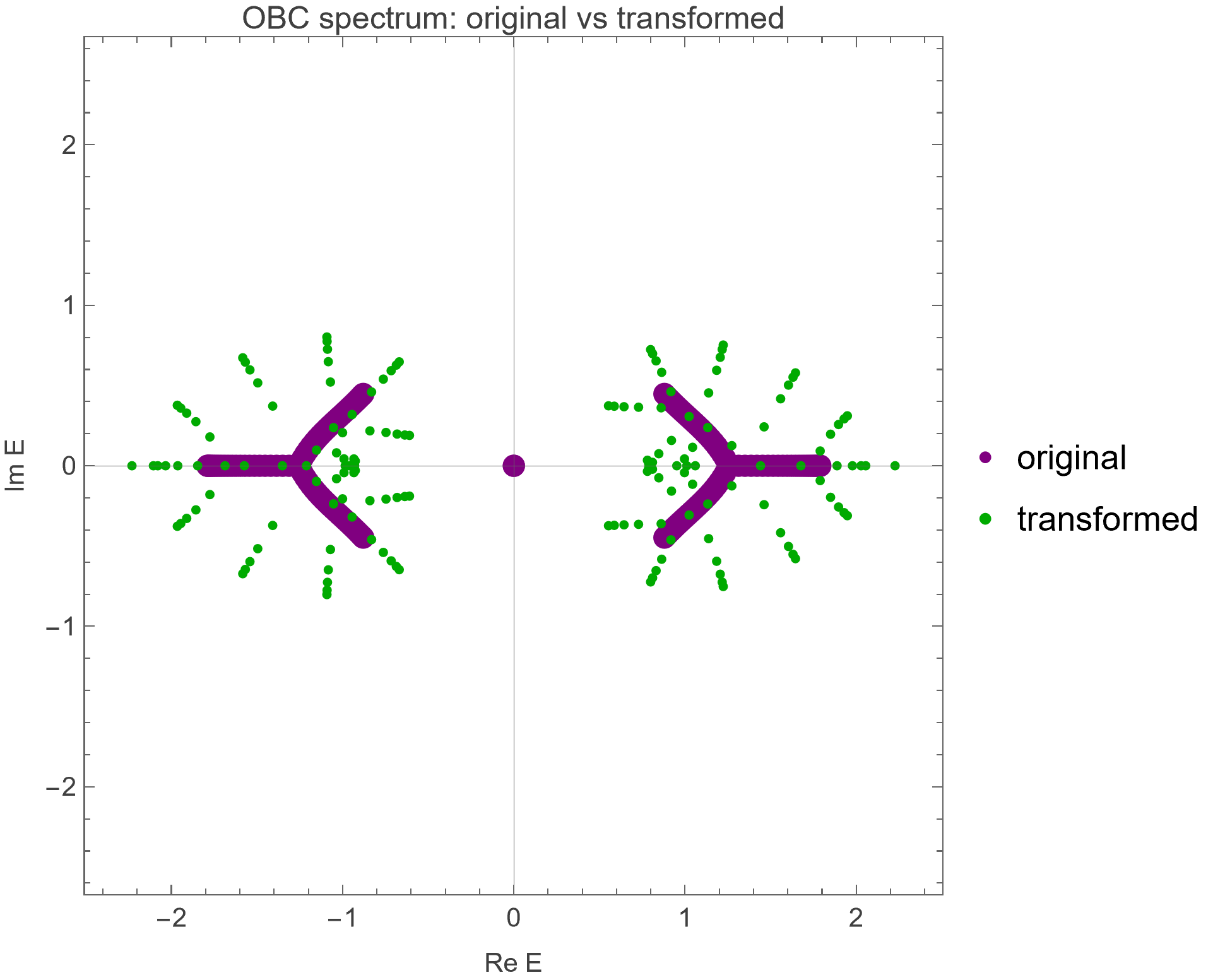}
\caption{OBC eigenvalues: original Hamiltonian (purple) versus transformed (green). The transformation, whose $\det S(z)$ has a zero inside the GBZ ($z_{\min}<r_{\mathrm{GBZ}}^{\max}$), fails to preserve the spectrum.}
\label{fig:nonconv_spec}
\end{figure}

Figure~\ref{fig:nonconv_spec} compares OBC eigenvalues of the original and transformed Hamiltonians. Significant discrepancies appear throughout the spectrum, confirming that transformation fails to preserve spectral properties when convergence conditions are violated.

\subsection{Momentum-dependent transformation: Converging case}
\label{sec:conv}

Consider now
\begin{equation}
S(z) = \begin{pmatrix}
z+2-\frac{6}{z} & 1\\[2pt]
1 & 1
\end{pmatrix},
\label{eq:Sgood}
\end{equation}
for which $\det S(z)=(z+3)(z-2)/z$ has zeros of modulus $\{2,3\}$, so $z_{\min}=2>r_{\mathrm{GBZ}}^{\max}=0.841$ and Theorem 4's radial condition is comfortably satisfied. Note that $z_{\min}$ need only exceed the GBZ radius, not unity.

\begin{figure}[htbp]
\centering
\includegraphics[width=0.9\linewidth]{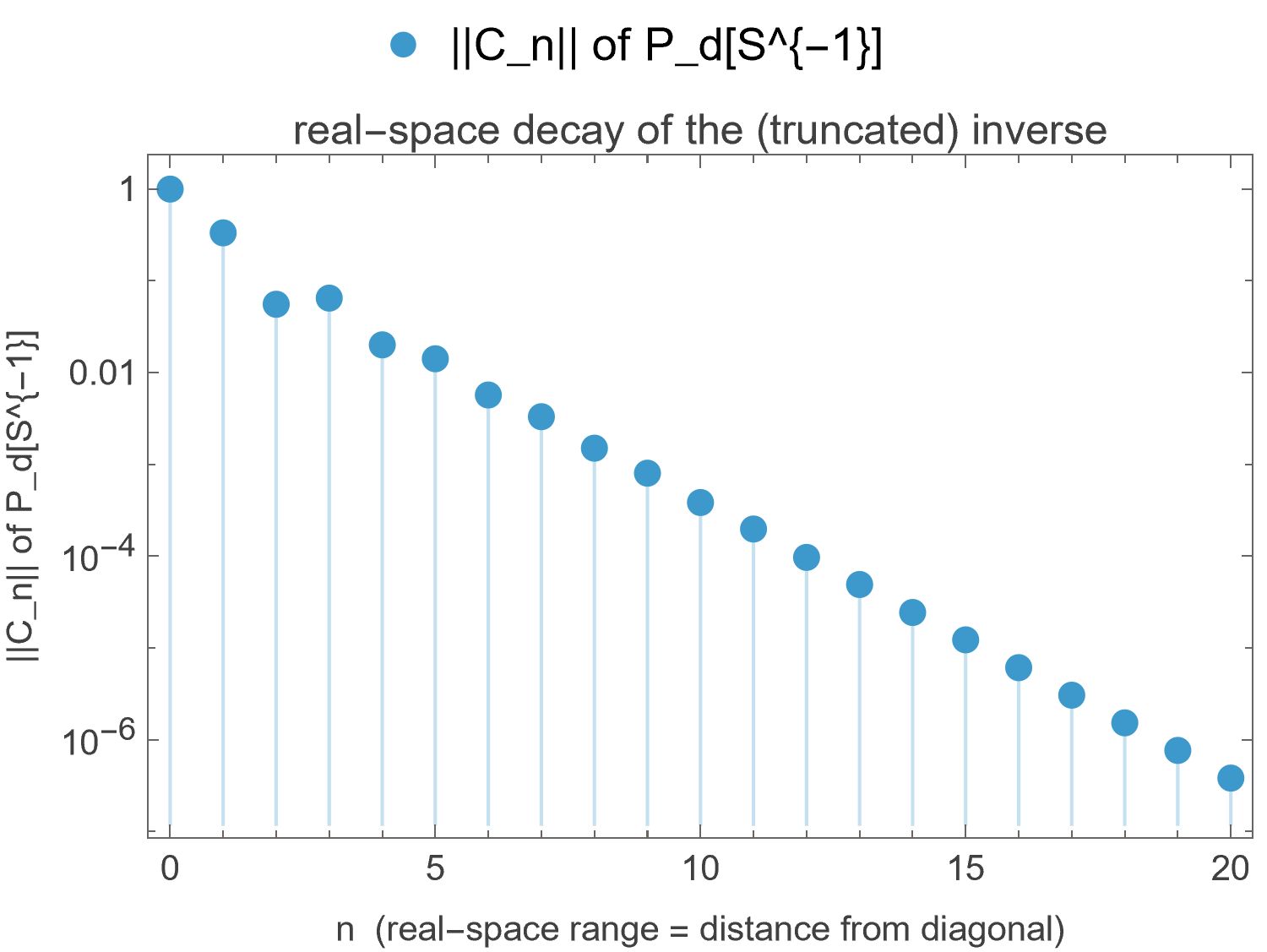}
\caption{Real-space Taylor coefficients $\|C_n\|$ of $P_d[S^{-1}]$. Linear horizontal axis, logarithmic vertical axis. The straight-line decay is exponential, ensuring good approximation in the bulk when Theorem 4's conditions hold.}
\label{fig:conv_inv}
\end{figure}

Figure~\ref{fig:conv_inv} shows rapid decay of $S^{-1}$ coefficients in real space, allowing good approximation with finite truncation. The product $S^{-1} \cdot S$ approximates identity well in the bulk, with errors confined to boundary sites.

\begin{figure}[htbp]
\centering
\includegraphics[width=0.9\linewidth]{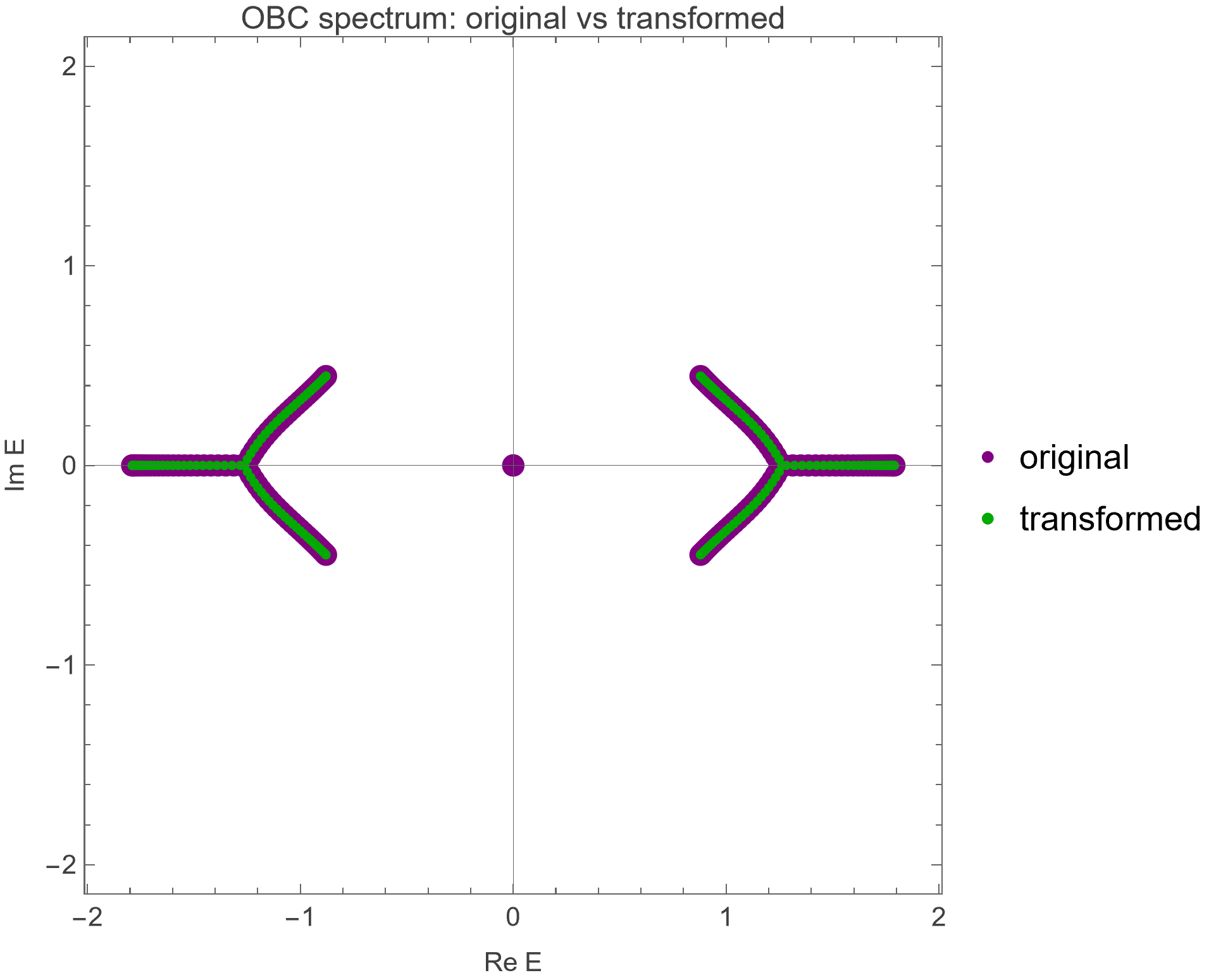}
\caption{OBC eigenvalues for a converging transformation. The bulk spectrum is preserved, but individual eigenvalues differ: most visibly, the near-zero state of the original (purple, at the origin) has no counterpart in the transformed spectrum, having been pushed to the band edge once the transform broke the chiral grading that protected it.}
\label{fig:conv_spec}
\end{figure}

Figure~\ref{fig:conv_spec} compares OBC eigenvalues. The bulk spectrum is preserved after transformation: at $N=80$ and $d_{S^{-1}}=12$ the median paired eigenvalue discrepancy is $4.8\times10^{-3}$, and only $2$ of the $160$ eigenvalues move by more than $0.1$. Those two are boundary states, reflecting the modification predicted by Theorem~5.

The most visible instance is the deletion of a near-zero state. Here $H(z)$ is skew-diagonal and therefore chiral, so its OBC eigenvalues come in $\pm$ pairs and states near $E=0$ are governed by the off-diagonal blocks alone. In real space the block built from $b(z)=z+\tfrac25+\tfrac{1}{10z}$ is tridiagonal Toeplitz, and its determinant obeys $D_N=\tfrac25D_{N-1}-\tfrac{1}{10}D_{N-2}$, whose characteristic roots have modulus $\sqrt{1/10}$; hence $\det\propto10^{-N/2}\to0$ and the chain carries a chirally protected edge mode, consistent with the winding $+1$ of $b(z)$ against $-1$ of $a(z)$. The state is exponentially close to zero rather than exactly zero, so no index theorem is at stake. Because $S(z)$ has nonzero off-diagonal entries, it mixes the two sublattices and the transformed symbol is no longer skew-diagonal: the chiral grading that protected the mode is gone, and the state is pushed out to the band edge at $|E|\approx0.99$. We verified that this is a property of the transform rather than of the truncation---the relocated eigenvalue is unchanged for $d_{S^{-1}}=8$ through $28$ and across $N=20$ to $100$---and confirmed it in exact rational arithmetic, where $|\det H_{\mathrm{OBC}}|$ falls from $2.7\times10^{-4}$ to $5.5\times10^{-8}$ over $N=6$ to $14$ for the original while \emph{growing} from $10.6$ to $271$ for the transformed operator. This is the boundary channel of Theorem~5 in its sharpest form, and it is the counterpart of Sec.~\ref{sec:constant}: a constant transform can \emph{reveal} a hidden chiral symmetry and its zero mode, while a momentum-dependent one can remove the protection of an existing one. The bulk, as the theorem guarantees, is untouched.

\section{Quantitative Accuracy of Spectral Predictions}

While the previous sections establish the topological protection of zero modes under similarity transformations, practical applications require understanding the accuracy with which transformed Hamiltonians predict the full eigenvalue spectrum. Here we develop a quantitative framework for this accuracy.

\subsection{The Spectral Prediction Problem}

Given a Hamiltonian $H(z)$ and its transformed version $\tilde{H}(z) = S^{-1}(z)H(z)S(z)$, the eigenvalue spectra under periodic boundary conditions (PBC) are identical. However, under open boundary conditions (OBC), the spectra differ due to boundary effects. We quantify this difference through the weighted spectral distance
\begin{equation}
\Delta E = \sqrt{\sum_{n}\frac{1}{n}|E_n^{\text{OBC}}[H] - E_n^{\text{OBC}}[\tilde{H}]|^{2}}
\label{eq:weightedspec}
\end{equation}
where the ordinal number $n$ is decided with optimal pairing given by the Hungarian algorithm\cite{Burkard2012}. In this section we discuss general $S(z)$ transforms, not only transforms related to skew-diagonal matrices. To facilitate calculations for discrete eigenvalues, this adopted distance is different from the Hausdorff distance stated in Theorem 3.

\subsection{Separate scaling in $N$ and $d_{S^{-1}}$}

We study the weighted spectral distance $\Delta E$ of Eq.~(\ref{eq:weightedspec}) as a
function of two control parameters: the chain length $N$, and the degree $d_{S^{-1}}$ to which
the Laurent series of $S^{-1}(z)$ is truncated. In the two limits that freeze one parameter and
vary the other, the error obeys the one-variable laws
\begin{align}
&\Delta E(N,d_{S^{-1}})-\Delta E_{\mathrm{floor}}(N)\nonumber\\
&\qquad=B\,e^{-\alpha_d\,d_{S^{-1}}}\bigl[1+b\cos(2\theta d_{S^{-1}}+\delta)\bigr],
\label{eq:approach_main}\\[2pt]
&\Delta E_{\mathrm{floor}}(N)=A\,N^{-\alpha_N}.
\label{eq:floor_main}
\end{align}
Here $\Delta E_{\mathrm{floor}}(N)=\lim_{d_{S^{-1}}\to\infty}\Delta E(N,d_{S^{-1}})$ is the
residual error at fully converged truncation; $A,B>0$ are amplitudes; $\alpha_N\in[\tfrac12,1]$
and $\alpha_d>0$ are the floor and approach exponents; and in the bracketed modulation $\theta$,
$b\in[0,1]$ and $\delta$ are respectively the argument of the nearest zero of $\det S$, a
depth-independent amplitude, and a phase, all fixed by the transform and derived in
Appendix~\ref{app:osc}. Equation~(\ref{eq:approach_main}) is measured at fixed large $N$ (the
\emph{approach}) and Eq.~(\ref{eq:floor_main}) at converged $d_{S^{-1}}$ (the \emph{floor}):
the approach is the symbol-level error of replacing $S^{-1}(z)$ by its degree-$d_{S^{-1}}$
Laurent polynomial, the floor the irreducible boundary-state modification of Theorem~5.

\begin{figure*}[htbp]
\centering
\includegraphics[width=0.9\linewidth]{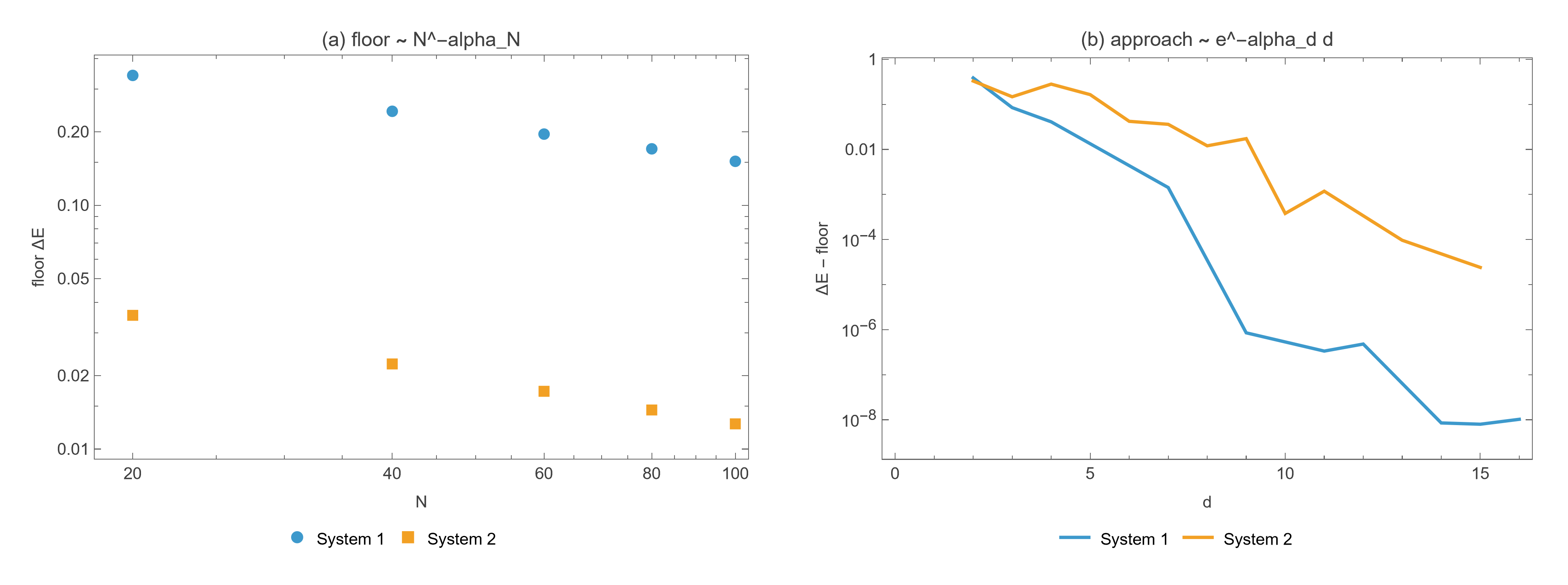}
\caption{The two independent scaling channels for the representative systems. (a) Floor $\Delta E_{\mathrm{floor}}(N)$ at converged $d_{S^{-1}}$ versus $N$, both axes logarithmic, giving a clean power law $N^{-\alpha_N}$. (b) Excess $\Delta E-\Delta E_{\mathrm{floor}}$ versus $d_{S^{-1}}$ at fixed $N$, linear horizontal axis and logarithmic vertical axis, giving an exponential $e^{-\alpha_d d_{S^{-1}}}$. Measured exponents: System~1 $\alpha_N\approx0.5,\ \alpha_d\approx1.3$; System~2 $\alpha_N\approx0.64,\ \alpha_d\approx0.76$.}
\label{fig:sepscale}
\end{figure*}

The two \emph{exponents} are independent: the approach rate $\alpha_d$ does not depend on $N$
(Appendix~G, Fig.~\ref{fig:sepscale}b), and the floor exponent $\alpha_N$ does not depend on
$d_{S^{-1}}$ once the truncation has converged. This is a statement about the two limiting
regimes, not about the full surface: at small $d_{S^{-1}}$ the approach has not yet reached the
floor, so there the two contributions are not simply separable, and the additive form below
should be read as the behaviour once each channel is in its own regime. The floor exponent is expected to lie in $\tfrac12\leq\alpha_N\leq1$, between the Ajtai--Koml\'os--Tusn\'ady optimal-matching value for a random density~\cite{AKT1984} and exact density matching (Appendix~G notes finite-size exceptions); the approach rate is controlled by the radial GBZ margin of Theorem~4---the symbol-level defect decays exactly as $\log(z_{\min}/r_{\mathrm{GBZ}}^{\max})$ (Appendix~G, Fig.~\ref{fig:corner}), and the RMS eigenvalue rate is of the same order but system-dependent (Appendix~G).

The bracketed factor in Eq.~(\ref{eq:approach_main}) is a genuine feature of the transform,
not fit scatter. Its frequency is not a fitted parameter: it is fixed at $2\theta$ with
$\theta=\arg z_\ast$, twice the argument of the zero $z_\ast$ of $\det S$ nearest the origin,
and its envelope decays at the rate $\log(z_{\min}/r_{\mathrm{GBZ}})$ set by the modulus of the
same zero. A real nearest zero gives no oscillation at all. The decay rate and the oscillation
frequency are therefore both read off a single zero of $\det S$, with nothing fitted, which
makes the ripple a parameter-free consistency check; the derivation and the perturbation-theory
account of the eigenvalue channel are given in Appendix~\ref{app:osc}.

\begin{figure*}[htbp]
\centering
\begin{subfigure}{0.48\linewidth}
  \centering
  \includegraphics[width=\linewidth]{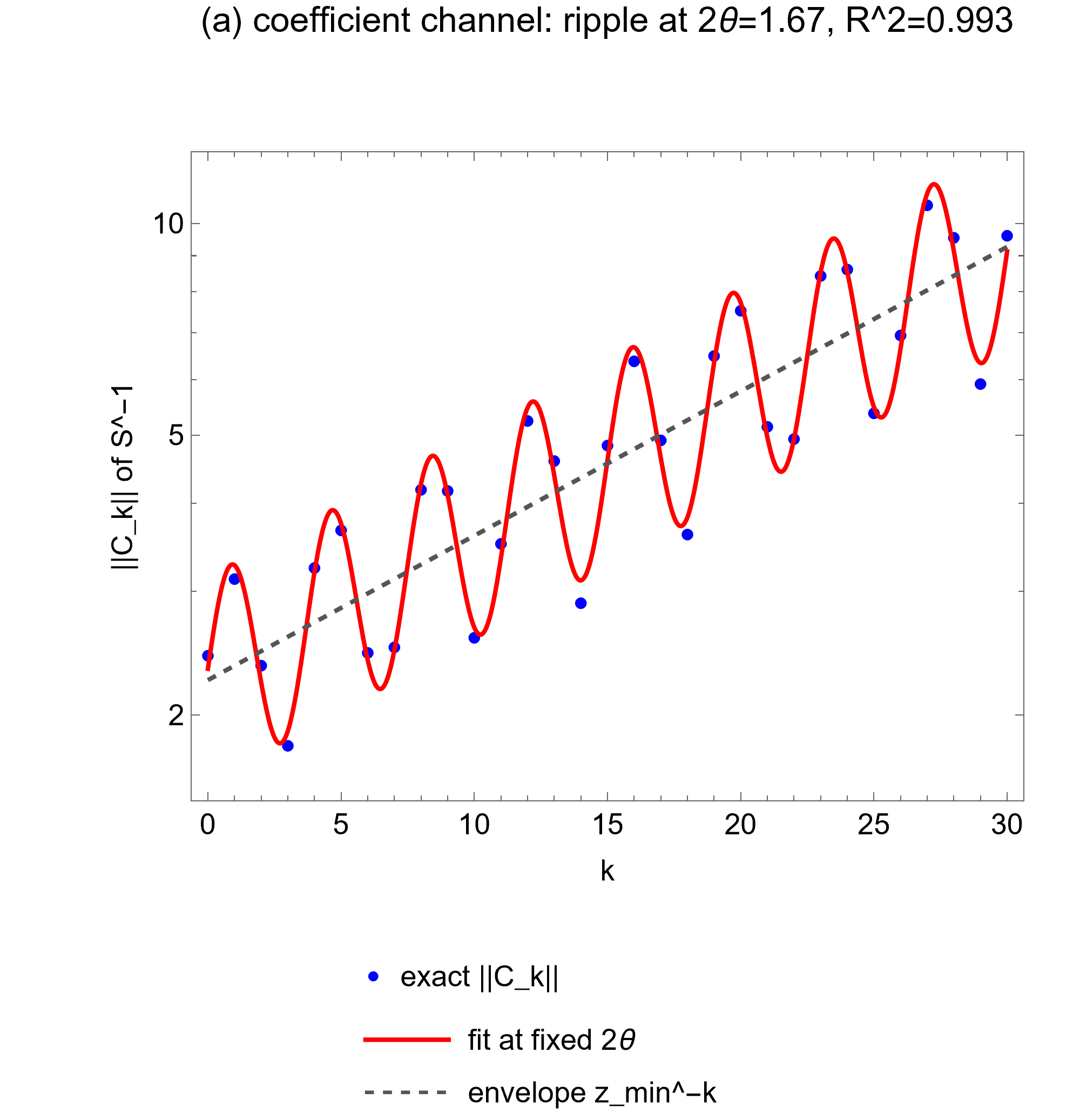}
  \caption{Coefficient channel (exact).}
  \label{fig:osccoeff}
\end{subfigure}\hfill
\begin{subfigure}{0.48\linewidth}
  \centering
  \includegraphics[width=\linewidth]{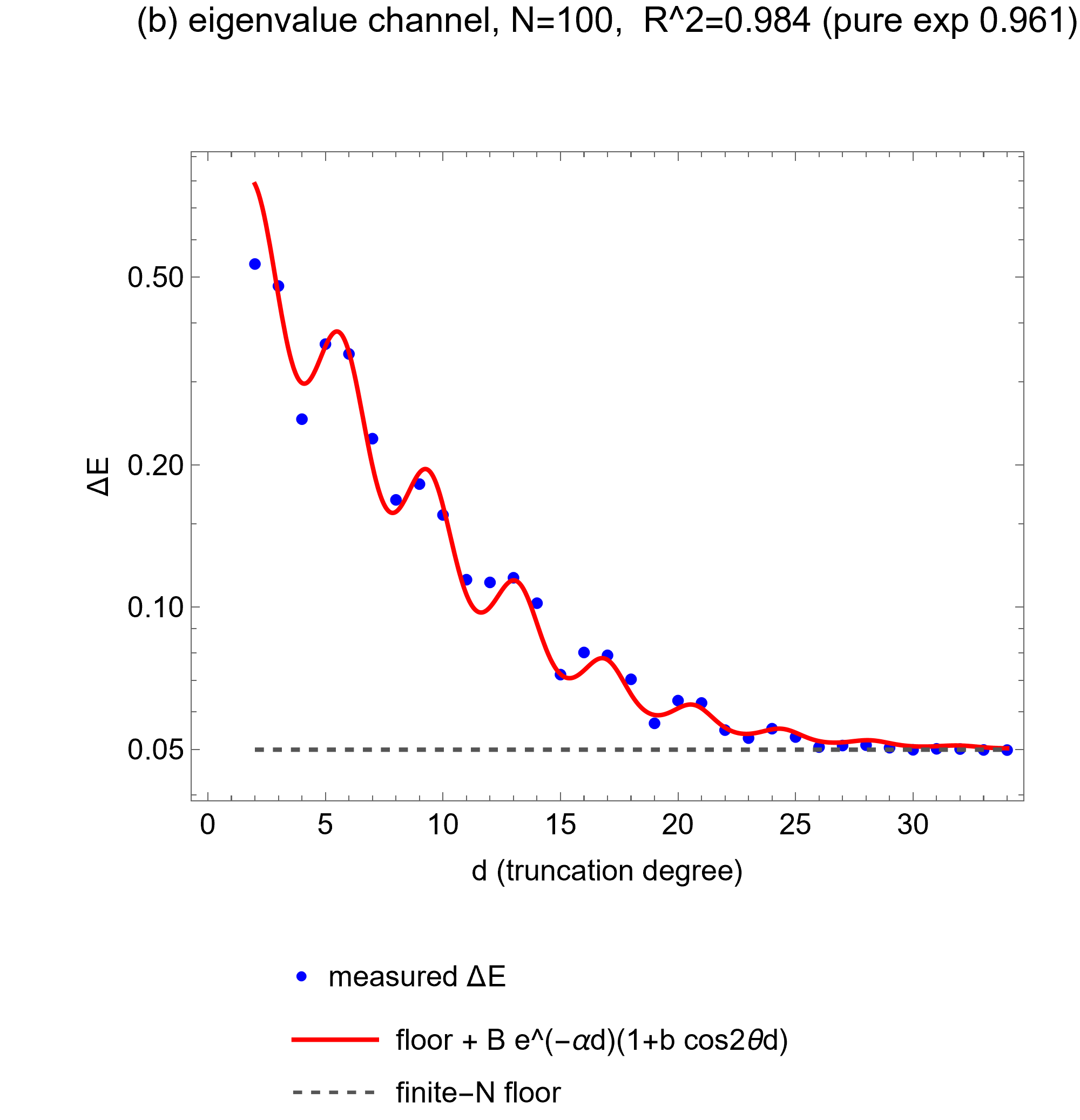}
  \caption{Eigenvalue channel, $N=100$.}
  \label{fig:osceig}
\end{subfigure}
\caption{Oscillatory fine structure for the transform $S_{\mathrm{osc}}$ of Eq.~(\ref{eq:Sosc}) (Appendix~\ref{app:osc}), whose $\det S$ has a single conjugate pair of zeros $z_\ast=0.640\pm0.707i$. Both panels use a linear horizontal axis and a logarithmic vertical axis. The ripple frequency is \emph{fixed} at the predicted $2\theta=1.670$; no frequency or phase is fitted. (a)~Coefficient channel: exact Frobenius norms $\|C_k\|_F$ of the Taylor coefficients of $S^{-1}$ (points), the global least-squares fit at $2\theta$ (solid), and the bare envelope $z_{\min}^{-k}$ (dashed); the envelope rises slowly because $z_{\min}=0.954<1$, which is immaterial---validity requires $z_{\min}>r_{\mathrm{GBZ}}^{\max}=0.841$, not $z_{\min}>1$. Fitted rate $-0.0473$ versus predicted $\log z_{\min}=-0.0474$, $R^2=0.992$. (b)~Eigenvalue channel at $N=100$: raw $\Delta E$ (points) against $\Delta E_{\mathrm{floor}}+Be^{-\alpha_dd_{S^{-1}}}[1+b\cos(2\theta d_{S^{-1}}+\delta)]$ (solid), with the finite-$N$ floor marked (dashed); plotting the raw error avoids subtracting the floor. The small margin $z_{\min}/r_{\mathrm{GBZ}}^{\max}=1.13$ keeps the excess above the floor for $24$ points, and the periodogram of the detrended residual peaks at $1.685$, within $0.9\%$ of $2\theta$. In both panels the same fit at the half frequency $\theta$ returns a null ripple. System~1, whose nearest zero is real, shows no ripple in either channel.}
\label{fig:oscfit}
\end{figure*}

To exhibit the ripple cleanly we use the transform
\begin{equation}
S_{\mathrm{osc}}(z)=zI-W,\qquad W=\begin{pmatrix}16/25 & -2\\[2pt] 1/4 & 16/25\end{pmatrix},
\label{eq:Sosc}
\end{equation}
whose $\det S_{\mathrm{osc}}=z^2-\tfrac{32}{25}z+\tfrac{1137}{1250}$ has the single conjugate
pair of zeros $z_\ast=\tfrac{16}{25}\pm\tfrac{\sqrt2}{2}i$, giving $z_{\min}=0.9537$ and
$\theta=0.8352$ ($47.85^\circ$), so the predicted frequency is $2\theta=1.6703$. Fixing the fit
at that value, the coefficient channel matches with $R^2=0.992$ and the eigenvalue channel at
$N=100$ with $R^2=0.984$, while the same fit at the half frequency $\theta$ returns a null in
both; independently, the periodogram of the detrended eigenvalue residual peaks at $1.685$,
within $0.9\%$ of $2\theta$ (Fig.~\ref{fig:oscfit}). System~1, whose nearest zero $z=2$ is real,
shows no ripple in either channel.

Once each channel is in its regime the total error is well described by the sum of the two,
$\Delta E\approx A\,N^{-\alpha_N}+B\,e^{-\alpha_d d_{S^{-1}}}$. This additive form is not an
independent fitting ansatz but the consequence of measuring each exponent in the limit that
freezes the other; we do not claim it holds uniformly at small $d_{S^{-1}}$, where the two
channels overlap. The full derivation and numerical tests are given in Appendix~G.

\subsection{Example: Two Representative Systems}
\label{sec:twosystems}

We illustrate with two systems having identical $H(z)$ but different $S(z)$:

\textbf{System 1 (Simple transformation):}
\begin{align}
H(z) &= \begin{pmatrix}
0 & z-\frac{1}{4}+\frac{3}{2z}\\
z+\frac{2}{5}+\frac{1}{10z} & 0
\end{pmatrix}\\
S(z) &= \begin{pmatrix}
z+2-\frac{6}{z} & 1\\
1 & 1
\end{pmatrix}
\end{align}

System 1 ($z_{\min}=2$) measured exponents (weighted distance $\Delta E$, Hungarian pairing): floor $\alpha_N\approx0.5$ ($R^2=0.9998$), approach $\alpha_d\approx1.3$.

\begin{figure*}[htbp]
    \centering
    \begin{subfigure}{0.47\linewidth}
        \centering
        \includegraphics[width=\linewidth]{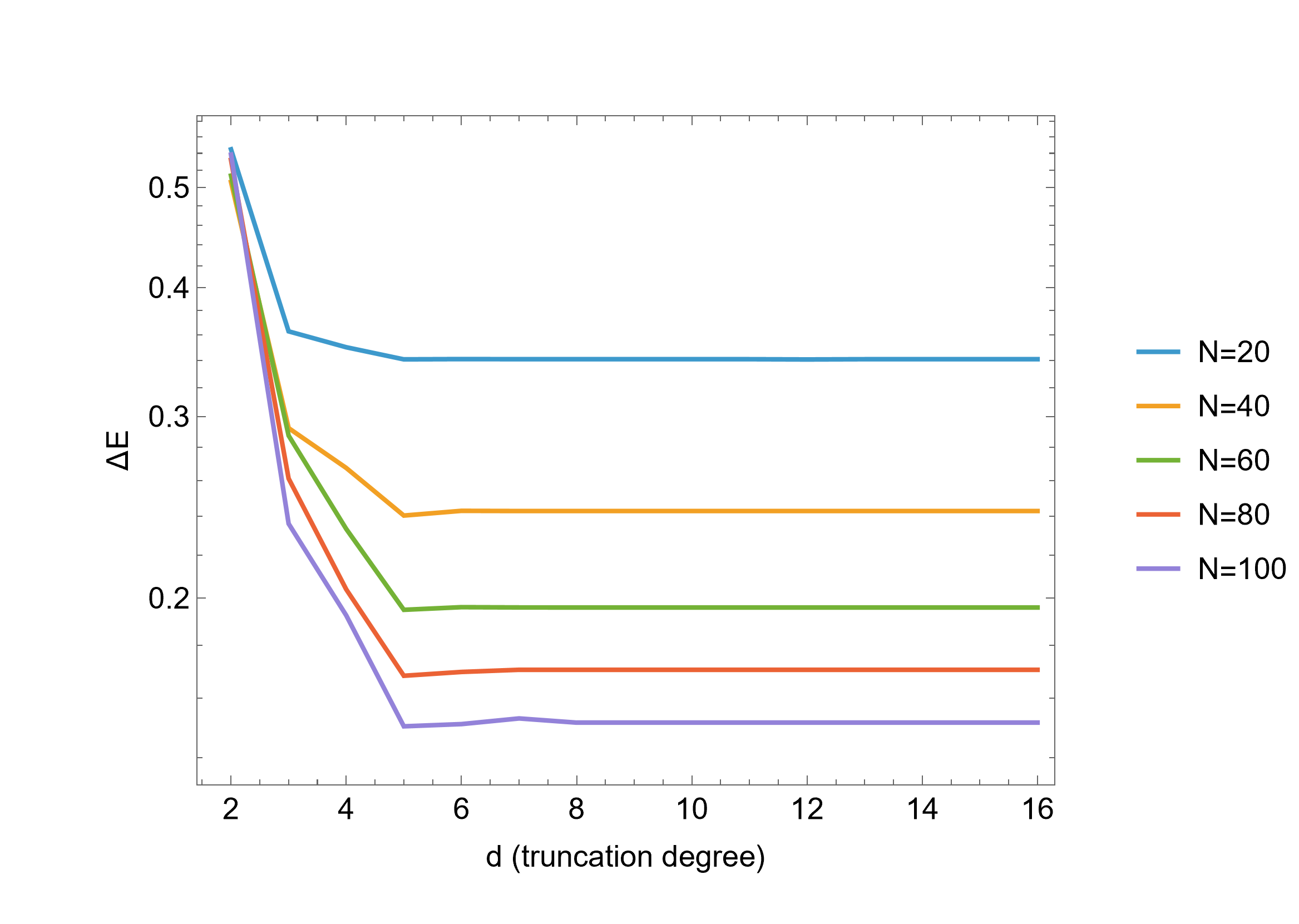}
        \caption{System 1, lin--log.}
        \label{fig:sub1-1}
    \end{subfigure}\hfill
    \begin{subfigure}{0.47\linewidth}
        \centering
        \includegraphics[width=\linewidth]{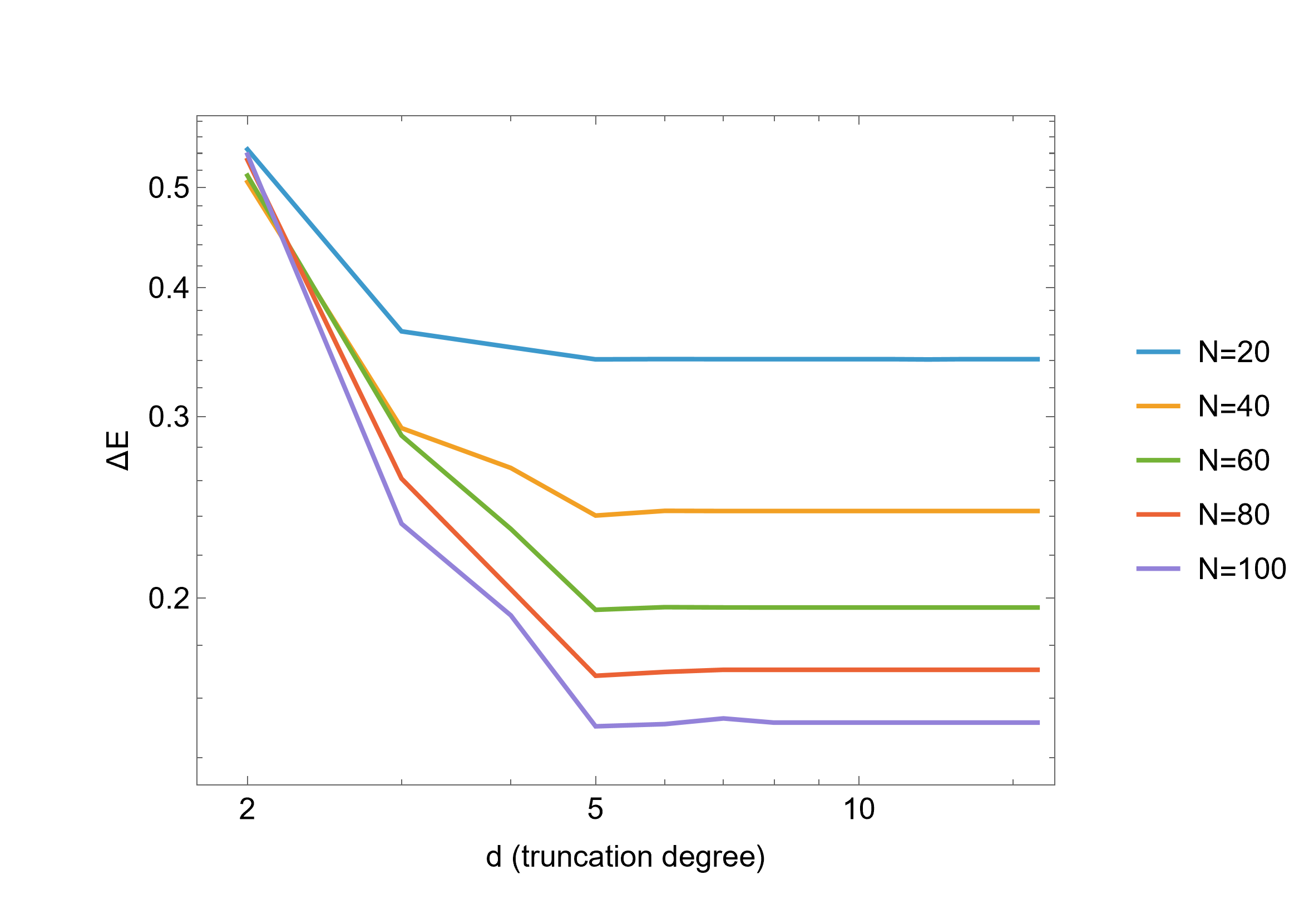}
        \caption{System 1, log--log.}
        \label{fig:sub2-1}
    \end{subfigure}

    \vspace{0.4cm}

    \begin{subfigure}{0.47\linewidth}
        \centering
        \includegraphics[width=\linewidth]{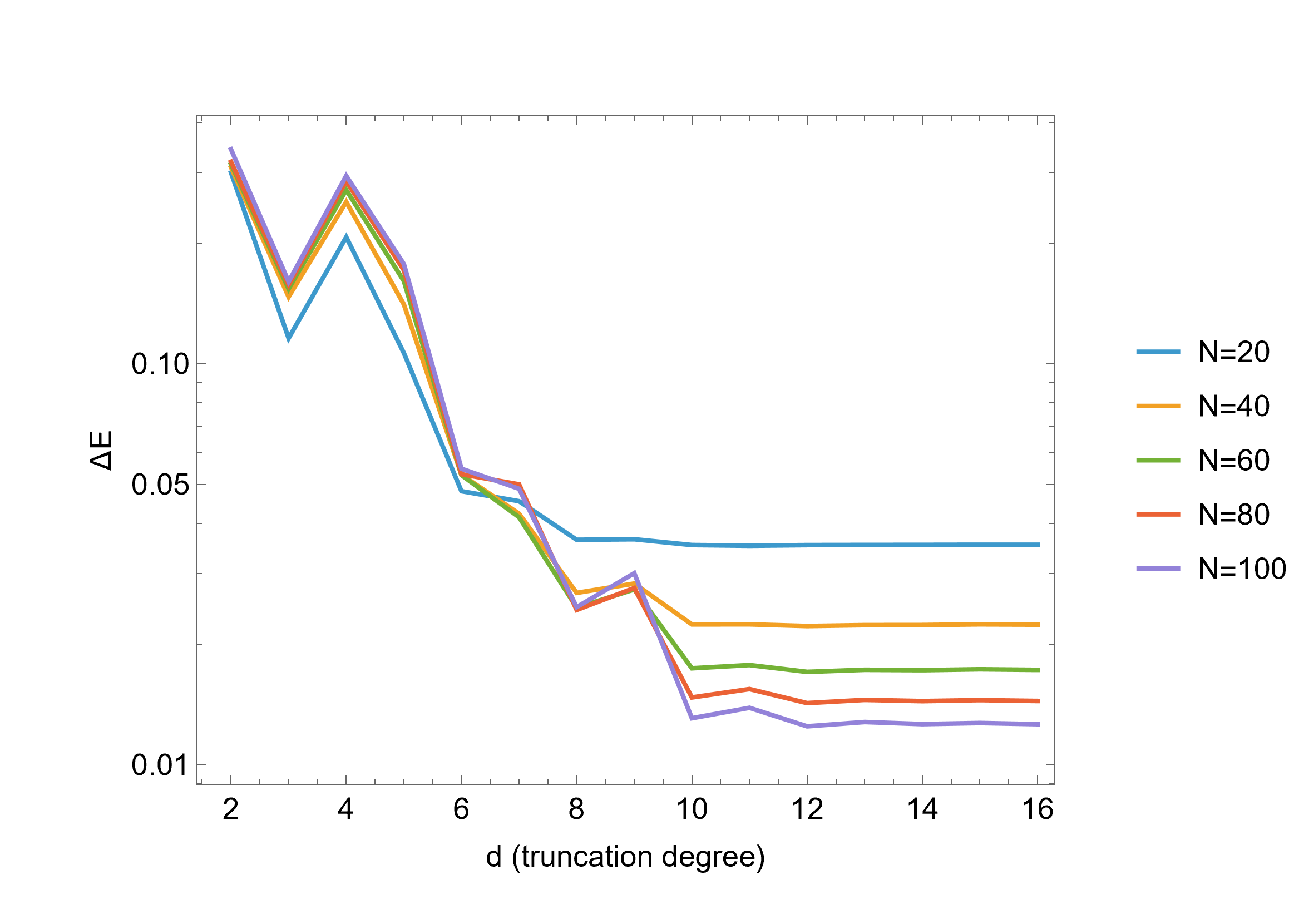}
        \caption{System 2, lin--log.}
        \label{fig:sub2-2}
    \end{subfigure}\hfill
    \begin{subfigure}{0.47\linewidth}
        \centering
        \includegraphics[width=\linewidth]{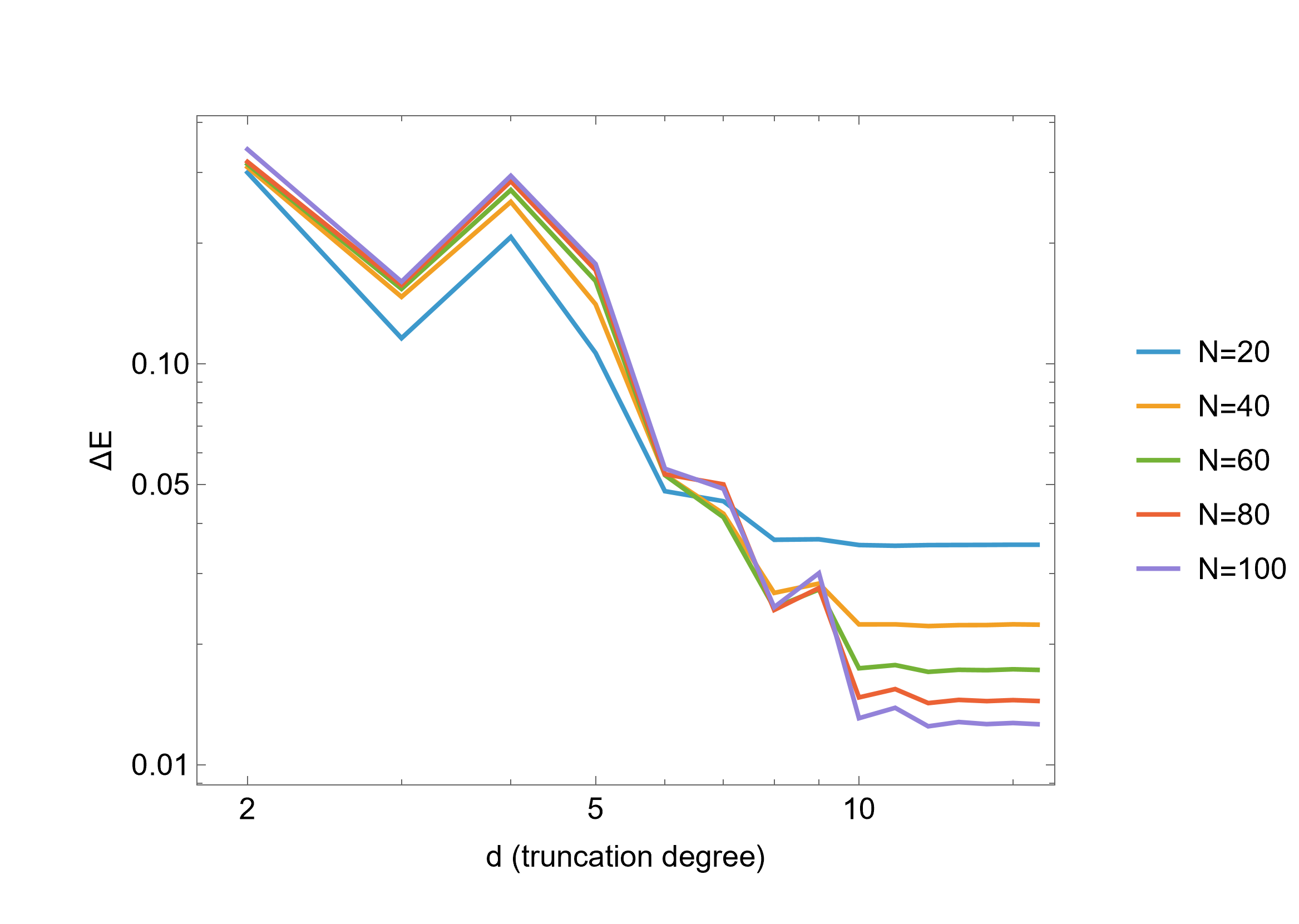}
        \caption{System 2, log--log; the region $d_{S^{-1}}<4$ has been cut for clarity.}
        \label{fig:sub2-3}
    \end{subfigure}

    \caption{Weighted spectral distance $\Delta E$ versus truncation degree $d_{S^{-1}}$, one curve per chain length $N$. (a),(b) System~1. (c),(d) System~2. The vertical axis is logarithmic throughout; the horizontal axis is linear in (a),(c) and logarithmic in (b),(d), so that the same data appear as lin--log and log--log. The simpler transform of System~1 gives the cleaner single-exponential approach.}
    \label{fig:errmain}
\end{figure*}

\textbf{System 2 (Complex transformation):}
Same $H(z)$, but different $S(z)$:
\begin{equation}
S(z) = \begin{pmatrix}
z+2-\frac{6}{z} & 0.1z+1\\
z+0.3/z & z-0.2+1.5/z
\end{pmatrix}
\end{equation}

System 2 ($z_{\min}\approx1.24$) measured exponents: floor $\alpha_N\approx0.64$ ($R^2=0.9995$), approach $\alpha_d\approx0.76$ ($N$-independent).

The calculated errors of the two systems are shown in Fig.~\ref{fig:errmain}.

\section{Discussion}

\subsection{Implications for topological analysis}

Our results establish precise criteria for reliable topological analysis using similarity transformations. When the linear dependence criterion Eq.~(\ref{eq:constraint}) is satisfied, constant transformations (Table \ref{tab:transform}) reveal hidden chiral symmetry without complications. Topological invariants computed in the transformed basis faithfully predict boundary phenomena, as demonstrated numerically.

For systems requiring momentum-dependent transformations, bulk topological invariants remain well-defined when Theorem 4's conditions hold ($r_{\mathrm{GBZ}}^{\max}<z_{\min}$ and spectral stability). However, boundary state counting may be unreliable since edge states probe boundary physics, which is modified by long-range transformation structure whenever $S$ is two-sided (Theorem 5). Bulk winding numbers computed after transformation therefore need not predict the correct number of edge states, even where the bulk-boundary correspondence of the untransformed model is well established~\cite{ZhangYangFang2020}.

The explicit formulas in Table \ref{tab:transform} demonstrate that momentum-dependent transformations arise naturally. Whenever transformation matrix elements contain nontrivial $z$-dependence—whether through direct appearance of the original Hamiltonian's momentum structure or through mathematical operations like square roots and ratios—the inverse $S^{-1}(z)$ becomes a rational function with potential poles. Our theorems provide the precise criterion: the poles of $S^{-1}(z)$ (the zeros of $\det S$) must lie strictly outside the GBZ of $H$ for bulk preservation.

\subsection{Experimental platforms}

These results have immediate relevance for several experimental platforms. In non-Hermitian photonic crystals, reciprocity breaking can obscure natural chiral bases \cite{ref8}. Our framework identifies when constant transformations can reveal hidden symmetry, enabling reliable prediction of topological edge modes in waveguide arrays with controlled gain and loss.

For dissipative quantum walks, chiral symmetry often emerges only after specific unitary transformations \cite{ref4}. Our explicit formulas provide a systematic procedure for constructing appropriate transformations and verifying that resulting topological predictions remain valid under experimental boundary conditions.

\subsection{Extensions and future directions}

Several natural extensions emerge from this work. First, higher-dimensional systems present additional subtleties since momentum space becomes multidimensional and boundary geometry affects truncation differently. Generalizing our framework to $d > 1$ dimensions while maintaining mathematical rigor remains an open challenge.

Second, certain experimental platforms may admit transformations with carefully controlled analytic properties beyond simple pole restrictions. Investigating whether weaker conditions than Theorem 4 can guarantee partial spectral preservation in specific physical contexts could expand the applicability of our framework.

Finally, extending beyond $2\times 2$ systems to general $N\times N$ Hamiltonians follows straightforwardly for our main theorems (conditions generalize to arbitrary dimension), but the explicit reduction procedures become considerably more involved. Developing systematic computational tools for higher-dimensional cases would benefit applications.

\section{Conclusion}

We have established a comprehensive framework for understanding spectral preservation under similarity transformations in lattice systems with open boundaries. For traceless $2\times 2$ Hamiltonians, reduction to skew-diagonal form via constant transformation is possible if and only if matrix elements satisfy a linear dependence relation. We provided explicit transformation formulas covering all cases, enabling systematic exploitation of hidden chiral symmetry.

For momentum-dependent transformations, bulk spectral preservation in finite systems requires spectral stability of $H$ together with the two-radius condition that the generalized Brillouin zone of $H$ lie inside the smallest zero of $\det S$ ($r_{\mathrm{GBZ}}^{\max}<z_{\min}$). This ensures that the truncated inverse transformation converges on the GBZ, permitting bulk spectral preservation despite boundary truncation. When $S$ is two-sided, boundary states are additionally modified even under optimal transformations, reflecting a fundamental incompatibility between momentum-space methods and real-space boundaries.

Our numerical verification confirms these predictions: constant transformations reliably reveal topological zero modes that exhibit expected robustness, while momentum-dependent transformations preserve bulk spectra only when convergence conditions hold, and always modify some boundary eigenvalues.

These results clarify when bulk topological invariants computed after transformation reliably predict boundary phenomena, with immediate implications for non-Hermitian systems, photonic platforms, and other contexts where symmetries emerge only after appropriate basis changes. The explicit formulas and rigorous conditions provide both theoretical understanding and practical guidance for experimental implementations.

\begin{acknowledgments}
We thank S. Yuan and C. H. Lee for relevant discussion.
\end{acknowledgments}

\newpage
\appendix

\section*{Appendix}

\section{Proof of Theorem 1: Reducibility to skew-diagonal form}

\subsection{Statement}

A traceless $2\times 2$ Hamiltonian $H(z)$ with entries
\begin{equation}
H(z) = \begin{pmatrix}
d(z) & q(z)\\
r(z) & -d(z)
\end{pmatrix}
\end{equation}
can be reduced to skew-diagonal form
\begin{equation}
S^{-1}H(z)S = \begin{pmatrix}
0 & a(z)\\
b(z) & 0
\end{pmatrix}
\end{equation}
by a constant similarity transformation $S$ (with matrix elements independent of $z$) if and only if there exist constants $k_1$, $k_2$, $k_3$ (not all zero) such that
\begin{equation}
k_1 d(z) + k_2 q(z) + k_3 r(z) = 0 \quad \forall z.
\label{eq:constraint_SM}
\end{equation}

\subsection{Proof of necessity}

Suppose there exists a constant matrix
\begin{equation}
S = \begin{pmatrix}
s_{11} & s_{12}\\
s_{21} & s_{22}
\end{pmatrix}, \quad s_{ij} \in \mathbb{C}
\end{equation}
such that $S^{-1}HS$ has vanishing diagonal elements.

Computing the similarity transformation:
\begin{equation}
S^{-1}HS = \frac{1}{\det S}\begin{pmatrix}
s_{22} & -s_{12}\\
-s_{21} & s_{11}
\end{pmatrix}\begin{pmatrix}
d & q\\
r & -d
\end{pmatrix}\begin{pmatrix}
s_{11} & s_{12}\\
s_{21} & s_{22}
\end{pmatrix}.
\end{equation}

The $(1,1)$ element of $S^{-1}HS$ is:
\begin{align}
(&S^{-1}HS)_{11} = \frac{1}{\det S}(s_{22},-s_{12})\begin{pmatrix}
d & q\\
r & -d
\end{pmatrix}\begin{pmatrix}
s_{11}\\
s_{21}
\end{pmatrix}\\
&= \frac{1}{\det S}(s_{22},-s_{12})\begin{pmatrix}
ds_{11} + qs_{21}\\
rs_{11} - ds_{21}
\end{pmatrix}\\
&= \frac{1}{\det S}[s_{22}(ds_{11} + qs_{21}) - s_{12}(rs_{11} - ds_{21})]\\
&= \frac{1}{\det S}[(s_{22}s_{11} + s_{12}s_{21})d + s_{22}s_{21}q - s_{12}s_{11}r].
\end{align}

Setting this equal to zero and multiplying by $\det S$:
\begin{equation}
(s_{22}s_{11} + s_{12}s_{21})d(z) + s_{22}s_{21}q(z) - s_{12}s_{11}r(z) = 0.
\end{equation}

Similarly, the $(2,2)$ element gives:
\begin{equation}
(-s_{21}s_{22} - s_{11}s_{12})d(z) - s_{21}s_{22}q(z) + s_{11}s_{12}r(z) = 0.
\end{equation}

Both equations are equivalent (the second is the negative of the first), giving the constraint:
\begin{equation}
k_1 d(z) + k_2 q(z) + k_3 r(z) = 0
\end{equation}
where
\begin{align}
k_1 &= s_{22}s_{11} + s_{12}s_{21}\\
k_2 &= s_{22}s_{21}\\
k_3 &= -s_{12}s_{11}.
\end{align}

Since $S$ is invertible, $\det S = s_{11}s_{22} - s_{12}s_{21} \neq 0$, so not all matrix elements vanish. Therefore, not all $k_i$ can be zero. This establishes necessity. $\square$

\subsection{Proof of sufficiency}

Conversely, suppose there exist constants $k_1$, $k_2$, $k_3$ (not all zero) satisfying Eq.~(\ref{eq:constraint_SM}). We construct $S$ explicitly.

Without loss of generality, assume $k_3 \neq 0$ (other cases are treated separately). The dependence relation is defined only up to an overall scale, so normalise it to $k_3=-1$ and set $s_{11} = s_{12} = 1$. The constraint equations then become:
\begin{align}
k_1 &= s_{21} + s_{22}\\
k_2 &= s_{21}s_{22}\\
k_3 &= -1.
\end{align}

From the first two equations:
\begin{equation}
s_{21} + s_{22} = -\frac{k_1}{k_3}, \quad s_{21}s_{22} = -\frac{k_2}{k_3}.
\end{equation}

These are the Vieta formulas for roots of the quadratic:
\begin{equation}
t^2 + \frac{k_1}{k_3}t - \frac{k_2}{k_3} = 0.
\end{equation}

The solutions are:
\begin{equation}
s_{21}, s_{22} = \frac{-k_1 \pm \sqrt{k_1^2 + 4k_2k_3}}{2k_3}.
\end{equation}

We can verify that with these values, $S^{-1}HS$ has vanishing diagonal elements. Computing explicitly:
\begin{align}
(S^{-1}HS)_{11} &= \frac{1}{\det S}[(s_{21}+s_{22})d + s_{21}s_{22}q - r]\\
&= \frac{1}{\det S}\left[-\frac{k_1}{k_3}d - \frac{k_2}{k_3}q - r\right]\\
&= -\frac{1}{k_3\det S}[k_1d + k_2q + k_3r]\\
&= 0
\end{align}
by the constraint. This establishes sufficiency \emph{provided} $S$ is invertible, i.e.\ $\det S=s_{22}-s_{21}=\sqrt{k_1^2+4k_2k_3}/k_3\neq0$.

\subsection{The degenerate (isotropic) case}

When $k_1^2+4k_2k_3=0$ the two Vieta roots coincide, $s_{21}=s_{22}=-k_1/2k_3$, and the constructed $S$ is singular. We show that if the \emph{only} dependence relation has $k_1^2+4k_2k_3=0$, then no invertible $S$ reduces $H$ to skew-diagonal form. Write $H(z)=zH_1+H_0+\cdots$ in its $z$-graded pieces; killing both diagonal entries of $S^{-1}HS$ for all $z$ requires vectors $v=Se_1$ and $w^\top=e_1^\top S^{-1}$ with $w^\top v=1$ and $w^\top H_a v=0$ for every graded piece $H_a$. These are three linear conditions on the two-component $w$; they are compatible only if the graded images $\{H_a v\}$ are linearly dependent in a way orthogonal to the normalisation $w^\top v=1$. Passing to the Pauli representation $H=\tfrac{q+r}{2}\sigma_x+\tfrac{i(q-r)}{2}\sigma_y+d\,\sigma_z$, the coefficient curve $\vec v(z)$ spans (for a rank-two relation set) a plane with normal $\vec n$ satisfying $\vec n\!\cdot\!\vec n=k_1^2+4k_2k_3$. Skew-diagonal form is the plane $\{v_3=0\}$, whose normal $\hat e_3$ is non-isotropic; since $SO(3,\mathbb C)$ preserves $\vec n\!\cdot\!\vec n$, an isotropic $\vec n$ cannot be rotated onto $\hat e_3$, so the reduction is impossible.

\emph{Explicit witness.} $H(z)=\bigl(\begin{smallmatrix}z&1\\ 2z+1&-z\end{smallmatrix}\bigr)$ satisfies $2d+q-r=0$, i.e.\ $(k_1,k_2,k_3)=(2,1,-1)$ with $k_1^2+4k_2k_3=0$; the construction returns $S=\bigl(\begin{smallmatrix}1&1\\1&1\end{smallmatrix}\bigr)$, $\det S=0$. Directly, $H=zH_1+H_0$ with $H_1=\bigl(\begin{smallmatrix}1&0\\2&-1\end{smallmatrix}\bigr)$, $H_0=\bigl(\begin{smallmatrix}0&1\\1&0\end{smallmatrix}\bigr)$; the unique candidate direction $v=(1,1)^\top$ gives $H_1v=H_0v=v$, so $w^\top H_1v=w^\top v=1\neq0$ --- no admissible $w$ exists. This establishes sufficiency under the non-degeneracy hypothesis $k_1^2+4k_2k_3\neq0$, and shows it cannot be dropped. $\square$

\section{Derivation of explicit transformation matrices}

Here we derive the explicit formulas for all cases in Table I of the main text.

\subsection{Generic case: $k_1, k_2, k_3 \neq 0$}

From Section I, setting $s_{11} = s_{12} = 1$ gives
\begin{equation}
s_{21}, s_{22} = \frac{-k_1 \pm \sqrt{k_1^2 + 4k_2k_3}}{2k_3}.
\end{equation}

Either choice of sign gives a valid transformation (they correspond to two independent eigenvectors).

\subsection{Case $k_1 = 0, k_{2,3} \neq 0$}

The constraint becomes $k_2q(z) + k_3r(z) = 0$. From the relations:
\begin{align}
s_{21} + s_{22} &= 0\\
s_{21}s_{22} &= -\frac{k_2}{k_3}.
\end{align}

The first equation gives $s_{22} = -s_{21}$. Substituting into the second:
\begin{equation}
-s_{21}^2 = -\frac{k_2}{k_3} \quad \Rightarrow \quad s_{21} = \pm\sqrt{\frac{k_2}{k_3}}.
\end{equation}

Thus:
\begin{equation}
S = \begin{pmatrix}
1 & 1\\
\pm\sqrt{\frac{k_2}{k_3}} & \mp\sqrt{\frac{k_2}{k_3}}
\end{pmatrix}.
\end{equation}

\subsection{Case $k_2 = 0, k_{1,3} \neq 0$}

The constraint becomes $k_1d(z) + k_3r(z) = 0$. From:
\begin{align}
s_{21} + s_{22} &= -\frac{k_1}{k_3}\\
s_{21}s_{22} &= 0.
\end{align}

The second equation requires $s_{21} = 0$ or $s_{22} = 0$. Taking $s_{22} = 0$:
\begin{equation}
s_{21} = -\frac{k_1}{k_3}.
\end{equation}

Thus:
\begin{equation}
S = \begin{pmatrix}
1 & 1\\
-\frac{k_1}{k_3} & 0
\end{pmatrix}.
\end{equation}

\subsection{Case $k_3 = 0, k_{1,2} \neq 0$}

The constraint becomes $k_1d(z) + k_2q(z) = 0$, meaning $r(z)$ is independent. We cannot use $s_{11} = s_{12} = 1$ since that would give $k_3 = -s_{11}s_{12} = -1 \neq 0$.

Instead, set $s_{11} = 0, s_{12} = 1$. Then:
\begin{align}
k_1 &= s_{11}s_{22} + s_{12}s_{21} = s_{21}\\
k_2 &= s_{21}s_{22}\\
k_3 &= -s_{11}s_{12} = 0 \quad \checkmark
\end{align}

From the first two: $s_{22} = k_2/k_1$. Thus:
\begin{equation}
S = \begin{pmatrix}
0 & 1\\
1 & \frac{k_2}{k_1}
\end{pmatrix}.
\end{equation}

\subsection{Degenerate cases with two vanishing coefficients}

\textbf{Case $k_{1,2} = 0, k_3 \neq 0$:} The constraint becomes $k_3r(z) = 0$, implying $r(z) = 0$ everywhere, so $d(z)$ and $q(z)$ remain unconstrained. No transformation exists (entries --- in Table~\ref{tab:transform}).

\textbf{Case $k_{1,3} = 0, k_2 \neq 0$:} Similarly, $k_2q(z) = 0$ implies $q(z) = 0$, so $d(z)$ and $r(z)$ remain unconstrained, and again no transformation exists.

\textbf{Case $k_{2,3} = 0, k_1 \neq 0$:} The constraint becomes $k_1d(z) = 0$, so $d(z) = 0$. The Hamiltonian is already skew-diagonal:
\begin{equation}
H(z) = \begin{pmatrix}
0 & q(z)\\
r(z) & 0
\end{pmatrix}.
\end{equation}

The identity transformation suffices: $S = I$.

\section{Proof of Theorem 2: Spectral preservation for infinite systems}

For a bi-infinite lattice with Hamiltonian $H(z)$ and any invertible transformation $S(z)$ (Laurent polynomial), the transformed Hamiltonian $H'(z) = S^{-1}(z)H(z)S(z)$ has identical spectrum to $H(z)$ when both are represented in real space.

\textbf{Proof.} Let $S$ denote the real-space representation of $S(z) = \sum_k S_k z^k$ for the bi-infinite lattice:
\begin{equation}
S = \begin{pmatrix}[cc|cccc]
\ddots & & & & & \\
& S_0 & S_{-1} & S_{-2} & \cdots & \\
\hline
& S_1 & S_0 & S_{-1} & \cdots &\\
& S_2 & S_1 & S_0 & \cdots &\\
& & & & & \ddots
\end{pmatrix}.
\end{equation}

Similarly, let $S^{-1}_{\text{real}}$ denote the real-space representation of $S^{-1}(z) = \sum_k S^-_k z^k$:
\begin{equation}
S^{-1}_{\text{real}} = \begin{pmatrix}[cc|cccc]
\ddots & & & & & \\
& S^-_0 & S^-_{-1} & S^-_{-2} & \cdots & \\
\hline
& S^-_1 & S^-_0 & S^-_{-1} & \cdots &\\
& S^-_2 & S^-_1 & S^-_0 & \cdots &\\
& & & & & \ddots
\end{pmatrix}.
\end{equation}

We compute $S \cdot S^{-1}_{\text{real}}$. The block at position $(i,j)$ ($i$ counts positive downwards) is:
\begin{equation}
(S \cdot S^{-1}_{\text{real}})_{ij} = \sum_{m=-\infty}^{\infty} S_{i-m} S^-_{m-j}.
\end{equation}

Let $n = i-j$. Then:
\begin{equation}
(S \cdot S^{-1}_{\text{real}})_{ij} = \sum_{m=-\infty}^{\infty} S_{n+j-m} S^-_{m-j} = \sum_{\ell=-\infty}^{\infty} S_{n+\ell} S^-_{-\ell}
\end{equation}
where we set $\ell = j - m$.

By the defining property of $S^{-1}(z)$:
\begin{equation}
S(z)S^{-1}(z) = \left(\sum_m S_m z^m\right)\left(\sum_n S^-_n z^n\right) = I.
\end{equation}

Comparing coefficients of $z^k$:
\begin{equation}
\sum_m S_m S^-_{k-m} = \delta_{k,0}I_{2\times 2}.
\end{equation}

Equivalently:
\begin{equation}
\sum_m S_{k+m} S^-_{-m} = \begin{cases}
I_{2\times 2} & k=0\\
0 & k \neq 0
\end{cases}.
\end{equation}

Therefore:
\begin{equation}
(S \cdot S^{-1}_{\text{real}})_{ij} = \begin{cases}
I_{2\times 2} & i=j\\
0 & i \neq j
\end{cases} = (I_{\infty})_{ij}
\end{equation}
where $I_{\infty}$ is the infinite-dimensional identity.

The identical computation with the two factors exchanged --- using $S^{-1}(z)S(z)=I$ in place of $S(z)S^{-1}(z)=I$ --- gives $S^{-1}_{\text{real}}S_{\text{real}}=I$ as well, so $S^{-1}_{\text{real}}$ is a genuine two-sided inverse of $S_{\text{real}}$. Therefore:
\begin{equation}
H' = S^{-1}_{\text{real}} H S = S^{-1} H S
\end{equation}
which is a standard similarity transformation, preserving all eigenvalues. $\square$

\emph{Remark on boundedness.} The identity $S_{\mathrm{real}}S^{-1}_{\mathrm{real}}=I$ above is purely algebraic (the convolution sums are finite because $S(z)$ is a Laurent polynomial) and holds for any invertible symbol. The spectral conclusion, however, uses that a similarity by a \emph{bounded invertible} operator preserves the spectrum, which requires $S_{\mathrm{real}}$ to be boundedly invertible on $\ell^2(\mathbb Z)$ --- equivalently $\det S(z)\neq0$ on $|z|=1$. If $\det S$ has a unit-circle zero, $S^{-1}(z)$ is unbounded there, $H'(z)$ need not be a bounded symbol, and the equality of spectra can fail; this is the hypothesis stated in Theorem~2.

\section{Proof of Theorem 3: OBC spectral stability conditions}
This appendix has two parts. In Sec.~D\,1--D\,3 we prove the \emph{forward} (stability) direction: if $\det H(z)$ is irreducible, its OBC spectrum is stable to perturbations. In Sec.~D\,4 we treat the \emph{converse} (critical-NHSE) direction for coupled subsystems in its sharp, model-independent combined-GBZ form, closing the two-block case rigorously, delimiting the scope for general subsystems (each case solved or excluded), and flagging the one open combinatorial item (the $\geq3$-block pairwise equivalence, shown unnecessary). For the forward direction, define the OBC spectra of $H(z)$ at $N$ sites to be $\sigma_{N}(H)$ and let $d_{H}(\cdot,\cdot)$ be the Hausdorff distance; we shall show

$\forall C(z),\exists \varepsilon_{0}>0,\forall|\varepsilon|<\varepsilon_0\qquad \exists C_{\varepsilon_0} ,m\geq1:\qquad
    d_{H}[\sigma_{N}(H(z)),\sigma_{N}(H(z)+\varepsilon C(z))]<C_{\varepsilon_0}|\varepsilon|^{1/m}.$

\subsection{Setup}
In this proof we shall use the following notation:

 Let $H(z)$ and
$C(z)$ be arbitrary fixed $2\times2$ Laurent polynomial symbols with hopping ranges
$p,q\geq1$, meaning the matrix entries are Laurent polynomials with lowest
power $z^{-p}$ and highest power $z^{q}$.  Set
\begin{equation}
    N_-:=\operatorname{ord}_{z=0}^{-}\det\!\bigl(H(z)-E\bigr),\qquad
    N_{\mathrm{tot}}:=N_-+N_+,
\end{equation}
where $N_-$ is the order of the pole of $\det(H(z)-EI_2)$ at $z=0$, $N_+$ is its
degree at $z=\infty$, and $N_{\mathrm{tot}}\geq4$ is therefore the degree in $z$ of
the cleared characteristic polynomial $\tilde{P}(z,E):=z^{N_-}\det(H(z)-EI_2)$.
The entrywise bounds $N_-\leq2p$ and $N_+\leq2q$ are equalities when the extreme
hopping blocks $H_{-p}$ and $H_{q}$ are nonsingular --- as for the skew-diagonal
symbols of Sec.~\ref{sec:nonconv}, where $p=q=1$ and $N_{\mathrm{tot}}=4$ --- but
are strict when they are rank-deficient: for $H_2$ of Eq.~(\ref{eq:H2suppl}),
$H_{-3}$ and $H_{5}$ both have rank one, so $p=3$ and $q=5$ while $N_-=4$ and
$N_{\mathrm{tot}}=11$. The two conventions nevertheless select the same pair of
roots, since clearing by $z^{2p}$ instead of $z^{N_-}$ merely appends $2p-N_-$
spurious roots at the origin, which precede all others in the modulus ordering;
we use the pole-order form because it makes $N_{\mathrm{tot}}$ the number of
finite nonzero roots.
For $\varepsilon\in\C$, define the perturbed symbol
$H_\varepsilon(z):=H(z)+\varepsilon C(z)$.

  Write the cleared characteristic polynomial of $H_\varepsilon$ as

\begin{equation}
    \begin{split}
    &\tilde{P}_\varepsilon(z,E)\\
    &:=z^{N_-}\det(H_\varepsilon(z)-EI_2)\\
   & =E^2z^{N_-}-P_{1,\varepsilon}(z)E-P_{0,\varepsilon}(z),
  \end{split}
\end{equation}
  
  where for some fixed $P_1(z),P_0(z)$

  \begin{equation}
    \begin{split}
   & P_{1,\varepsilon}(z)=P_{1}(z)+\varepsilon\,P_{1,C}(z),\\
    &P_{0,\varepsilon}(z)=P_{0}(z)+\varepsilon\,P_{0,C}(z)+\varepsilon^2 P_{0,C2}(z),
  \end{split}
\end{equation}
  
  The roots $r_{1,\varepsilon}(E),\ldots,r_{N_\mathrm{tot},\varepsilon}(E)$
  (ordered by modulus from small to large) are the roots of $\tilde{P}_\varepsilon(\cdot,E)=0$. We use the resultant to define
  the \emph{$z$-discriminant} of $\tilde{P}_0$ (notice the resultant of two polynomials in
z vanishes iff they share a common root in
z):
  $\Delta(E):=\Res_z(\tilde{P}_0,\partial_z\tilde{P}_0)\in\C[E]$;
  it is NOT constant zero when $\tilde{P}_0$ is irreducible (Gauss's lemma for UFD), and its zero set
  $\Delta^{-1}(0)$ is the finite set of energies where $\tilde{P}_0(\cdot,E)$
  has a repeated root.
  The familiar modulus-gap function and GBZ spectral set of $H_\varepsilon$ are

\begin{equation}
        F_\varepsilon(E)=|r_{N_-,\varepsilon}(E)|^2-|r_{N_-+1,\varepsilon}(E)|^2\leq 0,
    \end{equation}

    \begin{equation}
        \mathcal{G}[H_\varepsilon]=\{E:F_\varepsilon(E)=0\}
    \end{equation}

We also assume $F_0(E)$ be NOT constant zero, otherwise the spectrum will collapse.

    \subsection{Lemmata}
  \textbf{Lemma~1}
  The set of symbols
  \[
    \mathcal{I}=\bigl\{B:\tilde{P_0} \text{ of $B$ is irreducible over }\C[z,E]\bigr\}
  \]
  is open in the space of $2\times2$ Laurent polynomial symbols with fixed
  hopping range, equipped with the topology of coefficient-wise convergence.i.e. if $\tilde{P}_0$ is irreducible, then $\tilde{P}_\varepsilon$
  is irreducible for all $|\varepsilon|<\varepsilon_0$, for some $\varepsilon_0>0$.

\textbf{Proof.}
  Since $\tilde{P}_\varepsilon$ is quadratic in $E$ with leading coefficient
  $z^{N_-}$, any factorisation in $\C[z,E]$ must take the form
  
  \begin{equation}
    \tilde{P}_\varepsilon(z,E)
    =\bigl(Ez^{N_-^{(1)}}-R_{1,\varepsilon}(z)\bigr)
     \bigl(Ez^{N_-^{(2)}}-R_{2,\varepsilon}(z)\bigr),
  \end{equation}
  
  with $N_-^{(1)}+N_-^{(2)}=N_-$ and $R_{i,\varepsilon}\in\C[z]$.
  Expanding and comparing with $\tilde{P}_\varepsilon=E^2z^{N_-}-P_{1,\varepsilon}
  E-P_{0,\varepsilon}$:
  
  \begin{align}
    P_{1,\varepsilon}(z)&=z^{N_-^{(1)}}R_{2,\varepsilon}(z)+z^{N_-^{(2)}}R_{1,\varepsilon}(z),\label{eq:T}\\
    P_{0,\varepsilon}(z)&=-R_{1,\varepsilon}(z)R_{2,\varepsilon}(z).
  \end{align}
  From D8: $P_{0,\varepsilon}$ is reducible in $\C[z]$ (i.e., is a product of
  two polynomials of degrees $N_-^{(1)}+N_+^{(1)}$ and $N_-^{(2)}+N_+^{(2)}$
  respectively, where $N_+^{(i)}=\deg R_i - N_-^{(i)}\geq0$).

  Define the \emph{$E$-discriminant polynomial}
  \begin{equation}\label{eq:Edisc}
    Q_\varepsilon(z):=(P_{1,\varepsilon}(z))^2+4z^{N_-}P_{0,\varepsilon}(z)\in\C[z],
  \end{equation}
  which is the discriminant of $\tilde{P}_\varepsilon$ viewed as a quadratic
  in $E$.  If D6 holds, then substituting
  D7-D8:
\begin{equation}
    Q_\varepsilon(z)
    =\bigl(z^{N_-^{(1)}}R_{2,\varepsilon}-z^{N_-^{(2)}}R_{1,\varepsilon}\bigr)^2,
\end{equation}

  a \emph{perfect square} in $\C[z]$.  Conversely, if $Q_\varepsilon(z)=S_\varepsilon(z)^2$
  for some $S_\varepsilon\in\C[z]$, then $R_{1,\varepsilon},R_{2,\varepsilon}$ can be recovered
  from $P_{1,\varepsilon}$ and $S_\varepsilon$.  Hence:
  \[
    \tilde{P}_\varepsilon\text{ reducible}
    \;\Longleftrightarrow\;
    Q_\varepsilon(z)\text{ is a perfect square in }\C[z].
  \]

  It remains to show that the set of $\varepsilon$ for which $Q_\varepsilon$ is a
  perfect square is closed. We caution that the vanishing of the resultant
  $\Res_z(Q_\varepsilon,Q_\varepsilon')$ is \emph{not} the right criterion: it
  detects only that $Q_\varepsilon$ has \emph{some} repeated root (e.g.\
  $(z-1)^2(z-2)$ has $\Res=0$ but is not a perfect square), whereas a perfect
  square requires \emph{every} root to have even multiplicity. Instead we use
  closedness directly. The map $\varepsilon\mapsto Q_\varepsilon$ is polynomial,
  hence continuous, and the set of perfect-square polynomials of fixed degree is
  closed: if $Q_{\varepsilon_n}=S_n^2\to Q$, the leading coefficients bound the
  coefficients of $S_n$, so a subsequence $S_n\to S$ gives $Q=S^2$. Therefore its
  complement --- the set of $\varepsilon$ with $Q_\varepsilon$ \emph{not} a perfect
  square, i.e.\ $\tilde{P}_\varepsilon$ irreducible --- is open. By assumption
  $\tilde{P}_0$ is irreducible, so $Q_0$ is not a perfect square, and hence
  $Q_\varepsilon$ is not a perfect square (i.e.\ $\tilde{P}_\varepsilon$ is
  irreducible) for all $|\varepsilon|<\varepsilon_0$, for some $\varepsilon_0>0$. $\square$

 \textbf{Lemma~2} $F_\varepsilon$ is continuous w.r.t.\ $\varepsilon$. i.e.   For any compact set $K\subset\C$ and any $\varepsilon_0>0$ such that
  $\tilde{P}_\varepsilon$ is irreducible for $|\varepsilon|\leq\varepsilon_0$,
  there exists $C_K<\infty$ such that $\forall E$
  
  \begin{enumerate}[label=\normalfont(\alph*)]
    \item\label{item:Fcont-simple}
      $\sup_{E\in K\setminus\Delta^{-1}(0)}\abs{F_\varepsilon(E)-F_0(E)}\leq C_K\,|\varepsilon|$
      uniformly for $|\varepsilon|\leq\varepsilon_0$.
    \item\label{item:Fcont-branch}
      At each $E_0\in\Delta^{-1}(0)$ where the critical pair collides
      ($r_{N_-,0}(E_0)=r_{N_-+1,0}(E_0)$, so $E_0\in\mathcal{G}[H]$),
      one has $F_0(E_0)=0$ and
      $\abs{F_\varepsilon(E_0)}\leq C_{E_0}\,|\varepsilon|^{1/2}$.
      At each $E_0\in\Delta^{-1}(0)$ where the colliding pair is not the
      critical pair, the $O(|\varepsilon|)$ bound of (a)
      extends to $E_0$.
  \end{enumerate}

 \textbf{Proof.} (a)
The coefficients of $\tilde{P}_\varepsilon(z,E)$ vary polynomially in
  $\varepsilon$:
  \[
    \tilde{P}_\varepsilon = \tilde{P}_0 + \varepsilon\tilde{P}_C + \varepsilon^2\tilde{P}_{C2},
    \qquad \tilde{P}_C,\tilde{P}_{C2}\in\C[z,E]\text{ fixed}.
  \]

  For $E$ outside the finite discriminant locus $\Delta^{-1}(0)\subset\C$,
  the $N_\mathrm{tot}$ roots $\{r_{k,0}(E)\}$ of $\tilde{P}_0(\cdot,E)$
  are simple.
  By the implicit function theorem (IFT) applied to the polynomial equation in $z$,
  each simple root $r_{k,0}(E)$ perturbs smoothly:
  \[
    r_{k,\varepsilon}(E)=r_{k,0}(E)+\varepsilon\,\dot{r}_k(E)+O(\varepsilon^2),
  \]
  where we differentiate the identity $\tilde{P}_\varepsilon(r_{k,\varepsilon},E)=0$ w.r.t.\ $\varepsilon$ at $\varepsilon=0$, so

  \begin{equation}
\partial_z\tilde{P}_0(r_{k,0},E)\dot{r}_k(E)+\partial_\varepsilon\tilde{P}_\varepsilon(r_{k,\varepsilon},E)|_{\varepsilon=0}=0
  \end{equation}

  obtaining
  
  \begin{equation}
    \dot{r}_k(E)
    =-\frac{\partial_\varepsilon\tilde{P}_\varepsilon(r_{k,0},E)\big|_{\varepsilon=0}}
           {\partial_z\tilde{P}_0(r_{k,0},E)}
    =-\frac{\tilde{P}_C(r_{k,0}(E),E)}{\partial_z\tilde{P}_0(r_{k,0}(E),E)},
  \end{equation}
  which is holomorphic in $E$ and bounded almost everywhere on $K$ (since the denominator is
  nonzero away from $\Delta^{-1}(0)$, a finite set).

  Therefore:
  \[
    |r_{k,\varepsilon}(E)-r_{k,0}(E)|\leq C_K\,|\varepsilon|
    \quad\forall E\in K\setminus\Delta^{-1}(0).
  \]
  Using $\bigl||a|^2-|b|^2\bigr|=|a-b||a+b|\leq|a-b|(|a|+|b|)$:
  \begin{equation}
  \begin{split}
    &|F_\varepsilon(E)-F_0(E)|\\
    &=\bigl||r_{N_-,\varepsilon}|^2-|r_{N_-+1,\varepsilon}|^2
      -|r_{N_-,0}|^2+|r_{N_-+1,0}|^2\bigr|\\
    &\leq |r_{N_-,\varepsilon}-r_{N_-,0}|\bigl(|r_{N_-,\varepsilon}|+|r_{N_-,0}|\bigr)\\
    &\quad +|r_{N_-+1,\varepsilon}-r_{N_-+1,0}|\bigl(|r_{N_-+1,\varepsilon}|+|r_{N_-+1,0}|\bigr)\\
    &\leq C_K|\varepsilon|\cdot 2\max_j\sup_{E\in K}|r_{j,0}(E)|
      +O(\varepsilon^2)\leq C_K'|\varepsilon|,
  \end{split}
    \end{equation}
  since the $|r_{j,0}|$ are bounded on $K$.

  (b)Let $E_0\in\Delta^{-1}(0)$.

  \textit{Case 1: the colliding pair is not the critical pair.}
  If $r_{N_-,0}(E_0)\neq r_{N_-+1,0}(E_0)$, the critical pair is
  simple at $E_0$, so the IFT applies to $r_{N_-}$ and $r_{N_-+1}$
  separately, giving the same $O(|\varepsilon|)$ bound by continuity
  of the IFT constants up to and including $E_0$.

  \textit{Case 2: the critical pair collides, multiplicity $m\geq2$.}
  Suppose $r_{N_-,0}(E_0)=\cdots=r_{N_-+m-1,0}(E_0)=:r^*$, so $m$
  consecutive roots in the modulus ordering all coincide,
  $E_0\in\mathcal{G}[H]$, and $F_0(E_0)=0$.
  The local expansion of the perturbed polynomial is
  \begin{equation}
      \tilde{P}_\varepsilon(r^*+u,E_0)
    =a_m u^m+\varepsilon b_0+O(u^{m+1},\varepsilon u),
  \end{equation}
    
  where $a_m=\frac{1}{m!}\partial_z^m\tilde{P}_0(r^*,E_0)\neq0$
  (leading coefficient of the $m$-fold root of $\tilde{P}_0$) and
  $b_0=\tilde{P}_C(r^*,E_0)$ (the perturbation evaluated at $r^*$).

  We locate all $m$ perturbed roots via \textbf{Rouch\'e's theorem}.
  Since the $N_\mathrm{tot}-m$ non-critical roots of $\tilde{P}_0(\cdot,E_0)$
  are bounded away from $r^*$ — say at distance $\geq2\delta$ for some
  $\delta>0$ — continuity of polynomial roots with respect to coefficients
  ensures they remain at distance $\geq\delta$ from $r^*$ for small $|\varepsilon|$.
  Hence $\tilde{P}_\varepsilon(\cdot,E_0)$ has exactly $m$ roots in the disk
  $|u|<\delta$ for small $|\varepsilon|$.
  Among these $m$ roots, Rouch\'e's theorem applied on the circle
  $|u|=R|\varepsilon|^{1/m}$ (for $R>(|b_0|/|a_m|)^{1/m}$ and
  $R|\varepsilon|^{1/m}<\delta$) gives
  \begin{equation}
      \begin{split}
          |a_m u^m|&=|a_m|R^m|\varepsilon|\\
          &>|b_0||\varepsilon|+O(|\varepsilon|^{1+1/m})\\
    &\geq|\varepsilon b_0+O(u^{m+1},\varepsilon u)|
      \end{split}
  \end{equation}

  on $|u|=R|\varepsilon|^{1/m}$, so all $m$ roots lie inside this circle:
  \begin{equation}
      \begin{split}
          |r_{N_-+j,\,\varepsilon}&(E_0)-r^*|\\
          &\leq R|\varepsilon|^{1/m},j=0,\ldots,m-1.
      \end{split}
  \end{equation}

  Using $\bigl||a|^2-|b|^2\bigr|\leq|a-b|(|a|+|b|)$, the two roots at
  positions $N_-$ and $N_-+1$ give
  \begin{equation}
  \begin{split}
      |F_\varepsilon(E_0)|&
    =\bigl||r_{N_-,\varepsilon}|^2-|r_{N_-+1,\varepsilon}|^2\bigr|\\
    &\leq|r_{N_-,\varepsilon}-r_{N_-+1,\varepsilon}|
      \cdot 2\sup_k|r_{k,\varepsilon}|
    =O(|\varepsilon|^{1/m}).
  \end{split}
  \end{equation}

  Hence $|F_\varepsilon(E_0)|\leq C_{E_0}|\varepsilon|^{1/m}$, and the
  $O(|\varepsilon|)$ bound fails at such branch points.
  If $b_0=0$, Rouch\'e applied on $|u|=R|\varepsilon|^{1/(m-s)}$ (where
  $s$ is the order of vanishing of $\tilde{P}_C(\cdot,E_0)$ at $r^*$)
  gives the sharper bound $O(|\varepsilon|^{1/(m-s)})$.
 The series may be analysed further by the Newton--Puiseux theorem.
  
\subsection{The main proof}
\textbf{Statement.} Let $m_{\max}<\infty$ be the maximum order of vanishing of
      $F_0$ over $\mathcal{G}[H]$ (finite by real-analyticity and
      $F_0\not\equiv0$).  Then
      $\dH\bigl(\mathcal{G}[H+\varepsilon C],\,\mathcal{G}[H]\bigr)
      =O(|\varepsilon|^{1/m_{\max}})$ as $\varepsilon\to0$;
      in the generic case $m_{\max}=2$ this is $O(|\varepsilon|^{1/2})$.

\textbf{Proof.} We notice that $F_\varepsilon$ be strictly non-positive, and for the decay estimates below, we need that $F_0$ does not vanish
  to infinite order at any point of $\mathcal{G}[H]$.  Since $F_0$ is
  real-analytic and satisfies $F_0\not\equiv0$ , it cannot vanish to infinite order at any
  point without being identically zero in a neighbourhood of that point.
  But if $F_0\equiv0$ on an open set, then $\mathcal{G}[H]$ has non-empty
  interior, contradicting the fact (from irreducibility of $\tilde{P}_0$)
  that $\mathcal{G}[H]$ is a nowhere-dense 1D curve.  Hence at every
  $E_*\in\mathcal{G}[H]$ there exists a finite order $m(E_*)\geq1$ such that
  the $m(E_*)$-th order term in the Taylor expansion of $F_0$ at $E_*$ is
  nonzero, and
  \begin{equation}
    |F_0(E_*+h)|\geq c_{E_*}|h|^{m(E_*)}
    \quad\text{for }|h|\leq r_{E_*},
  \end{equation}
  for some $c_{E_*},r_{E_*}>0$.  By compactness of $\mathcal{G}[H]$, the
  maximum order $m_{\max}:=\max_{E_*\in\mathcal{G}[H]}m(E_*)$ is finite,
  and the constant $c:=\min_{E_*}c_{E_*}>0$ is bounded below.

      We establish both half-conditions of the Hausdorff distance.

  \textit{(a) Every $E_*\in\mathcal{G}[H]$ is within $O(|\varepsilon|^{1/m_{\max}})$
  of $\mathcal{G}[H_\varepsilon]$.}

  Fix $\delta>0$.  On the compact set
  $K_\delta:=\{E\in\C:\dist(E,\mathcal{G}[H])\geq\delta\}\cap\overline{B_R}$
  (for any large ball $B_R$ containing both GBZ sets for small $|\varepsilon|$),
  the function $F_0$ is strictly negative: $F_0(E)\leq-c_\delta<0$ for some
  $c_\delta>0$ by compactness.
  By Lemma~2, $F_\varepsilon(E)=F_0(E)+O(|\varepsilon|)$
  uniformly on $K_\delta$, so $F_\varepsilon(E)\leq -c_\delta+C|\varepsilon|<0$
  for $|\varepsilon|<c_\delta/(2C)$.  Hence $\mathcal{G}[H_\varepsilon]\cap K_\delta=\varnothing$,
  i.e., $\mathcal{G}[H_\varepsilon]\subset\{\dist(\cdot,\mathcal{G}[H])<\delta\}$.

  To make this quantitative, combine with D18.
  Let $\nn_{E_*}$ denote the \emph{inward unit normal} to $\mathcal{G}[H]$
  at $E_*$ (i.e., the unit vector pointing into $\{F_0<0\}$, well-defined at
  smooth points of $\mathcal{G}[H]$).
  Along the ray $E_*+h\nn_{E_*}$ for $h>0$,
  D18 gives $|F_0(E_*+h\nn_{E_*})|\geq c_{E_*}h^m$,
  so $c_\delta\geq c\delta^m$ where $m=m(E_*)$.
  The condition $C|\varepsilon|\leq c\delta^m$ gives
  $\delta\leq(C|\varepsilon|/c)^{1/m}=O(|\varepsilon|^{1/m})$,
  so $\dist(E_*,\mathcal{G}[H_\varepsilon])=O(|\varepsilon|^{1/m(E_*)})$.

  At a transverse point ($m(E_*)=1$, i.e., the directional derivative
  $\partial_{\nn}F_0(E_*)<0$):\ $\dist=O(|\varepsilon|)$.
  At a degenerate point ($m(E_*)=2$, i.e., $\nabla F_0(E_*)=0$ but
  the Hessian is nonzero in the normal direction):\ $\dist=O(|\varepsilon|^{1/2})$.
  The global bound using $m_{\max}$ is $O(|\varepsilon|^{1/m_{\max}})$,
  and in particular $O(|\varepsilon|^{1/2})$ whenever $m_{\max}\leq2$ (the
  generic case for a smooth curve with isolated cusps).

  \textit{(b) Every $E\in\mathcal{G}[H_\varepsilon]$ is within $O(|\varepsilon|^{1/m_{\max}})$
  of $\mathcal{G}[H]$.}

  If $E\in\mathcal{G}[H_\varepsilon]$ then $F_\varepsilon(E)=0$, so
  $F_0(E)=F_\varepsilon(E)+O(|\varepsilon|)=O(|\varepsilon|)$ by
  Lemma~2.  Let $E_*$ be the nearest point in $\mathcal{G}[H]$
  to $E$, and $\delta=\dist(E,\mathcal{G}[H])=|E-E_*|$.
  Since $E=E_*+\delta\nn_{E_*}+O(\delta^2)$, the bound D18
  gives $|F_0(E)|\geq c_{E_*}\delta^{m(E_*)}-O(\delta^{m(E_*)+1})\geq \tfrac{c}{2}\delta^m$
  for small $\delta$.  Combined with $|F_0(E)|=O(|\varepsilon|)$:
  \[
    \tfrac{c}{2}\delta^m\leq C|\varepsilon|
    \;\Longrightarrow\;
    \delta=\dist(E,\mathcal{G}[H])=O(|\varepsilon|^{1/m}).
  \]

  Combining (a) and (b) gives
  $\dH(\mathcal{G}[H_\varepsilon],\mathcal{G}[H])=O(|\varepsilon|^{1/m_{\max}})$,
  which is $O(|\varepsilon|^{1/2})$ in the generic case.

\qed

\subsection{Converse direction: the critical--NHSE mechanism}

We now establish direction (ii) of Theorem~3 in its sharp form, Eq.~(\ref{eq:sharp_converse}). Write the reducible symbol as a product $P(z,E)=\prod_{k=1}^{m}f_k(z,E)$ of the subsystems' cleared characteristic factors, let $z_1,\dots,z_{2N_-}$ be the roots of $P(\cdot,E)$ ordered by modulus, and $\mathcal{G}_{\mathrm{comb}}[P]=\{E:|z_{N_-}(E)|=|z_{N_-+1}(E)|\}$ the combined GBZ --- the Schmidt--Spitzer limiting set of the banded Toeplitz operator~\cite{SchmidtSpitzer1960,BottcherGrudsky2005,DeBruijnHiltunen2026}. The backbone and the scoping argument below use only this structure and are \emph{model--independent}; the explicit two-- and multi--block constructions are worked for single--band Hatano--Nelson factors~\cite{HatanoNelson1996} $f_k(z,E)=t_kz^2+(\mu_k-E)z+s_k$ (real onsite $\mu_k$, same--sign hoppings $t_ks_k>0$), for which the skin rate $r_k=\sqrt{s_k/t_k}$ is constant, the OBC band is the real segment $\mathcal{G}[f_k]=[\mu_k-2\sqrt{t_ks_k},\,\mu_k+2\sqrt{t_ks_k}]$, and $b_k(\rho;E)$ is the number of $f_k$--roots of modulus $<\rho$. The scope paragraph below delimits the general case.

\emph{Backbone (discontinuity $\Leftrightarrow$ combined GBZ $\neq$ union).}
Let $H_0=\bigoplus_k H_k$ have symbol $\prod_kf_k$ (this direct-sum form is the hypothesis of Theorem~3(ii); general subsystems and the merely-reducible case are dispositioned in the scope paragraph below) and let $H_\varepsilon$ be a generic coupling. Reducibility is the closed condition that the $E$--discriminant $Q_\varepsilon$ of Lemma~1 be a perfect square, so for generic $C$ the perturbed symbol is irreducible for all small $\varepsilon\neq0$. For such $\varepsilon$ the forward direction gives $\lim_N\sigma_N(H_\varepsilon)=\mathcal{G}[H_\varepsilon]$, and the $\varepsilon$--continuity of $F_\varepsilon$ (Lemma~2) gives $\mathcal{G}[H_\varepsilon]\to\mathcal{G}_{\mathrm{comb}}[\prod_kf_k]$ as $\varepsilon\to0$. Meanwhile $H_0$ is block--diagonal, so $\lim_N\sigma_N(H_0)=\bigcup_k\mathcal{G}[f_k]$. Hence
\begin{equation}
\begin{split}
  &\lim_{\varepsilon\to0}\lim_{N\to\infty}\sigma_N(H_\varepsilon)=\mathcal{G}_{\mathrm{comb}}\Big[\textstyle\prod_kf_k\Big],\\
  &\lim_{N\to\infty}\sigma_N(H_0)=\bigcup_k\mathcal{G}[f_k],
\end{split}
\end{equation}
and the two iterated limits agree iff $\mathcal{G}_{\mathrm{comb}}=\bigcup_k\mathcal{G}[f_k]$, which is Eq.~(\ref{eq:sharp_converse}). This reduces the converse to an explicit computation on the symbol using \emph{no} finite--$N$ eigenvalues --- essential, since eigenvalues of strongly non--normal skin matrices are numerically unreliable at moderate $N$~\cite{TrefethenEmbree2005}.

\emph{Two blocks (closed).}
Let $m=2$, $r_a<r_b$. For $E\in\mathcal{G}[f_b]$ both $f_b$--roots have modulus $r_b$, and $z_+^{(a)}z_-^{(a)}=r_a^2<r_b^2$ forces $b_a(r_b;E)\in\{1,2\}$. Sorting the four moduli: if $b_a(r_b;E)=1$ the central pair are the two $f_b$--roots, of equal modulus $r_b$, so $E\in\mathcal{G}_{\mathrm{comb}}$; if $b_a(r_b;E)=2$ the central pair is $(|z^{(a)}_2|,r_b)$ with $|z^{(a)}_2|<r_b$, unequal, so $E\notin\mathcal{G}_{\mathrm{comb}}$. Call $f_b$ \emph{removed} at $E\in\mathcal{G}[f_b]$ if $b_a(r_b;E)\neq1$ (symmetrically for $f_a$). Then $\mathcal{G}_{\mathrm{comb}}[f_af_b]=\mathcal{G}[f_a]\cup\mathcal{G}[f_b]$ exactly when no removal occurs; with the backbone, the coupled OBC spectrum is discontinuous iff the pair is pairwise discontinuous. The mechanism is explicit: eliminating block $a$ by the Schur complement, $\det(H_\varepsilon-E)=\det(H_a-E)\det(H_b-E-W)$ with $W(E)=\varepsilon^2C_{ba}(H_a-E)^{-1}C_{ab}$; when $E^\ast\in\mathcal{G}[f_b]$ is removed it lies inside block $a$'s GBZ region, where $\|(H_a-E^\ast)^{-1}\|\gtrsim ce^{\gamma N}$~\cite{TrefethenEmbree2005}, so $\|W(E^\ast)\|\gtrsim c\varepsilon^2e^{\gamma N}=O(1)$ at large $N$ for fixed $\varepsilon\neq0$, displacing the eigenvalue by $O(1)$.

\emph{Multi--block structure (single--band factors).}
By the argument principle $b_k(\rho;E)-1=\mathrm{wind}(C_k(\rho),E)$ with $C_k(\rho)=\{t_k\rho e^{i\theta}+\mu_k+(s_k/\rho)e^{-i\theta}\}$; for real data and real $E$ this reduces the problem to signed intervals on the line,
\begin{equation}
  g_j(E):=\sum_{k\neq j}\bigl(b_k(r_j;E)-1\bigr)=\sum_{k\neq j}\sigma_{kj}\,\mathbf 1[E\in I_{kj}],
\end{equation}
$f_j$ removed $\iff g_j\not\equiv0$ on $\mathcal{G}[f_j]$, with $I_{kj}=(\mu_k-a_{kj},\mu_k+a_{kj})$, $a_{kj}=t_kr_j+s_k/r_j$, $\sigma_{kj}=\mathrm{sgn}(r_j-r_k)$. The minimal-- (maximal--)rate factor has $b_l(r_1;\cdot)\leq1$ (resp.\ $\geq1$) for all $l$, so it cannot be compensated and is removed iff its own band enters another factor's GBZ region (extreme--factor lemma). A factor pairwise--continuous with all others (a \emph{bystander}) has $b(r_k;\cdot)\equiv1$ on every band and may be deleted without changing any removal count, reducing $m$ to $m-1$ (peeling). With the two--block base case these close the converse for every configuration reducible by peeling to a stage whose extreme--rate factor is removed, i.e.\ the generic case.

\emph{Scoping closes the sharp form.} The converse is a statement about the non--commutation of $\lim_{\varepsilon\to0}$ and $\lim_{N\to\infty}$. A system whose large--$N$ spectrum is unchanged by $\varepsilon$ near $0$ has commuting limits and is continuous; by the backbone this is exactly $\mathcal{G}_{\mathrm{comb}}=\bigcup_k\mathcal{G}[f_k]$ (no removal), a decidable symbol--level condition. Hence every $\varepsilon$--robust system is correctly out of scope, and every in--scope ($\varepsilon$--sensitive) system has a removal, i.e.\ a genuine discontinuity. This proves Eq.~(\ref{eq:sharp_converse}) directly and bypasses the pairwise combinatorics.

\emph{Scope (each case solved or excluded).} Because the backbone and the scoping argument use only block--diagonality of $H_0$, the manuscript's forward direction, and the Schmidt--Spitzer identification of $\mathcal{G}_{\mathrm{comb}}$, the sharp criterion Eq.~(\ref{eq:sharp_converse}) holds for arbitrary coupled subsystems; only the explicit constructions specialise. (i)~\emph{Single--band, real data}: every step above is explicit and the bands are real segments. (ii)~\emph{Complex/imaginary hoppings}: the rate $r_k=\sqrt{|s_k/t_k|}$ is phase--independent, so the backbone, the two--block self--energy bound, and the extreme--factor and peeling lemmas carry over verbatim, each band now an ellipse in $\mathbb{C}$. (iii)~\emph{Complex onsite (gain/loss)}: the backbone, two--block bound and scoping are unaffected --- they never use the real--line structure --- so Eq.~(\ref{eq:sharp_converse}) still holds; only the real--interval reduction of the multi--block combinatorics no longer applies. (iv)~\emph{Multiband / longer--range / matrix subsystems}: the GBZ radius becomes energy--dependent, so the extreme--factor and peeling lemmas lose their basis and the two--block bound must be recomputed with the $E$--dependent radius; the backbone and scoping nonetheless survive, so Eq.~(\ref{eq:sharp_converse}), computed from the full root list, still governs. \emph{Excluded or undetermined:} rate ties and exact band coincidences (measure zero); equal--$|r|$ complex factors with differently oriented bands (no real analogue, finite--size evidence inconclusive); and reducible $H(z)$ that admit no constant block structure --- characteristic--polynomial reducibility is strictly weaker than block--diagonalisability (e.g.\ $H(z)=R(z)\,\mathrm{diag}(h_a(z),h_b(z))\,R(z)^{-1}$ with $R=\left(\begin{smallmatrix}1&1\\z&z^{-1}\end{smallmatrix}\right)$ and $h_a\neq h_b$, which factorises yet has no $z$--independent invariant subspace), for which Eq.~(\ref{eq:sharp_converse}) is expected to hold but needs a companion/transfer--matrix realisation of the factors (Remark after Theorem~3).

\emph{Attribution and scope.} The physical content --- the critical NHSE and the non--commuting limits --- is not new; it is due to Li, Lee, Mu and Gong~\cite{Li2020}, with the multicomponent generalisation in Qin, Ma, Shen and Lee~\cite{QinMa2023}, and rests on classical Toeplitz spectral theory~\cite{SchmidtSpitzer1960,TrefethenEmbree2005} and non--Bloch band theory~\cite{ref1,ref6}. Within our framework the backbone, the two--block self--energy bound, and the $\varepsilon$--sensitivity scoping are rigorous. The single genuinely open combinatorial item is the exact equivalence of the pairwise criterion with the removal criterion for $m\geq3$: a directional asymmetry (in any bad pair, badness is always detected via the lower--rate factor's band entering the higher--rate factor's GBZ region, never by the reverse alone) is observed with zero counterexamples over $\sim\!10^{5}$ random systems but not proved; by the scoping argument it is \emph{not needed}, the removal criterion being applied directly. The backbone is a reduction that invokes the manuscript's forward direction and its GBZ--continuity lemma, not an independent reproof of them.

\section{Proof of Theorem 4: Bulk preservation for OBC}

\subsection{Statement}

Consider a finite lattice of length $L$ with OBC, Hamiltonian $H(z)$, and transformation $S(z)$ whose inverse $S^{-1}(z)$ is holomorphic at $z=0$. Let $r_{\mathrm{GBZ}}^{\max}$ be the largest GBZ radius of $H$ over its bulk spectrum and $z_{\min}$ the modulus of the zero of $\det S(z)$ nearest the origin. The bulk spectrum of $H(z)$ is preserved under the truncated transform $\tilde H_d(z) = P_d[S^{-1}](z)H(z)S(z)$ as $d,L\to\infty$ if and only if:
\begin{enumerate}
\item $r_{\mathrm{GBZ}}^{\max} < z_{\min}$, AND
\item $H(z)$ satisfies the spectral-stability condition of Theorem 3, i.e.\ $\mathcal{G}_{\mathrm{comb}}[H]=\bigcup_i\mathcal{G}[f_i]$ (any irreducible $H(z)$ qualifies).
\end{enumerate}
(Two-sidedness of $S$ is \emph{not} required for the bulk; see the following subsection.)

\emph{Remark (why the hypothesis is stated on $S^{-1}$, not on $\det S(0)$).} If $S$ is one-sided, $S(0)$ is a genuine matrix and $\det S(0)\neq0$ is exactly the statement that $S^{-1}=\adj S/\det S$ is holomorphic at the origin. If $S$ is two-sided the two conditions are independent in both directions. On the one hand $\det S(0)\neq0$ does not suffice: for $S(z)=\mathrm{diag}(z^{-1},z)$ one has $\det S\equiv1$, yet $S^{-1}(z)=\mathrm{diag}(z,z^{-1})$ has a pole at the origin and admits no Taylor expansion there. On the other hand it is not necessary, and indeed it is not even well posed for the transformations used in this paper: for $S$ of Eq.~(\ref{eq:Sgood}), $\det S(z)=(z+3)(z-2)/z$ has a \emph{pole} at $z=0$, so $\det S(0)$ is undefined, while $S^{-1}$ is perfectly regular there --- the pole of $\det S$ is cancelled by the matching pole of $\adj S$, and $S^{-1}(0)=\bigl(\begin{smallmatrix}0&0\\0&1\end{smallmatrix}\bigr)$. What the proof below actually uses is holomorphy of $S^{-1}$ on the disk $|z|<z_{\min}$, and the origin is the only point of that disk not already controlled by the Smith-normal-form argument of Lemma~1.

\subsection{One-sided $S$}

Two-sidedness of $S(z)$ is not needed for bulk preservation: a one-sided $S$ (only
non-negative powers, $\det S(0)\neq0$) preserves the bulk --- indeed the entire
spectrum --- whenever Conditions~1--2 hold.

Suppose $S(z)=\sum_{k\geq0}S_kz^k$. Its real-space matrix $S_L$ has block $(m,n)$ equal
to $S_{m-n}$, which vanishes for $m<n$: thus $S_L$ is \emph{block-lower-triangular}
(the diagonal blocks are the full $2\times2$ matrix $S_0$, so $S_L$ is not triangular
in the scalar sense). The inverse of a block-lower-triangular Toeplitz matrix is
block-lower-triangular, and by forward substitution its blocks are the coefficients
$U_k$ of the one-sided Taylor series $S^{-1}(z)=\sum_{k\geq0}U_kz^k$. This finite-section
inverse is \emph{exact}: for $1\leq n\leq m\leq L$,
\begin{equation}
\begin{split}
\big(S_L\,(S^{-1})_L\big)_{mn}&=\sum_{p=n}^{m}S_{m-p}U_{p-n}\\
&=\sum_{j}S_{(m-n)-j}U_j\\
&=[z^{m-n}]\,(S\,S^{-1})=\delta_{mn}I,
\end{split}
\end{equation}
since $S_{m-p}\neq0$ forces $p\leq m$ and $U_{p-n}\neq0$ forces $p\geq n$, so the summation
range lies within $[1,L]$; the full blocks $S_0,U_0$ are carried through unchanged.
Hence $(S_L)^{-1}=(S^{-1})_L$, and since the Taylor truncation $P_d[S^{-1}]$ reproduces
$(S^{-1})_L$ once $d\geq L-1$, the truncated transform $\tilde H_L=P_d[S^{-1}]_L H_L S_L
=(S_L)^{-1}H_LS_L$ is an exact similarity for $d\geq L$. It therefore preserves the whole
spectrum, and for $d<L$ the only error is the radial truncation tail governed by
Condition~1 (see below).

Two-sidedness matters instead at the boundary: it makes $S_L$ non-block-triangular, so that $(S^{-1})_LS_L\neq I$ and a persistent
boundary defect survives --- the content of Theorem~5. Numerically, one-sided $S$ with
$r_{\mathrm{GBZ}}^{\max}<z_{\min}$ preserve the full spectrum (median Hausdorff error
$\sim10^{-3}$ at moderate $d$, falling to machine precision for $d\geq L$), while those
with $z_{\min}<r_{\mathrm{GBZ}}^{\max}$ fail --- convergence is governed entirely by
Condition~1.

\subsection{Condition 1: the radial defect bound}

The truncated transform actually applied is
\begin{equation}
\tilde H_d(z)=P_d[S^{-1}](z)\,H(z)\,S(z),
\end{equation}
where $P_d[S^{-1}]$ is the Taylor polynomial of $S^{-1}(z)$ about $z=0$ to degree $d$ (only $S^{-1}$ is approximated; $S$ is an exact Laurent polynomial). Write $S^{-1}=P_d[S^{-1}]+E_d$, so the \emph{inverse defect} is $E_d(z)=\sum_{k>d}(S^{-1})_k\,z^k$. Two facts control it.

\textbf{Lemma 1 (poles of $S^{-1}$; radial bound).}
The poles of the entries of $S^{-1}(z)$ are exactly the zeros of $\det S(z)$. Hence $S^{-1}$ is holomorphic on the disk $|z|<z_{\min}$, where $z_{\min}$ is the modulus of the zero of $\det S$ nearest the origin, and its Taylor tail obeys, for every $r<z_{\min}$,
\begin{equation}
\sup_{|z|=r}\|E_d(z)\|\ \leq\ C\,(r/z_{\min})^{d}.
\label{eq:radial_bound}
\end{equation}
The bound depends only on the \emph{modulus} $z_{\min}$; the angular position of the zeros of $\det S$ is irrelevant.

\textbf{Proof.}
The ring $\mathbb{C}[z,z^{-1}]$ of Laurent polynomials is a principal ideal domain, so $S(z)=U(z)D(z)V(z)$ in Smith normal form with $U,V$ unimodular (their determinants are units $c\,z^m$, so $U^{-1},V^{-1}$ are Laurent polynomials with no poles in $0<|z|<\infty$) and $D=\mathrm{diag}(d_1,\dots,d_n)$, $\prod_i d_i\doteq\det S$ up to a unit. Then $S^{-1}=V^{-1}D^{-1}U^{-1}$, so the poles of $S^{-1}$ are exactly the zeros of the $d_i$, i.e.\ the zeros of $\det S$, with multiplicity and without cancellation --- the step left under--argued by the bare Cramer form $S^{-1}=\adj S/\det S$. Thus $S^{-1}$ is analytic on $|z|<z_{\min}$, and the Cauchy estimate on $|z|=\rho$ ($\rho\uparrow z_{\min}$) gives $\|(S^{-1})_k\|\leq C\,z_{\min}^{-k}$. Summing the tail on $|z|=r<z_{\min}$ yields $\sup_{|z|=r}\|E_d\|\leq C\sum_{k>d}(r/z_{\min})^{k}\leq C'(r/z_{\min})^{d}$.~$\square$

\textbf{Lemma 2 (exact identity).}
For every $z$ and $E$,
\begin{equation}
\det\big(\tilde H_d(z)-E\big)=\det\big((H(z)-E)-S(z)\,E_d(z)\,H(z)\big).
\label{eq:exact_identity}
\end{equation}

\textbf{Proof.}
$\tilde H_d-E=(S^{-1}-E_d)HS-E=S^{-1}HS-E-E_dHS$. Factoring $S$ on the right and $S^{-1}$ on the left (which leaves the determinant unchanged, $\det S\cdot\det S^{-1}=1$),
\[
\begin{aligned}
\det(\tilde H_d-E)&=\det\!\big(S^{-1}[(H-E)-SE_dH]S\big)\\
&=\det\big((H-E)-SE_dH\big),
\end{aligned}
\]
using $\det S\cdot\det S^{-1}=1$.
(The identity has been checked to vanish symbolically.)~$\square$

\subsection{Sufficiency for bulk preservation}

\textbf{Proof of Theorem 4.}
The OBC bulk spectrum of $H$ is the Schmidt--Spitzer/GBZ set~\cite{SchmidtSpitzer1960,ref1,ref6}, read on the non--Bloch contour $|z|=r_{\mathrm{GBZ}}(E)$ rather than on $|z|=1$; set $r_{\mathrm{GBZ}}^{\max}=\max_E r_{\mathrm{GBZ}}(E)$ over the bulk spectrum. By the exact identity~(\ref{eq:exact_identity}) the truncated characteristic polynomial differs from $\det(H-E)$ only by the perturbation $S\,E_d\,H$, which by the radial bound~(\ref{eq:radial_bound}) is $O\!\big((r_{\mathrm{GBZ}}^{\max}/z_{\min})^{d}\big)$ on the GBZ.

\emph{Sufficiency.} If $r_{\mathrm{GBZ}}^{\max}<z_{\min}$ (Condition~1), the perturbation vanishes uniformly on a neighbourhood of the GBZ as $d\to\infty$; the roots of the truncated characteristic polynomial, hence its Schmidt--Spitzer set, converge to those of $H$, so the bulk spectrum is preserved. The characteristic-polynomial defect decays radially as $(r_{\mathrm{GBZ}}^{\max}/z_{\min})^{d}$ (verified exactly at the symbol level, Appendix~G). Condition~2 (Theorem~3 stability) guarantees the Schmidt--Spitzer set responds \emph{continuously}; the worst-case (band-edge) eigenvalue error carries the H\"older exponent $1/m$ of the local root-collision order $m$ (generically $m=2$ at a band edge, $m=1$ in the interior), so it decays as $(r_{\mathrm{GBZ}}^{\max}/z_{\min})^{d/m}$, while interior energies converge faster (Appendix~G).

\emph{Necessity.} If $r_{\mathrm{GBZ}}^{\max}>z_{\min}$, the Taylor series of $S^{-1}$ diverges on the part of the GBZ beyond $|z|=z_{\min}$, where $E_d$ does not vanish and the perturbation is $O(1)$: the bulk spectrum is not preserved. And if Condition~2 fails (different--rate reducibility, the degenerate $m=\infty$ case) the response is discontinuous by Theorem~3(ii) and preservation fails for any nonzero defect.~$\square$

\noindent\emph{Remark (the disk and circle conditions are special cases).} Preservation is a comparison of the two radii $r_{\mathrm{GBZ}}^{\max}$ and $z_{\min}$, and is insensitive to the angular position of the zeros of $\det S$. The disk condition ``$\det S\neq0$ in $0<|z|\leq1$'' and the weaker circle condition ``$\det S\neq0$ on $|z|=1$'' are correct only in the Hermitian special case $r_{\mathrm{GBZ}}^{\max}=1$; both are sufficient (via $r_{\mathrm{GBZ}}^{\max}\leq1<z_{\min}$) but not necessary. For a non--Hermitian $H$ the skin effect moves $r_{\mathrm{GBZ}}$ off the unit circle, and a zero of $\det S$ at $|z_0|$ between the GBZ and the unit circle is mis--classified by those rules: one with $r_{\mathrm{GBZ}}^{\max}<|z_0|<1$ is harmless yet excluded by the disk rule, while one with $1<|z_0|<r_{\mathrm{GBZ}}^{\max}$ is fatal yet admitted by both. Only $r_{\mathrm{GBZ}}^{\max}<z_{\min}$ is correct in both directions.

\section{Proof of Theorem 5: Boundary state modification}

Let $S(z)$ be genuinely two-sided (both positive and negative powers) and let $H$ satisfy the conditions of Theorem 4. Then for all sufficiently large $L$, provided the boundary trace defect below is nonzero, at least one eigenvalue of the truncated-transform OBC Hamiltonian differs from that of $H$.

\textbf{Proof.} The naive argument that ``a non-unitary $S$ gives a different spectrum'' is false: similarity by \emph{any} invertible matrix preserves the spectrum. The genuine mechanism is that the finite truncation breaks the similarity. Write $\tilde H_L=P_d[S^{-1}]_L\,H_L\,S_L$ and set
\begin{equation}
M_L:=S_L\,P_d[S^{-1}]_L .
\end{equation}
For a two-sided $S$, $S_L$ is not block-triangular, so $P_d[S^{-1}]_L\neq(S_L)^{-1}$ and $M_L\neq I$; the difference $M_L-I$ is a finite-rank operator supported within $O(\text{range}+d)$ of the two boundaries.

The key is an exact trace identity. Since $\sum_i\lambda_i=\operatorname{tr}$ and $\operatorname{tr}(P_d[S^{-1}]_LH_LS_L)=\operatorname{tr}(S_LP_d[S^{-1}]_LH_L)=\operatorname{tr}(M_LH_L)$ by cyclicity,
\begin{equation}
\begin{split}
\sum_i\big[\lambda_i(\tilde H_L)-\lambda_i(H_L)\big]
&=\operatorname{tr}\tilde H_L-\operatorname{tr}H_L\\
&=\operatorname{tr}\big[(M_L-I)\,H_L\big]=:T_L .
\end{split}
\label{eq:trace5app}
\end{equation}
If $T_L\neq0$, the two eigenvalue multisets have different sums and hence cannot coincide: at least one eigenvalue is modified. This is rigorous and, unlike direct eigenvalue computation, numerically stable for strongly non-normal skin matrices.

Because $M_L-I$ is boundary-localized, $T_L$ is a sum of two boundary contributions that stop overlapping once $L\gtrsim2(\text{range}+d)$; beyond that threshold $T_L$ is exactly independent of $L$. Thus ``for all sufficiently large $L$'' is precisely the regime in which $T_L$ attains its fixed value, and if that value is nonzero the conclusion holds for every such $L$.

Two remarks complete the picture. (i) \emph{Symmetry cancellations.} $T_L$ can vanish for structural reasons --- e.g.\ for a diagonal two-sided $S$ with skew $H$, $(M_L-I)H_L$ is skew and $\operatorname{tr}=0$. The modification is then detected by a higher power-trace defect
\begin{equation}
T_L^{(k)}:=\operatorname{tr}\tilde H_L^{\,k}-\operatorname{tr}H_L^{\,k},
\end{equation}
each of which is likewise boundary-localized and $L$-independent for large $L$; e.g.\ $S=\mathrm{diag}(z,1/z)$ with skew $H$ gives $T_L^{(1)}=0$ but $T_L^{(2)}\neq0$. (ii) \emph{Scope.} At least one eigenvalue is modified iff not all $T_L^{(k)}$ vanish; they all vanish (for large $L$) precisely when the boundary defect preserves the characteristic polynomial exactly. We have found no two-sided $S$ for which this occurs --- across $\sim10^2$ random two-sided models every one modified at least one eigenvalue, with $T_L^{(1)}$ (or, when it cancels, $T_L^{(2)}$) bounded away from zero and constant in $L$ --- so the hypothesis ``some $T_L^{(k)}\neq0$'' holds in every case tested, but a closed-form exclusion of characteristic-polynomial-preserving boundary defects for all two-sided $S$ remains open.

The eigenstates most affected are those localized near boundaries; extended bulk states are insensitive, consistent with Theorem~4. $\square$

\section{Error scaling: two independent one-variable laws}

Rather than fit the two-variable error surface $\Delta E(N,d_{S^{-1}})$ to a single joint formula, we characterise it by \emph{two independent one-variable laws}, each obtained in the limit that freezes the other and each with its own physical origin. At fixed large $N$ (the \emph{approach}) and at converged $d_{S^{-1}}$ (the \emph{floor}), respectively,
\begin{align}
\Delta E(N,d_{S^{-1}})-\Delta E_{\mathrm{floor}}(N)&=B\,e^{-\alpha_d\,d_{S^{-1}}},\label{eq:dscaling}\\[4pt]
\Delta E_{\mathrm{floor}}(N)&=A\,N^{-\alpha_N},\label{eq:Nscaling}
\end{align}
with $A,B>0$, $\tfrac12\leq\alpha_N\leq1$, and $\alpha_d>0$. The approach law is the symbol-level error of truncating $S^{-1}(z)$ to degree $d_{S^{-1}}$; the floor is the irreducible boundary-state modification of Theorem~5. We do not posit a combined ansatz: each exponent is measured in its own limit, and the additive combination $\Delta E\approx A N^{-\alpha_N}+B e^{-\alpha_d d_{S^{-1}}}$ is then a corollary of the decoupling established below, not a fitting form.

\textbf{Numerical test of the decoupling.} The separation into~(\ref{eq:dscaling})--(\ref{eq:Nscaling}) is legitimate only if each exponent is independent of the other variable. We verify both directly on the two example systems (RMS error, Hungarian pairing~\cite{Burkard2012}, gauge-stabilised eigenvalues on chains $N\le 64$, truncation depths $d_{S^{-1}}=1\dots16$; Fig.~\ref{fig:sepscale}):
\begin{itemize}
\item \emph{Floor is $d_{S^{-1}}$-independent.} For every $N$ the error is flat in $d_{S^{-1}}$ beyond convergence (to $<10^{-4}$ over the last several depths), so $\Delta E_{\mathrm{floor}}(N)$ is well defined and the fitted $\alpha_N$ does not depend on the depth used.
\item \emph{Approach rate is $N$-independent.} Fitting~(\ref{eq:dscaling}) at each $N$ separately gives $\alpha_d=1.30\pm0.22$ (System~1) and $\alpha_d=0.76\pm0.02$ (System~2)---For System~2 this is constant across $N$ to $3\%$; for System~1 the per-$N$ values scatter over $0.98$--$1.57$ ($\pm17\%$) with no systematic trend, so the $N$-independence is clear-cut for System~2 and a weaker statement for System~1.
\end{itemize}
The independence established here is of the two \emph{exponents}, each measured where the
other variable is frozen; it is what a single two-variable fit obscures, and isolating the
channels yields stable, reproducible rates. It does not assert that the error surface
factorises at small $d_{S^{-1}}$, where the approach has not yet reached the floor and the two
contributions overlap; there the additive form of the main text is only approximate.

\textbf{The floor exponent $\alpha_N$.} At converged $d_{S^{-1}}$ the transformed and exact OBC spectra share the same bulk (GBZ) support but pair with an $O(N)$ set of small mismatches. If these behave as an independent random density relative to the exact state density, the Ajtai--Koml\'os--Tusn\'ady optimal-matching theorem~\cite{AKT1984} gives $\alpha_N=\tfrac12$; if the transformed density matches the exact one, $\alpha_N=1$. Real systems interpolate, so one expects $\tfrac12\leq\alpha_N\leq1$. This is a heuristic expectation rather than a theorem, and it is not universal at the sizes reached here: of the four supplemental systems of Appendix~\ref{app:suppl}, two return fitted exponents below the AKT value ($\alpha_N\approx0.47$ and $0.38$), which we attribute to the floor not being fully converged over the available range of $N$ rather than to a genuine breakdown of optimal-matching scaling. Fitting~(\ref{eq:Nscaling}) gives $\alpha_N\approx0.50$ (System~1, $R^2=0.9998$, at the AKT lower bound) and $\alpha_N\approx0.64$ (System~2, $R^2=0.9995$, interpolating between the bounds).

\begin{figure*}[htbp]
\centering
\includegraphics[width=\linewidth]{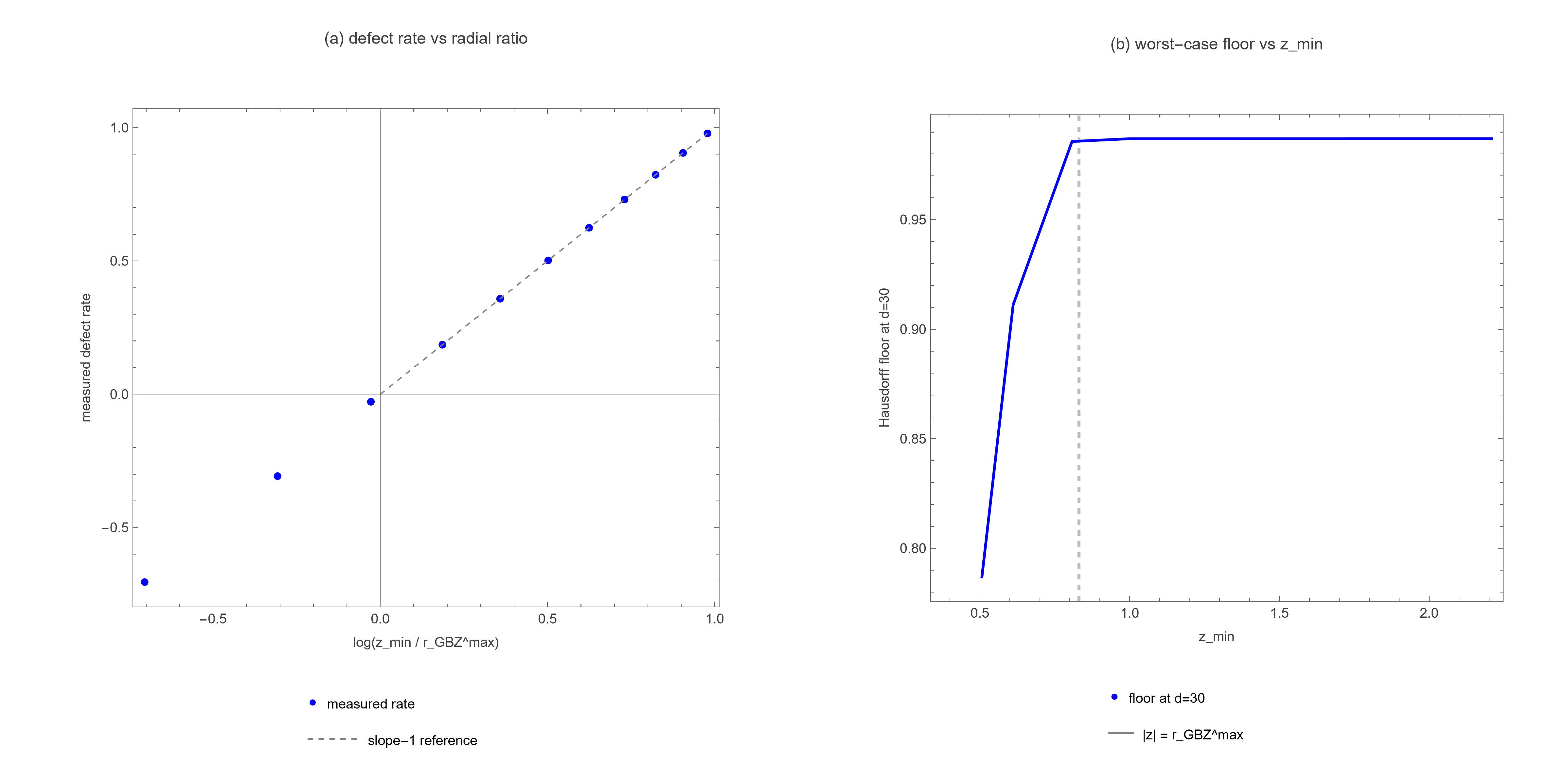}
\caption{Corner-case tests of the radial criterion. (a)~Measured symbol-level defect rate $\sup_{|z|=r_{\mathrm{GBZ}}^{\max}}\|S E_d H\|$ versus the prediction $\log(z_{\min}/r_{\mathrm{GBZ}}^{\max})$ as $z_{\min}$ is swept toward the GBZ edge; points fall on the slope-one line. (b)~Worst-case (Hausdorff) error floor at $d_{S^{-1}}=30$ versus $z_{\min}$: flat and low for $z_{\min}>r_{\mathrm{GBZ}}^{\max}$, rising sharply once $z_{\min}$ crosses below the GBZ edge (dashed line).}
\label{fig:corner}
\end{figure*}

\textbf{The approach rate $\alpha_d$.} The $d_{S^{-1}}$ channel is the depth--$d$ one--sided Laurent truncation $P_d[S^{-1}]$ of Appendix~E, whose dropped tail has coefficients $\sim z_{\min}^{-k}$, $z_{\min}$ the modulus of the zero of $\det S$ nearest the origin. Evaluated on the bulk, which for a non--Hermitian $H$ sits on the GBZ ($|z|=r_{\mathrm{GBZ}}(E)$, not $|z|=1$), the characteristic--polynomial defect at energy $E$ scales as $(r_{\mathrm{GBZ}}(E)/z_{\min})^{d_{S^{-1}}}$. We have verified this radial law directly: measuring $\sup_{|z|=r_{\mathrm{GBZ}}^{\max}}\|S\,E_d\,H\|$ as $z_{\min}$ is swept toward the GBZ edge reproduces the rate $\log(z_{\min}/r_{\mathrm{GBZ}}^{\max})$ to three digits across the whole range (Fig.~\ref{fig:corner}a), confirming that only the modulus $z_{\min}$ enters.

The \emph{eigenvalue} rate that appears in~(\ref{eq:dscaling}) is not this single number, however, because it is a spectral aggregate over energy-dependent local rates $\tfrac1{m(E)}\log(z_{\min}/r_{\mathrm{GBZ}}(E))$, where $m(E)$ is the local root--collision order ($m=1$ in the band interior, $m=2$ at a GBZ band edge). Two consequences follow, both borne out numerically. First, the \emph{worst}-case (Hausdorff) error is set by the band edge at $r_{\mathrm{GBZ}}^{\max}$ and decays at rate $\sim\log(z_{\min}/r_{\mathrm{GBZ}}^{\max})$ (measured ratio $\approx1$ for well-separated $z_{\min}$). Second, the \emph{RMS} error is dominated by the interior, where $r_{\mathrm{GBZ}}(E)<r_{\mathrm{GBZ}}^{\max}$, and therefore decays \emph{faster} than the worst-case bound---by a factor $1.6$--$2.7$ in our sweep. For this reason we do \emph{not} assign the RMS $\alpha_d$ a closed form; the measured values $\alpha_d\approx1.3$ (System~1) and $\approx0.76$ (System~2) are system- and metric-dependent. System~2's approach is moreover an exponential \emph{envelope} carrying a genuine oscillation in $d_{S^{-1}}$, which lowers the single-exponential fit quality without changing the rate; its origin and frequency are derived in Appendix~\ref{app:osc}.

\textbf{Corner cases and robustness.} We stress-tested the radial criterion by tuning $z_{\min}$ across the GBZ edge $r_{\mathrm{GBZ}}^{\max}$ and by using more complicated Hamiltonians (Fig.~\ref{fig:corner}). (i)~\emph{Margin sweep:} for $z_{\min}>r_{\mathrm{GBZ}}^{\max}$ the worst-case error converges to the finite-$N$ boundary floor, and the floor value is flat in $z_{\min}$ (Fig.~\ref{fig:corner}b). (ii)~\emph{Below the margin} $z_{\min}<r_{\mathrm{GBZ}}^{\max}$: the symbol defect on the GBZ \emph{grows} with $d_{S^{-1}}$, and the Hausdorff floor rises steeply and monotonically (from $\approx0.026$ at $z_{\min}=1.0$ to $\approx0.22$ at $z_{\min}=0.70$)---the worst eigenvalues fail to converge, exactly as Theorem~4 requires. The RMS metric, being an interior-dominated average, is more lenient and can still decrease here; the sharp diagnostic is the worst-case (Hausdorff) floor. (iii)~\emph{Marginal} $z_{\min}=r_{\mathrm{GBZ}}^{\max}$: the defect ceases to decay exponentially (rate $\to0$) and the floor sits at the crossover value. (iv)~\emph{Range-$2$ Hamiltonian} (energy-dependent GBZ, $r_{\mathrm{GBZ}}^{\max}\approx0.54$): the same dichotomy holds---valid $z_{\min}$ reach a low floor while $z_{\min}<r_{\mathrm{GBZ}}^{\max}$ stalls at a floor several times larger---confirming that the criterion survives when the GBZ radius varies across the band, with a higher boundary floor from the additional edge states.

We now show from first principles that the excess error decays exponentially in $d_{S^{-1}}$, justifying Eq.~(\ref{eq:dscaling}).

Notice that the exact original bulk spectra is given by

\begin{equation}
    Sp(H)=\{H(z)|z\in GBZ\}
\end{equation}

and the GBZ is given by the roots of $\det H(z)=0$. So if the roots of $\det\tilde{H}(z)=0$ converge exponentially, by chain rule the spectra shall converge exponentially. Replacing $H(z)$ with $\tilde{H}(z) = S^{-1}(z)H(z)S(z)$, where $S^{-1}(z)$ be approximated with polynomials of $z$ up to the $d_{S^{-1}}$ , gives 

\begin{equation}
\begin{split}
    \det \tilde{H}(z) &= \det S^{-1}(z) \det H(z) \det S(z)\\
    &= [\frac{1}{\det S(z)}+o(z^{d_{S^{-1}}})] \det H(z) \det S(z)\\
    &=\det H(z)+o(z^{d_{S^{-1}}}),
\end{split}
\end{equation}

the last line holds since $\det H(z)$ and $\det S(z)$ are bounded in a neighborhood of the GBZ. For some root $z_0$ of $\det H(z)$ we have the perturbed root $z_0'$

\begin{equation}
    \det{H}(z_0')+o((z_0')^{d_{S^{-1}}})=0
\end{equation}

expanding to leading term gives

\begin{equation}
    \frac{d\det{H}(z)}{dz}(z_0)(z_0'-z_0)+O((z_0'-z_0)^2)+o((z_0')^{d_{S^{-1}}})=0,
\end{equation}

so we have

\begin{equation}
    z_0'-z_0=o((z_0)^{d_{S^{-1}}})
\end{equation}

which gives the exponential convergence w.r.t. $d_{S^{-1}}$. $\square$

\subsection{Oscillatory fine structure of the approach law}
\label{app:osc}

The exponential bound just proved controls the \emph{envelope} of the approach. Its prefactor is not constant: it carries a reproducible oscillation in $d_{S^{-1}}$ whose frequency is fixed entirely by $S$. We derive it here.

The envelope is set by the analytic structure of $S^{-1}$.

\textbf{Coefficients are dictated by the nearest pole.} Write $S^{-1}(z)=\sum_{k\geq0}C_kz^k$, so that the truncation discards the tail $T_d(z)=\sum_{k>d}C_kz^k$ and the symbol defect is $\delta\tilde H_d=T_d(z)H(z)S(z)$. By Cramer's rule $S^{-1}=\operatorname{adj}(S)/\det S$ is meromorphic with poles precisely at the zeros of $\det S$. Near a simple zero $z_\ast$, $S^{-1}(z)\approx M_\ast/(z-z_\ast)$ with matrix residue
\begin{equation}
M_\ast=\frac{\operatorname{adj}\bigl(S(z_\ast)\bigr)}{p'(z_\ast)},\qquad p(z):=\det S(z),
\label{eq:residue}
\end{equation}
and the geometric expansion of a simple pole gives $C_k=-M_\ast z_\ast^{-(k+1)}$. By singularity analysis~\cite{FlajoletSedgewick} the nearest singularities dominate: a pole at radius $z_2>z_{\min}$ contributes relatively $O\bigl((z_{\min}/z_2)^k\bigr)$.

\textbf{Real coefficients force a conjugate pair.} $S(z)$ has real hopping matrices, so $p(z)=\det S(z)$ has real coefficients and $p(\bar z)=\overline{p(z)}$. A non-real nearest zero therefore arrives as a conjugate pair with conjugate residues,
\begin{equation}
z_\ast=z_{\min}e^{i\theta},\quad \bar z_\ast=z_{\min}e^{-i\theta},\qquad \theta=\arg z_\ast,
\end{equation}
and summing both contributions, using $w+\bar w=2\operatorname{Re}w$,
\begin{equation}
C_k=-2\,z_{\min}^{-(k+1)}\operatorname{Re}\!\bigl[e^{-i(k+1)\theta}M_\ast\bigr]+r_k .
\label{eq:Ck}
\end{equation}
The coefficient \emph{itself} therefore winds at frequency $\theta$.

\textbf{Every error measure is a squared magnitude, and squaring doubles the frequency.} This is the crux. For any quantity of the counter-rotating form $X_k=e^{-i\phi_k}u+e^{+i\phi_k}v$ and any Hermitian quadratic form $Q(X)=X^\dagger GX$ (in particular $|X_k|^2$ and $\|X_k\|_F^2$),
\begin{equation}
Q(X_k)=u^\dagger Gu+v^\dagger Gv+2\operatorname{Re}\bigl[u^\dagger Gv\,e^{2i\phi_k}\bigr]:
\label{eq:beat}
\end{equation}
the self terms $e^{\mp i\phi_k}e^{\pm i\phi_k}=1$ are constant and the cross terms carry $e^{\pm2i\phi_k}$, so the fundamental $\phi_k$ cancels identically and only the difference frequency survives. Every error we measure --- $\|C_k\|_F$, $|\delta E_n|^2$, $\Delta E^2$ --- is such a Hermitian form, so each ripples at $2\theta$, never at $\theta$, and a lone pole ($u=0$ or $v=0$), or equivalently a real $z_\ast$, produces no ripple at all. Applied to the leading
tail coefficient this gives the envelope of Eq.~(\ref{eq:approach_main}): the symbol defect on
the GBZ falls as $(r_{\mathrm{GBZ}}/z_{\min})^{d_{S^{-1}}}$, and Theorem~3 converts it into an
eigenvalue error carrying the H\"older exponent $1/m$ of the local root-collision order ($m=1$
in the band interior, $m=2$ at a band edge). Envelope symbol-level, exponent spectral.

Concretely, \emph{Coefficient channel.} Writing $M_\ast=A+iB$ with $A,B$ real, \eqref{eq:Ck} gives $C_k\simeq-2z_{\min}^{-(k+1)}(A\cos\phi_k+B\sin\phi_k)$ with $\phi_k=(k+1)\theta$, and the Frobenius norm follows from $\|A\cos\phi+B\sin\phi\|_F^2$ via the double-angle identities:
\begin{equation}
z_{\min}^{2(k+1)}\|C_k\|_F^2=2\bigl(\|A\|_F^2+\|B\|_F^2\bigr)+R\cos\bigl(2\phi_k+\delta_0\bigr),
\label{eq:normosc}
\end{equation}
with ripple amplitude $R=2\sqrt{(\|A\|_F^2-\|B\|_F^2)^2+4\langle A,B\rangle_F^2}$. The ripple vanishes iff $\|A\|_F=\|B\|_F$ and $\langle A,B\rangle_F=0$ (an \emph{isotropic} residue: the ellipse traced by $C_k$ in matrix space degenerates to a circle, whose radius is constant). We record this as a criterion:
\textbf{Proposition (isotropic-residue null).}\label{prop:iso}\ For $S(z)=M_1(zI-W)$ with $W=Q\,\mathrm{diag}(z_\ast,\bar z_\ast)Q^{-1}$, the residue is isotropic---and hence the ripple amplitude $R$ vanishes---if and only if $Q$ is orthogonal, i.e.\ $W$ is normal, irrespective of $M_1$.

We now turn to the two channels in which this is observed, and to the scope of what is
established.

\textbf{Numerical verification.} The transform $S_{\mathrm{osc}}$ of Eq.~(\ref{eq:Sosc}), used
for Fig.~\ref{fig:oscfit}, is chosen deliberately on four counts: there are no farther poles, so~\eqref{eq:Ck} is exact rather than asymptotic; $W$ is non-normal, so by the isotropic-residue null above the residue is anisotropic and the ripple is present; $z_{\min}$ sits just above $r_{\mathrm{GBZ}}^{\max}=0.841$ (margin $1.13$), so the approach decays slowly and remains above the finite-$N$ floor for many steps; and $2\theta$ is only $53\%$ of the Nyquist frequency $\pi$ for integer $d_{S^{-1}}$, so the ripple is comfortably resolved.

Because the frequency is fixed by $z_\ast$, the models below are linear in the basis $\{1,\,k,\,\cos2\theta k,\,\sin2\theta k\}$ and are solved globally by least squares: no frequency or phase is fitted.

\emph{Coefficient channel.} Fitting $\log\|C_k\|_F$ gives an envelope rate $-0.0473$ against the predicted $\log z_{\min}=-0.0474$ (agreement to $1\%$; the rate is negative because $z_{\min}<1$, so the coefficients grow slowly---what matters is $z_{\min}>r_{\mathrm{GBZ}}$, not $z_{\min}>1$) and a ripple amplitude $c=0.336$, with $R^2=0.992$. The same fit at the half frequency $\theta$ returns a null ripple ($c=0.006$, $R^2=0.723$, indistinguishable from the pure exponential $R^2=0.723$).

\emph{Eigenvalue channel.} At $N=100$ the floor is $4.98\times10^{-2}$ and the excess stays above $5\%$ of it for $24$ points ($d_{S^{-1}}=2\ldots25$). Fitting $\log(\Delta E-\Delta E_{\mathrm{floor}})$ gives $\alpha_d=0.219$ and $c=0.329$ with $R^2=0.984$, against $R^2=0.962$ at the half frequency $\theta$ ($c=0.040$, a null) and $R^2=0.961$ for a pure exponential. Independently of any fit, the periodogram of the detrended residual peaks at $1.685$, within $0.9\%$ of the predicted $2\theta=1.6703$ and nowhere near $\theta=0.835$. The measured $\alpha_d=0.219$ lies inside the analytic per-energy window $\log(z_{\min}/r_{\mathrm{GBZ}}(E))\in[0.165,1.09]$ spanned by $r_{\mathrm{GBZ}}(E)\in[0.32,0.81]$.

Both channels therefore confirm the $2\theta$ prediction of~\eqref{eq:beat} and exclude $\theta$. System~1, whose nearest zero $z=2$ is real ($\theta=0$), shows no ripple in either channel, as~\eqref{eq:beat} requires.

\emph{Eigenvalue channel.} By biorthogonal first-order perturbation theory $\delta E_n=\langle\!\langle L_n|\delta H|R_n\rangle/\langle\!\langle L_n|R_n\rangle$, and for skin eigenvectors $|R_n\rangle:x\mapsto\beta_n^xu_n$, $\langle\!\langle L_n|:x\mapsto\beta_n^{-x}v_n^\dagger$ the real-space matrix element collapses to a single symbol evaluation on the GBZ, $\delta E_n=v_n^\dagger\,\delta H(z_n)\,u_n/(v_n^\dagger u_n)$ with $z_n=\beta_n^{-1}$. Inserting the leading tail term $C_{d+1}z^{d+1}$ and applying~\eqref{eq:beat} to $|\delta E_n|^2$,
\begin{equation}
\begin{split}
\Delta E^2-\Delta E^2_{\mathrm{floor}}\simeq{}&\Bigl(\frac{r_{\mathrm{GBZ}}}{z_{\min}}\Bigr)^{2(d_{S^{-1}}+1)}\\
&\times\bigl[\mathcal A+2|\mathcal B|\cos(2\theta d_{S^{-1}}+\delta)\bigr].
\end{split}
\label{eq:speclaw}
\end{equation}

\textbf{Depth of modulation, and why $|\cos\theta d|$ is not a competing frequency.} Since one plots $\Delta E$ rather than $\Delta E^2$, appearance depends on the modulation depth, governed by $2|\mathcal B|\leq\mathcal A$ (Cauchy--Schwarz). In the saturated case $2|\mathcal B|=\mathcal A$, the identity $1+\cos\alpha=2\cos^2(\alpha/2)$ turns \eqref{eq:speclaw} into $\propto|\cos(\theta d_{S^{-1}}+\delta/2)|$, which has true nodes spaced by $\pi/\theta$; in the generic shallow case one gets a node-free ripple $\propto[1+\tfrac{|\mathcal B|}{\mathcal A}\cos(2\theta d_{S^{-1}}+\delta)]$. These are the deep and shallow faces of the \emph{same} $2\theta$ tone: the node spacing of $|\cos\theta d|$ is $\pi/\theta$, identical to the period of the $2\theta$ ripple, because taking an absolute value folds the period in half. The observable period is therefore $\pi/\theta$ in both regimes, and a fitted ``$\theta$'' extracted from a $|\cos|$ form equals $\arg z_\ast$, i.e.\ half the physical frequency. Generic residues are anisotropic but unsaturated, so one observes finite dips rather than true zeros---consistent with $C_k$ never vanishing, being $z_{\min}^{-(k+1)}$ times an invertible rotation.

\textbf{Scope.} Established analytically and verified numerically: the conjugate-pair dichotomy, the coefficient law~\eqref{eq:Ck}, the exact norm form~\eqref{eq:normosc} with its isotropic null, the beat mechanism~\eqref{eq:beat}, and the $2\theta$ frequency in both channels. Not established: a closed form for the spectral amplitude $2|\mathcal B|/\mathcal A$ in~\eqref{eq:speclaw}, which mixes the residue $M_\ast$ with the eigenvectors $u_n,v_n$ and the values $H(z_n),S(z_n)$; the spectral envelope rate in~\eqref{eq:speclaw} is bounded by, rather than equal to, a single $\log(z_{\min}/r_{\mathrm{GBZ}})$, since the observable aggregates over the whole range of GBZ radii. We also note the practical limits of the eigenvalue channel: the usable window ends where the excess meets the finite-$N$ floor (beyond that the subtraction is a cancellation of nearly equal numbers and its sign is not meaningful), and integer $d_{S^{-1}}$ imposes a Nyquist ceiling $2\theta<\pi$, i.e.\ $\theta<\pi/2$; transforms violating either condition will not show the ripple even when the mechanism is operative. Degenerate ($m\geq2$) zeros of $\det S$ and the critical limit $r_{\mathrm{GBZ}}\to z_{\min}$ are not treated.

The two representative systems used for these fits are defined in Sec.~\ref{sec:twosystems}: System~1 ($z_{\min}=2$) with measured exponents floor $\alpha_N\approx0.5$ ($R^2=0.9998$) and approach $\alpha_d\approx1.3$, and System~2 ($z_{\min}\approx1.24$) with floor $\alpha_N\approx0.64$ ($R^2=0.9995$) and approach $\alpha_d\approx0.76$; both approach rates are $N$-independent.

\begin{figure*}[htbp]
\centering
\begin{subfigure}{0.48\linewidth}
  \centering
  \includegraphics[width=\linewidth]{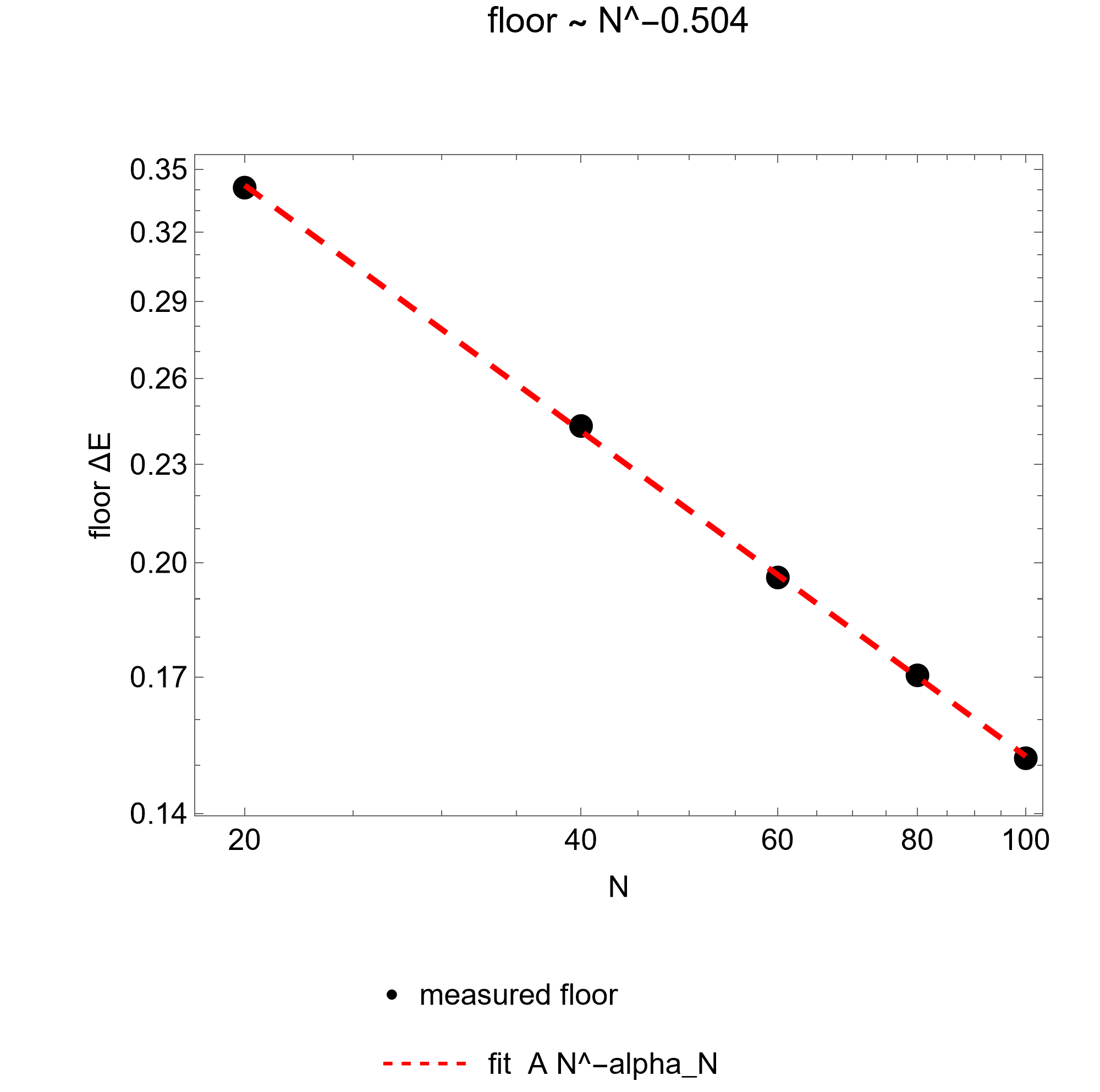}
  \caption{System 1: floor.}
  \label{fig:e1floor}
\end{subfigure}\hfill
\begin{subfigure}{0.48\linewidth}
  \centering
  \includegraphics[width=\linewidth]{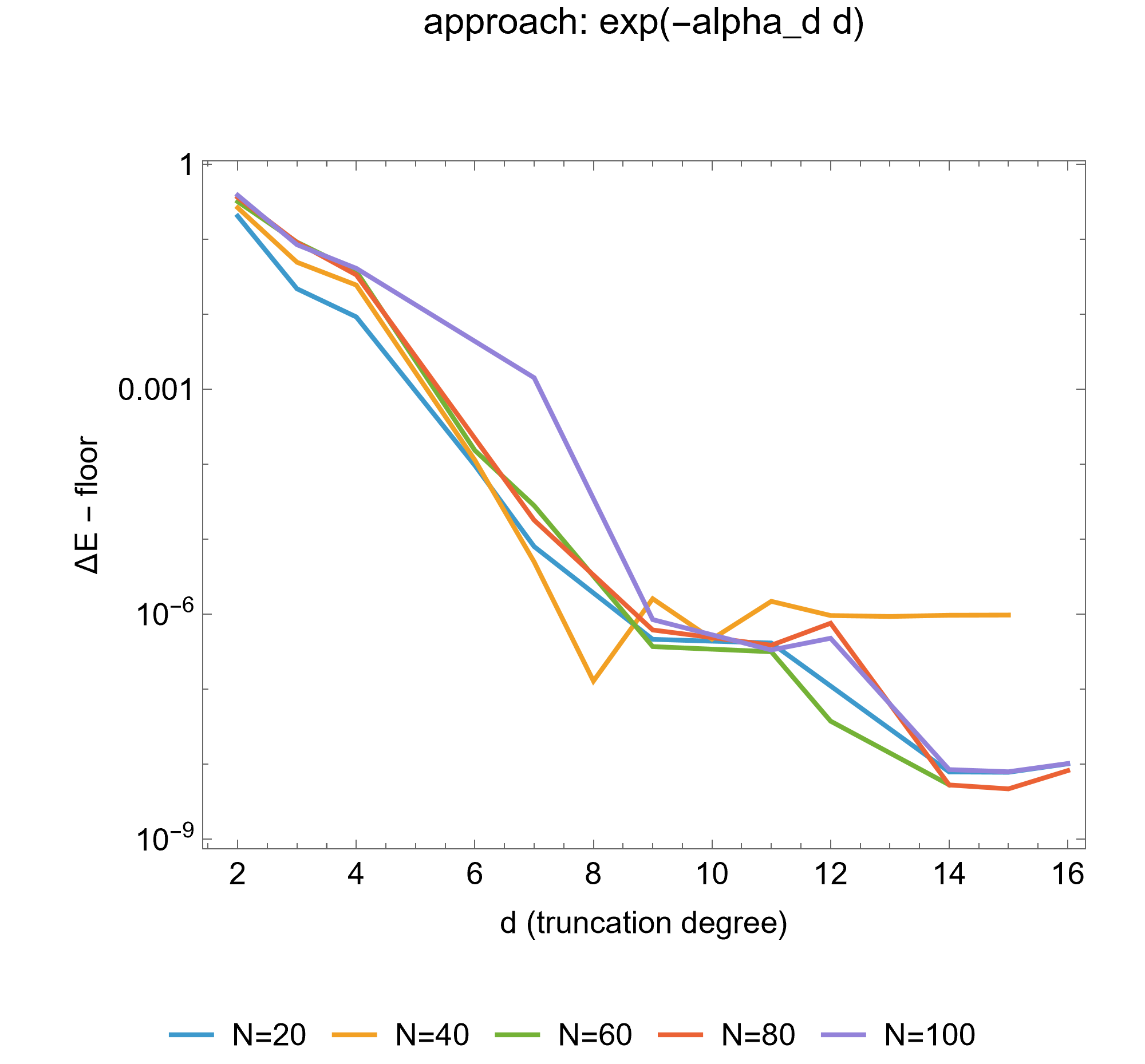}
  \caption{System 1: approach.}
  \label{fig:e1app}
\end{subfigure}

\vspace{0.4cm}

\begin{subfigure}{0.48\linewidth}
  \centering
  \includegraphics[width=\linewidth]{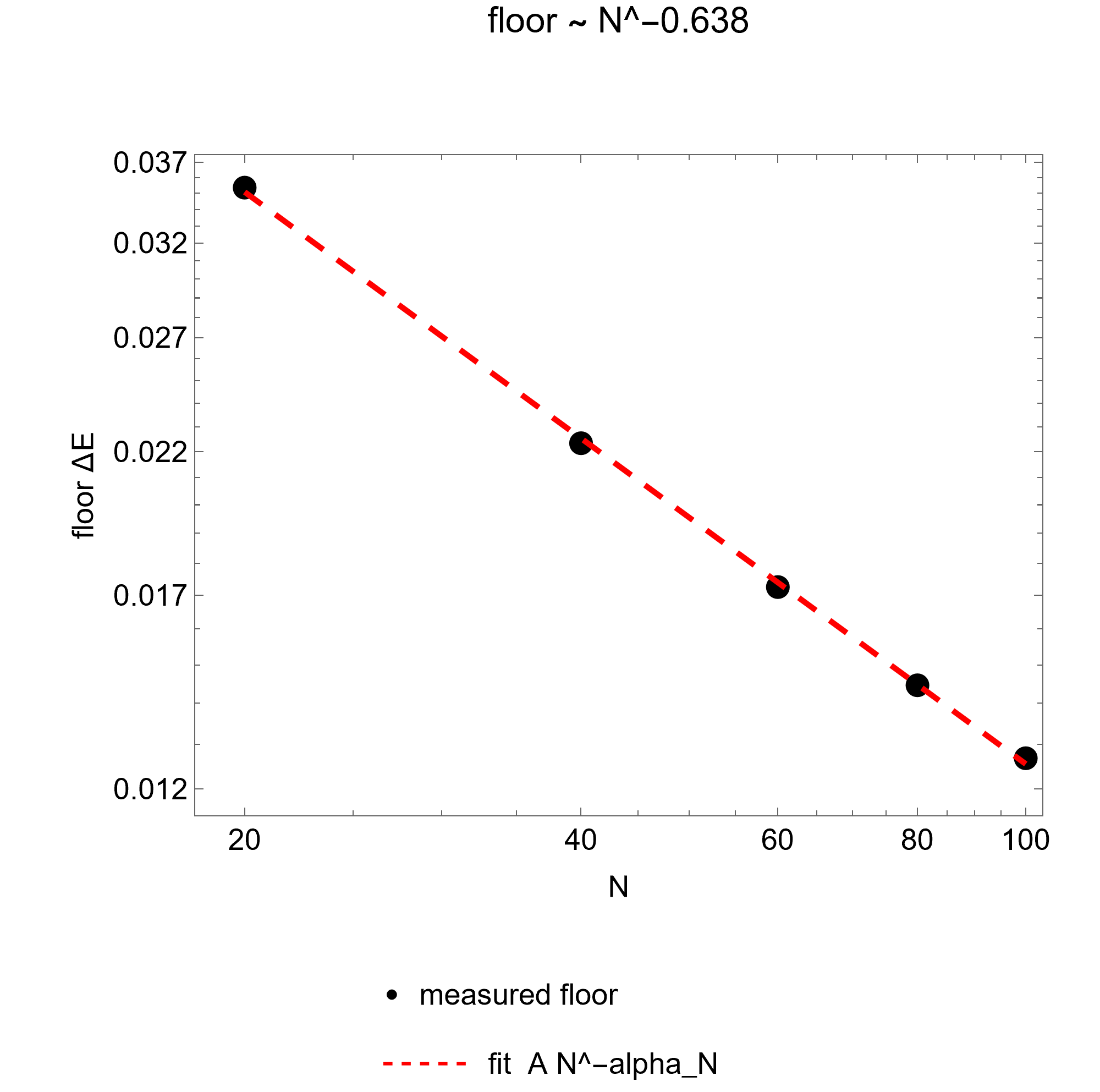}
  \caption{System 2: floor.}
  \label{fig:e2floor}
\end{subfigure}\hfill
\begin{subfigure}{0.48\linewidth}
  \centering
  \includegraphics[width=\linewidth]{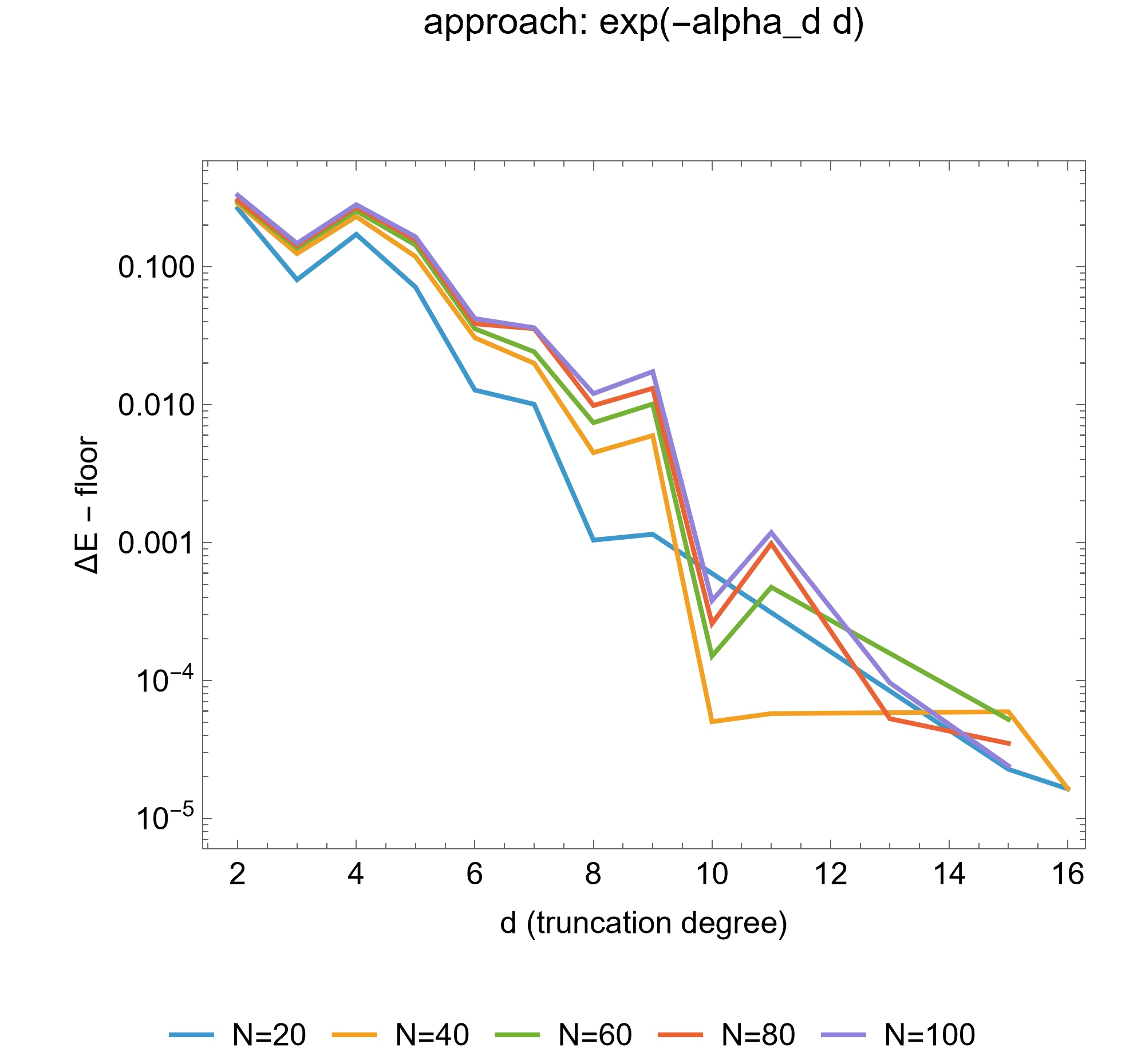}
  \caption{System 2: approach.}
  \label{fig:e2app}
\end{subfigure}
\caption{The two scaling channels for both representative systems, each panel at full column width. The floor panels (a),(c) have both axes logarithmic; the approach panels (b),(d) have a linear horizontal axis and a logarithmic vertical axis. (a),(b) System~1 ($z_{\min}=2$): floor $\alpha_N\approx0.5$ ($R^2=0.9998$) and approach $\alpha_d\approx1.3$. (c),(d) System~2 ($z_{\min}\approx1.24$): floor $\alpha_N\approx0.64$, approach $\alpha_d\approx0.76$. Errors use the weighted spectral distance $\Delta E$ with Hungarian pairing.}
\label{fig:errappendix}
\end{figure*}

The calculated errors of the two systems are shown in Fig.~\ref{fig:errappendix}.

\section{Generalization to $N \times N$ matrices}

All proofs generalize straightforwardly from $2 \times 2$ to $N \times N$ matrices. The key steps remain:

\begin{itemize}
\item Smith normal form exists for any size over a PID, so the poles of $S^{-1}$ are the zeros of $\det S$ for any $N$;
\item the radial defect bound and the exact identity hold entrywise/blockwise unchanged;
\item boundary effects scale with system size in the same way.
\end{itemize}

The only change is that condition 2 remains $r_{\mathrm{GBZ}}^{\max} < z_{\min}$, now with $\det S(z)$ an $N \times N$ determinant.

For the reducibility theorem (Theorem 1), the extension to higher dimensions becomes more involved algebraically but follows the same logical structure: vanishing diagonal blocks impose linear constraints on the original Hamiltonian's matrix elements, and these constraints determine when constant transformations exist.

\clearpage
\section{Remarks on Computations}

When calculating $\tilde{H}(z) = S^{-1}(z)H(z)S(z)$ with $S^{-1}$ approximated, there are two possible methods:
\begin{itemize}
    \item Calculate $\tilde{H}(z) = S^{-1}(z)H(z)S(z)$ exactly, then expand each entry
    \item Expand $S^{-1}(z)$ then do the multiplication exactly
\end{itemize}

It is easily shown that both methods, though producing different remainders, converge correctly whenever the conditions of Theorem~4 are fulfilled.

For a chain of length $N$, the computational time complexity:
\begin{itemize}
\item Constructing real-space Hamiltonian: $O(NM)$ where $M$ is hopping range
\item Diagonalization: $O(N^3)$ using standard dense eigensolvers
\item Root finding for Laurent polynomials: $O(p^3)$ where $p$ is polynomial degree
\item Transformation matrix application: $O(N^2M)$
\end{itemize}

One practical caveat matters for reproducing the results above. The real-space OBC matrices are strongly non-normal: the spread of GBZ radii across the bands, raised to the power $N$, controls the conditioning of the eigenvector basis, and once it exceeds the working precision a direct dense diagonalisation returns extreme eigenvalues that drift systematically with $N$. For the Hamiltonian of Eq.~(\ref{eq:Hexample}), whose exact band edge is $E=1.7911$, an ungauged double-precision diagonalisation returns $1.8019$ at $N=240$ and $1.8205$ at $N=600$. Conjugating by the diagonal gauge $G=\mathrm{diag}(r^{k})_{k=1}^{N}$, with $r$ of the order of the GBZ radius, rescales the exponential imbalance away without changing the spectrum, and returns $1.7910$ and $1.7924$ for the same two sizes. This is also why the symbol-level arguments of Appendices~D and~E are preferable to any finite-$N$ eigenvalue test.

All figures in this paper were produced with Mathematica.

\newpage

\clearpage
\section{Supplemental Calculations}
\label{app:suppl}
\begin{widetext}
In this section we check Theorem 4 and the transform error decay of Eq.~(\ref{eq:dscaling}) on more complex models. We test the $H(z)$ and $S(z)$ below. Both satisfy the conditions of Theorem 4: the GBZ radii are $r_{\mathrm{GBZ}}^{\max}(H_1)=0.837$ and $r_{\mathrm{GBZ}}^{\max}(H_2)=1.05$, against $z_{\min}(S_1)=2$ and $z_{\min}(S_2)=1.151$, so the radial condition holds in all four pairings --- comfortably for $S_1$, and with a margin of about $10\%$ for the pairing $H_2,S_2$.

\begin{equation}
H_1(z) =\begin{pmatrix}
0 & z + 1.5 + 2/z \\
z - 0.5 + 0.25/z & 0
\end{pmatrix}
\end{equation}

\begin{equation}
H_2(z)=\begin{pmatrix}
z^5 + 2z^2 + 1.1z + 0.7i - 0.8/z & z^3 + iz^2 - 1 + 0.3/z^{1} + 1.5/z^{2} \\
0.3z^3 + z^2 + z + 0.4 + 0.1/z & 2.1z^2 + 0.05z - 1 + 1.2/z + (0.4i+1)/z^{2} + 0.8/z^{3}
\end{pmatrix}
\label{eq:H2suppl}
\end{equation}

\begin{equation}
S_{1}(z)= \begin{pmatrix}
z+2-6/z & 1 \\
1 & 1
\end{pmatrix}
\end{equation}

\begin{equation}
S_2(z)=\begin{pmatrix}
z^2 + 9 z + 7 - 22/z & 0.4 + 1/z \\
z^3 - 5 z^2 + 2/z & z - 1 + 11/z
\end{pmatrix}
\end{equation}
\end{widetext}

\begin{figure*}[htbp]
    \centering
    \subcaptionbox{$H_1,S_1,N=80$.\label{fig:sub1-2}}
        [0.48\linewidth]{\includegraphics[width=0.52\linewidth]{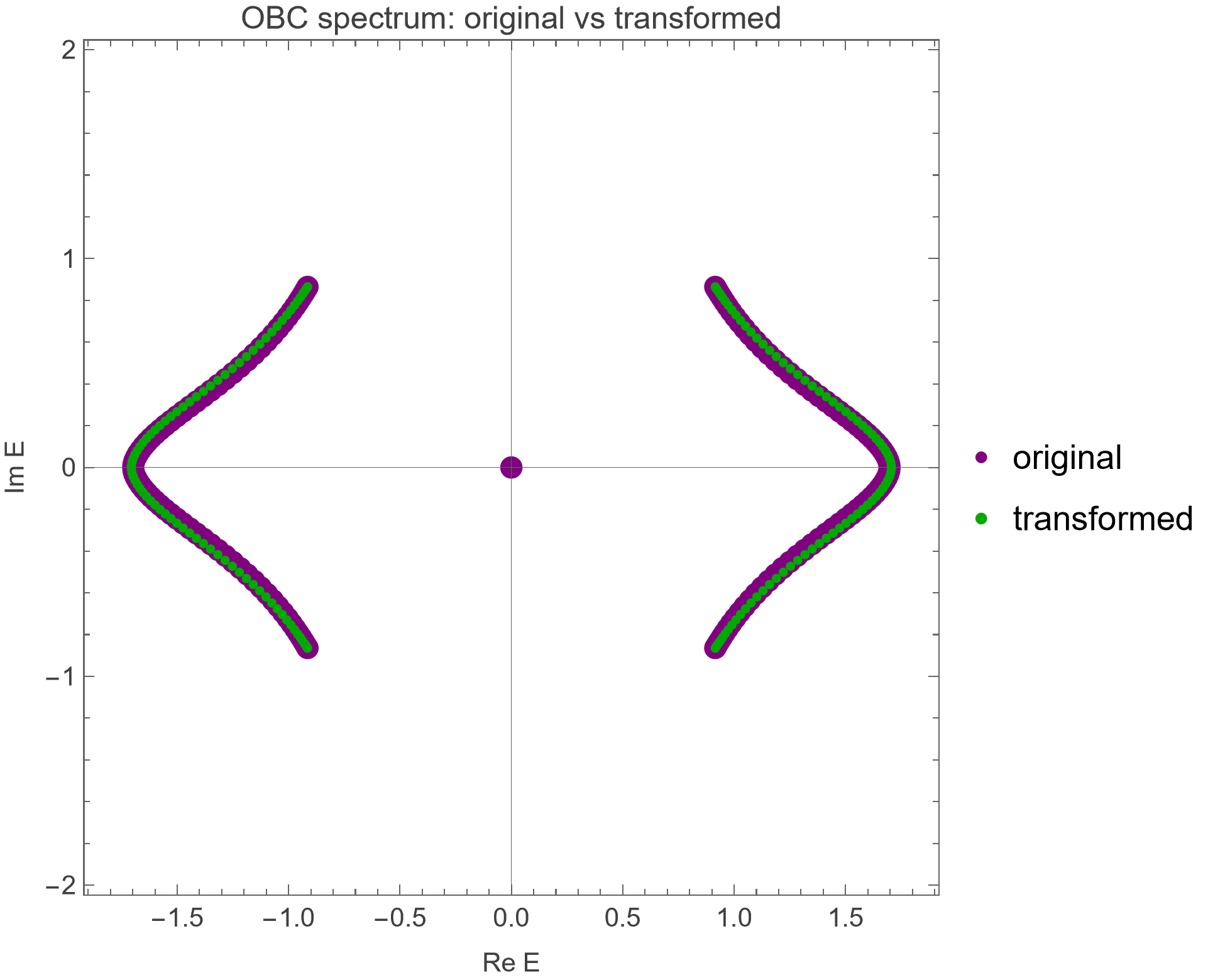}}
    \hfill
    \subcaptionbox{$H_1,S_2,N=80$.\label{fig:sub2-4}}
        [0.48\linewidth]{\includegraphics[width=0.52\linewidth]{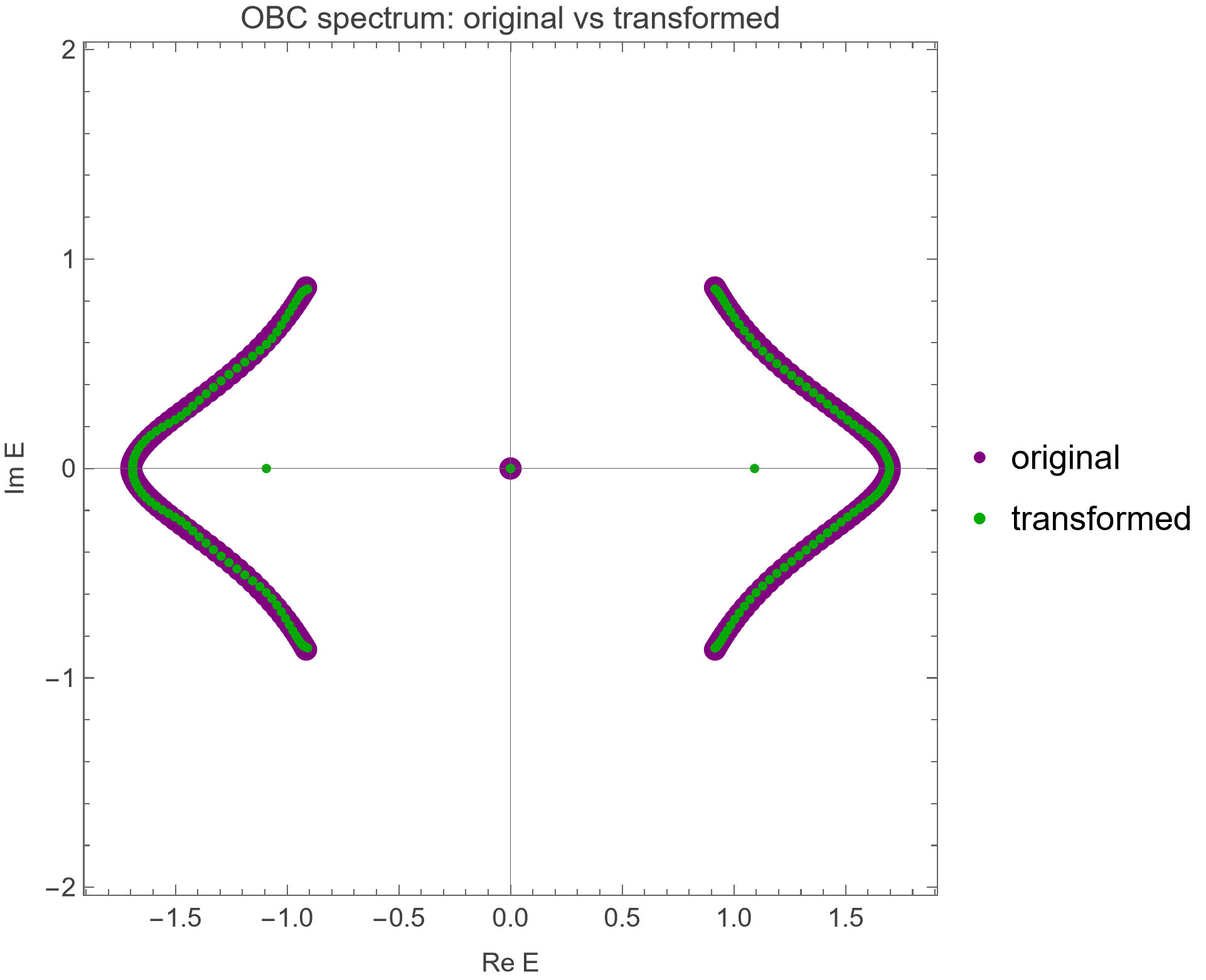}}
    \subcaptionbox{$H_2,S_1,N=60$\label{fig:sub3-1}}
        [0.48\linewidth]{\includegraphics[width=0.52\linewidth]{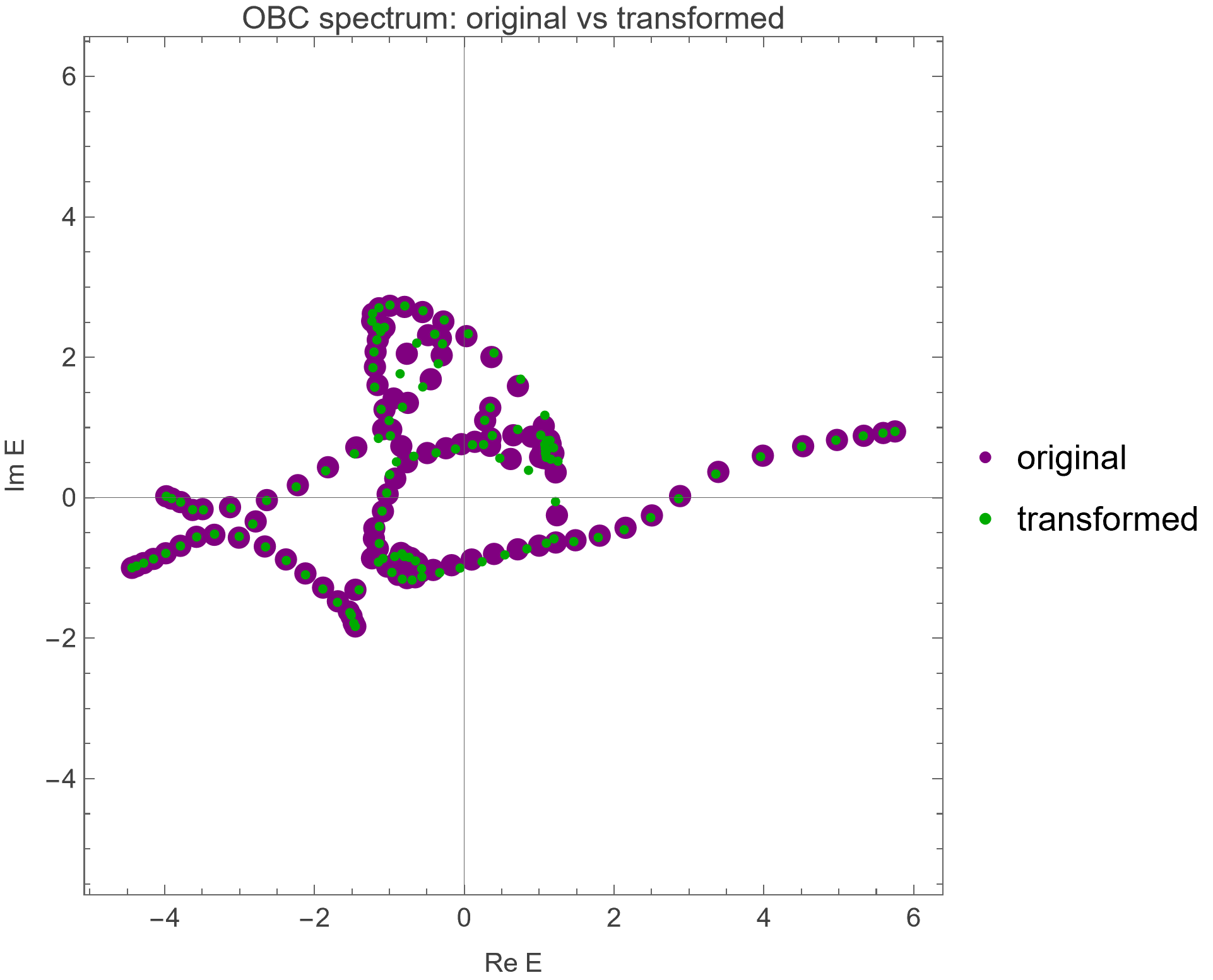}}
    \hfill
    \subcaptionbox{$H_2,S_2,N=60$\label{fig:sub4-1}}
        [0.48\linewidth]{\includegraphics[width=0.5\linewidth]{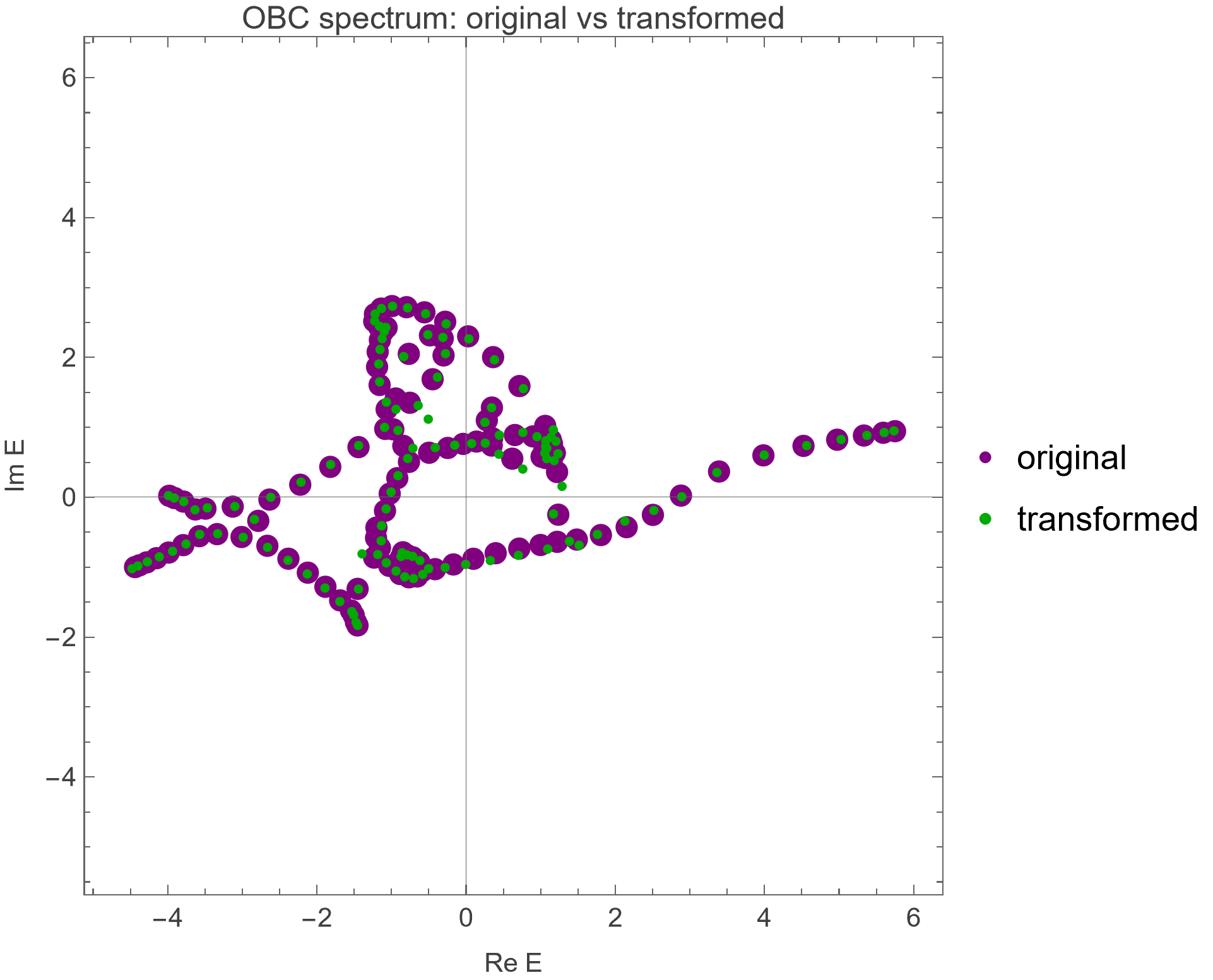}}
    \caption{The eigenspectra before and after transform with the given $H,S$ and finite size $N$. After transform while the bulk is preserved correctly, some isolated eigenvalues differ significantly. The isolated discrepancies are boundary states; we verified that no additional zero mode is created, the winding of $\det H$ about the origin being unchanged by the transform.}
    \label{fig:main-1}
\end{figure*}

\begin{figure*}[htbp]
    \centering
    \subcaptionbox{$H_1,S_1$ error.\label{fig:sub1-3}}
        [0.48\linewidth]{\includegraphics[width=0.52\linewidth]{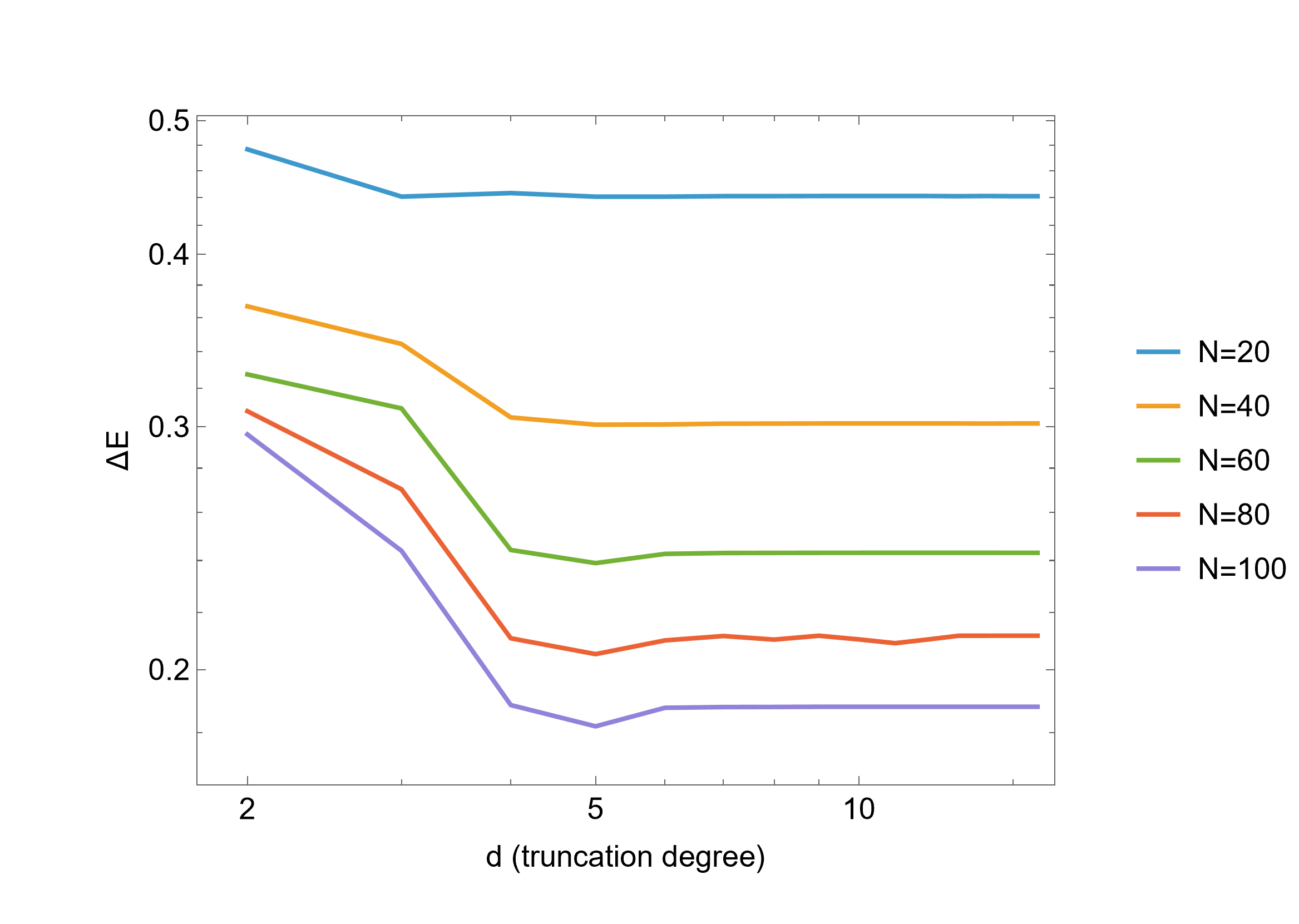}}
    \hfill
    \subcaptionbox{$H_1,S_1$ log of decay term.\label{fig:sub2-5}}
        [0.48\linewidth]{\includegraphics[width=0.52\linewidth]{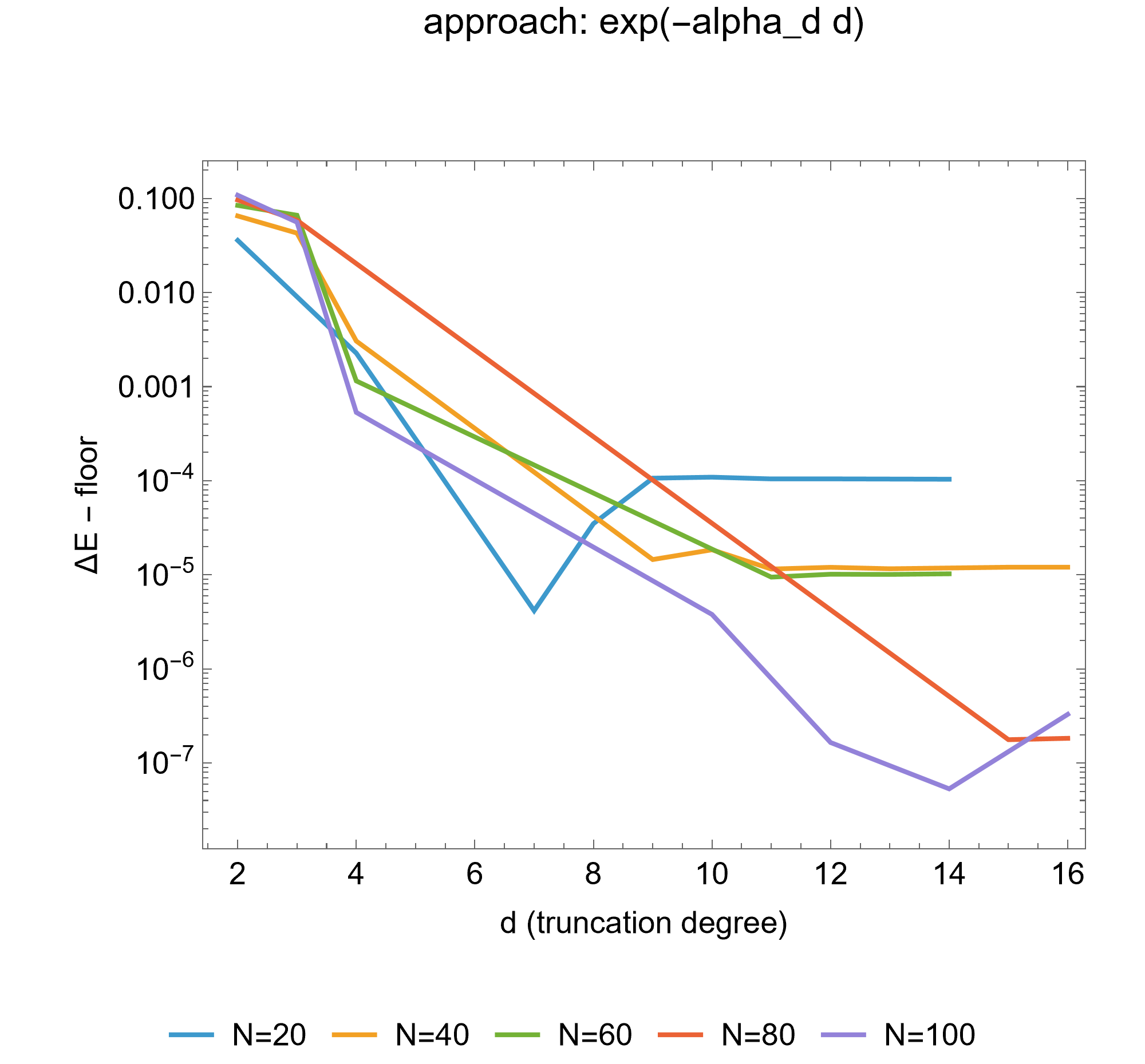}}
    \subcaptionbox{$H_1,S_2$\label{fig:sub3-2}}
        [0.48\linewidth]{\includegraphics[width=0.52\linewidth]{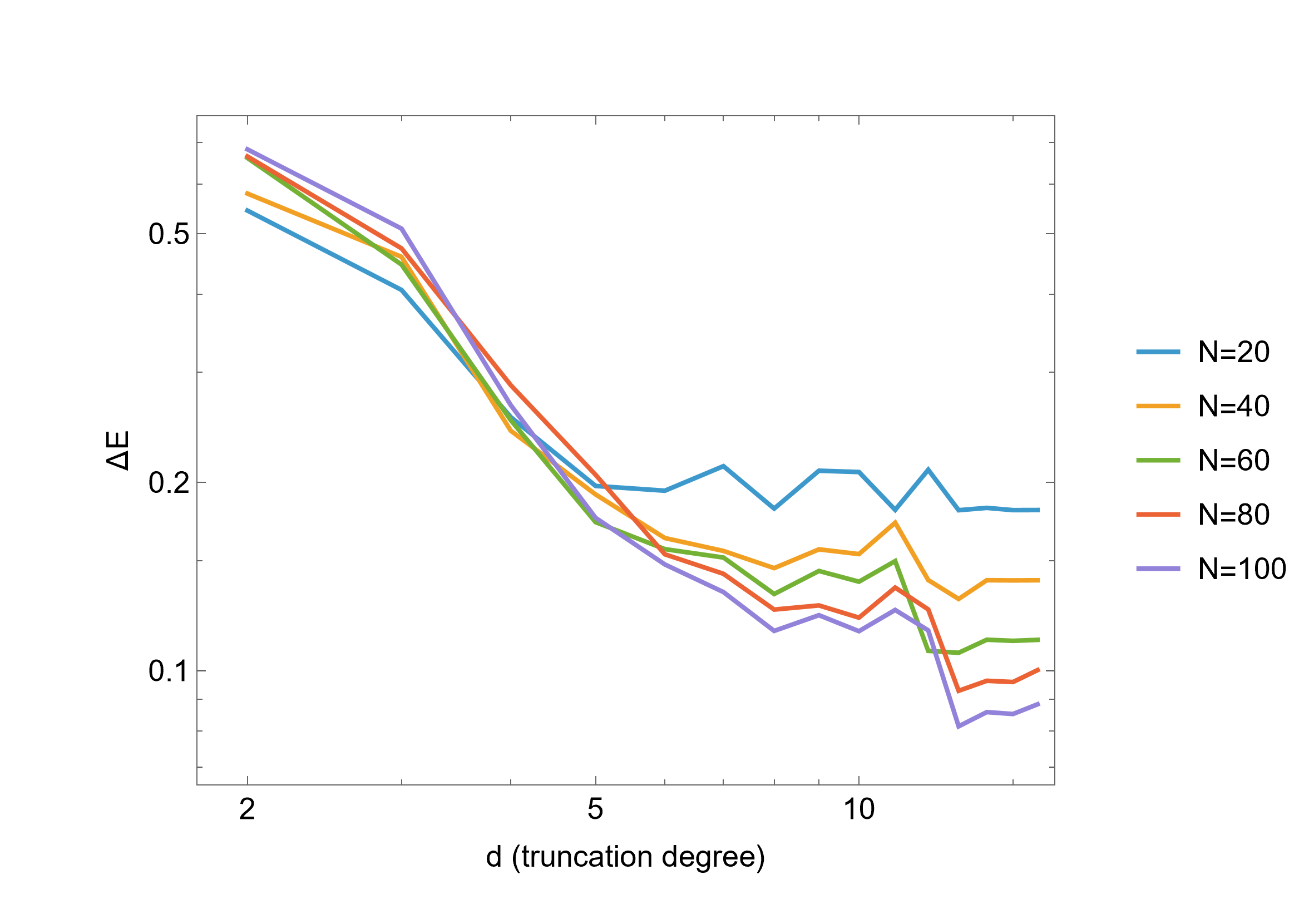}}
    \hfill
    \subcaptionbox{$H_1,S_2$ log of decay term\label{fig:sub4-2}}
        [0.48\linewidth]{\includegraphics[width=0.5\linewidth]{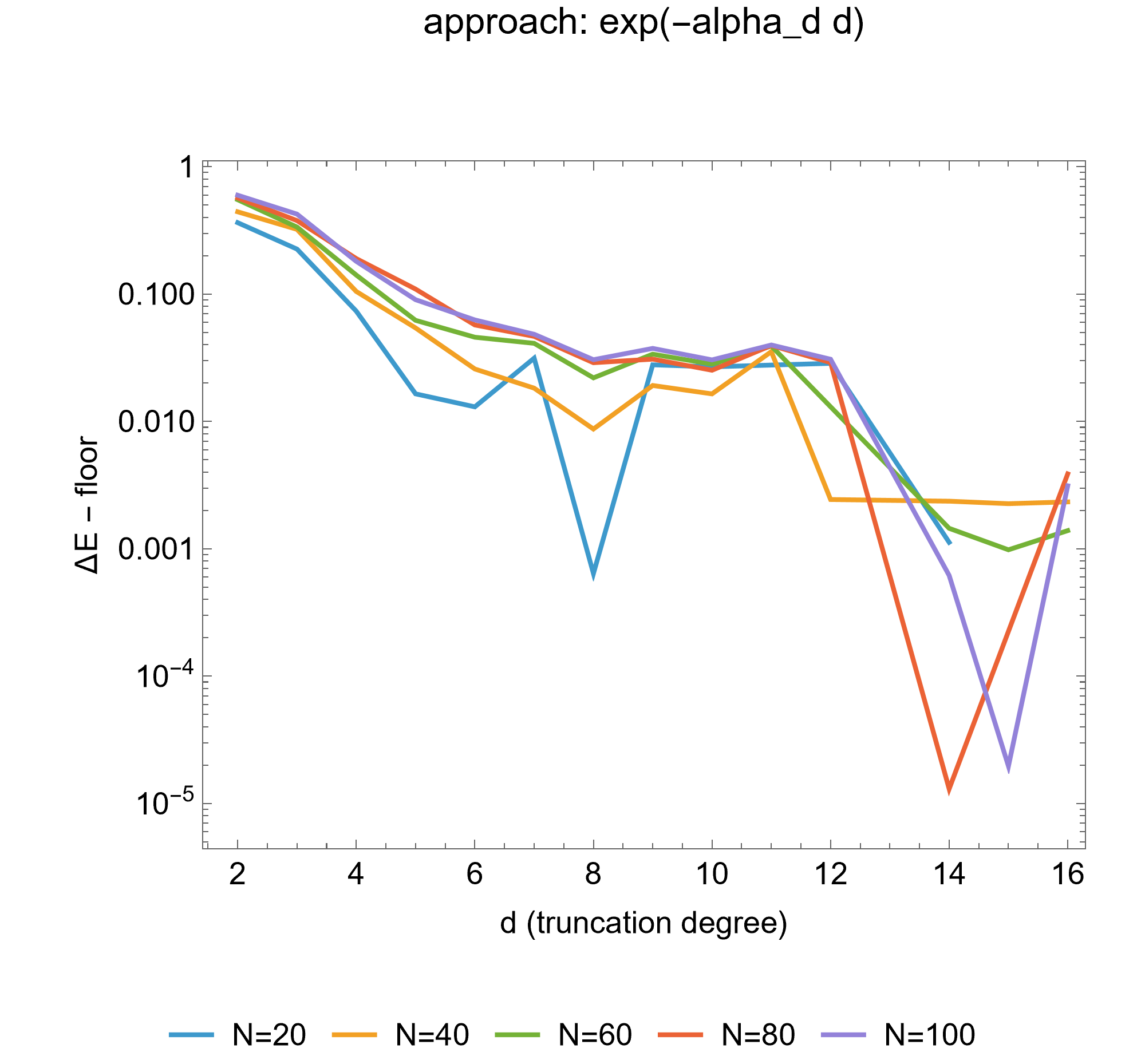}}
        \subcaptionbox{$H_2,S_1$ error.\label{fig:sub1-4}}
        [0.48\linewidth]{\includegraphics[width=0.52\linewidth]{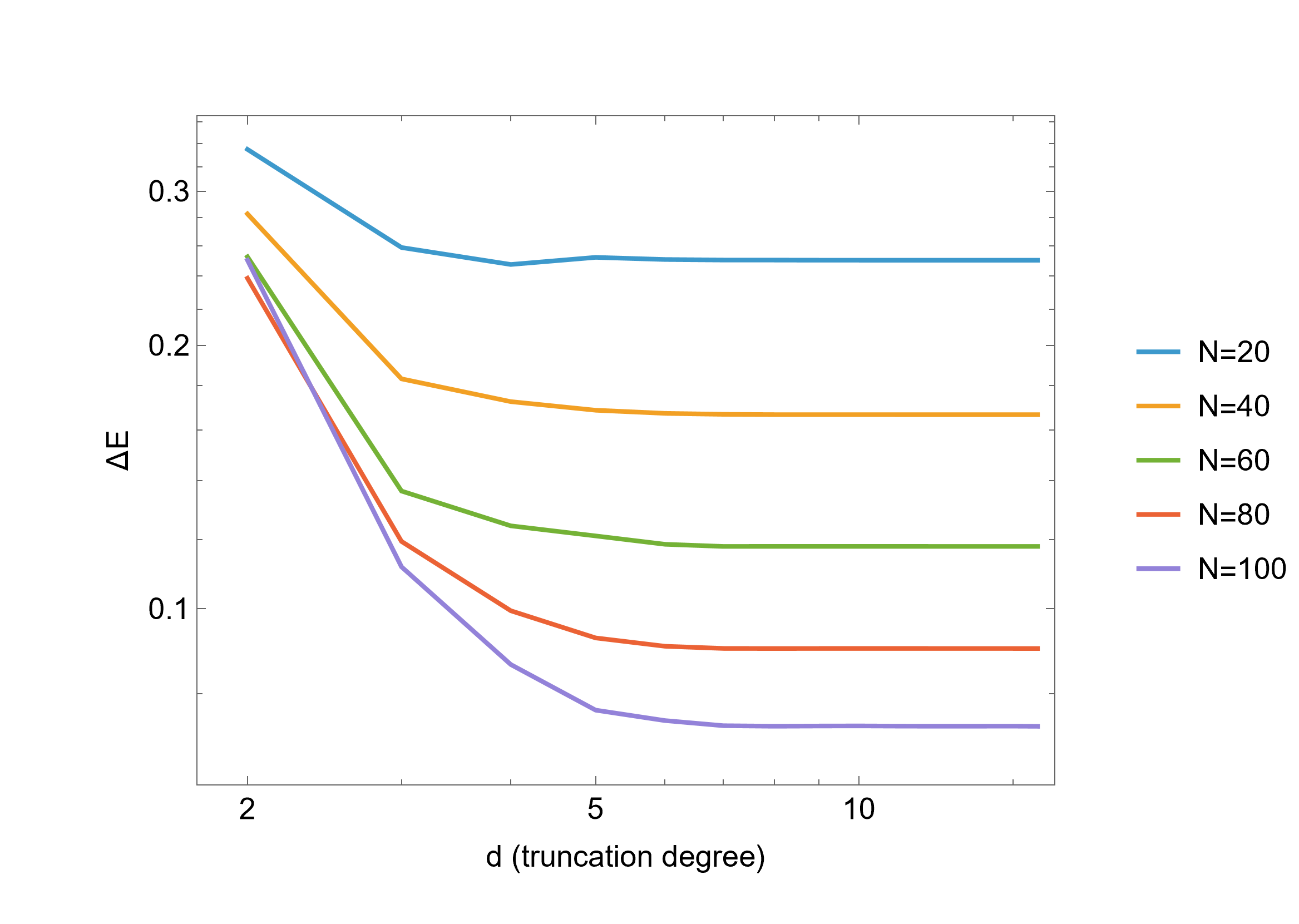}}
    \hfill
    \subcaptionbox{$H_2,S_1$ log of decay term.\label{fig:sub2-6}}
        [0.48\linewidth]{\includegraphics[width=0.52\linewidth]{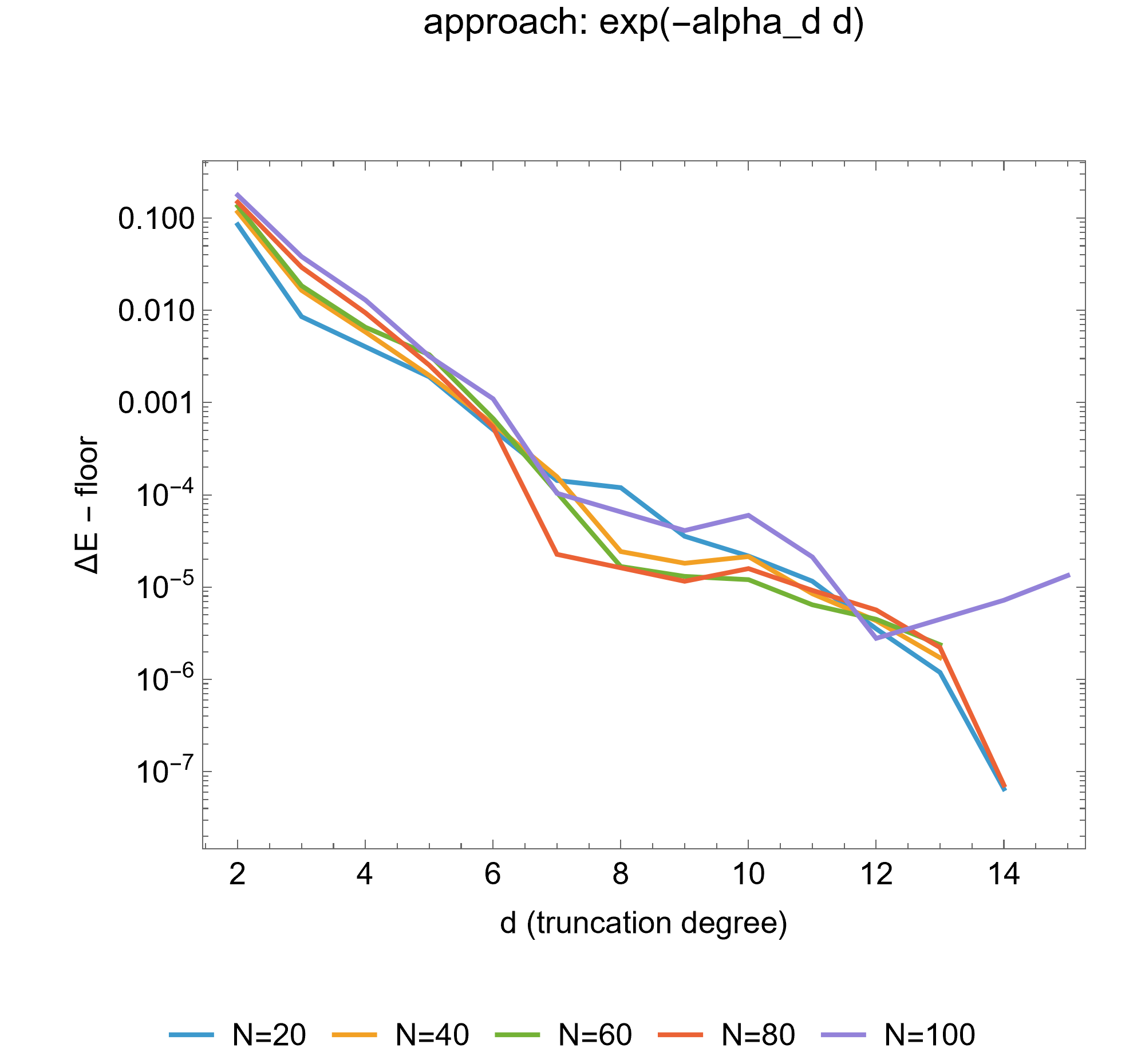}}
    \subcaptionbox{$H_2,S_2$ error\label{fig:sub3-3}}
        [0.48\linewidth]{\includegraphics[width=0.52\linewidth]{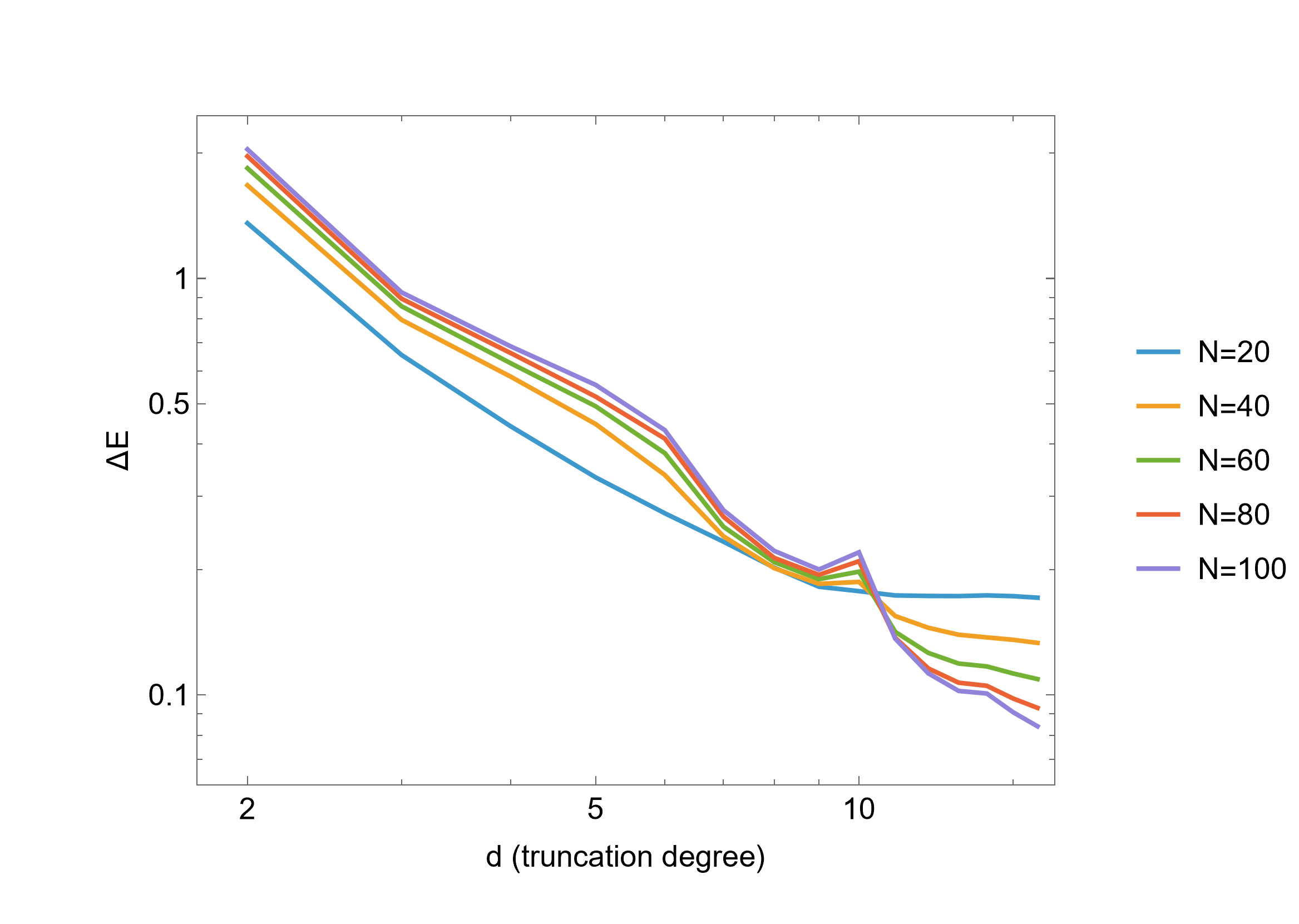}}
    \hfill
    \subcaptionbox{$H_2,S_2$ log of decay term\label{fig:sub4-3}}
        [0.48\linewidth]{\includegraphics[width=0.5\linewidth]{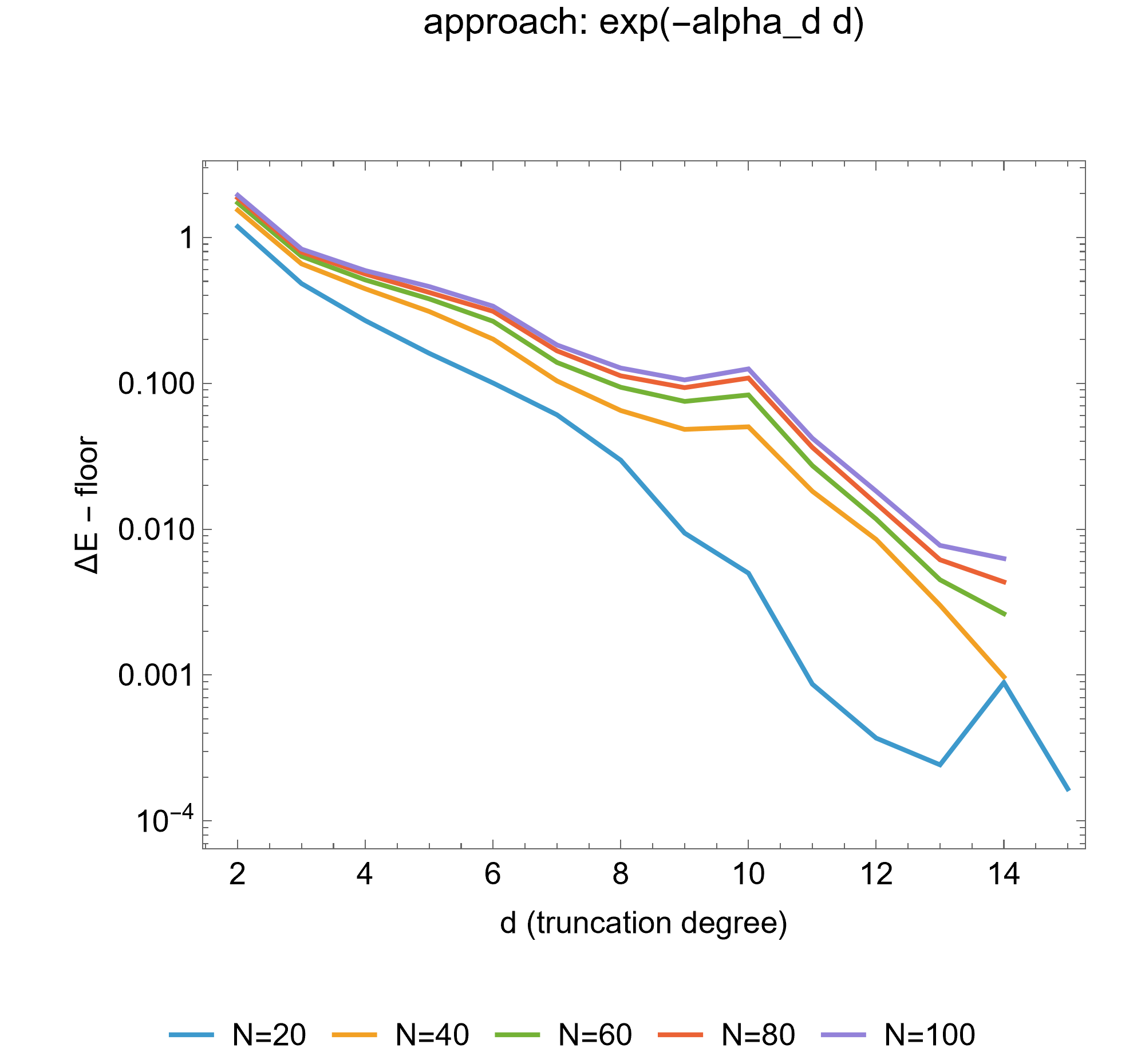}}
    \caption{Error scaling for the supplemental systems. The left column has both axes logarithmic; the right column has a linear horizontal axis and a logarithmic vertical axis. Left column: the weighted spectral distance $\Delta E$ versus truncation degree $d_{S^{-1}}$, one curve per chain length $N$. Right column: the excess $\Delta E-\Delta E_{\mathrm{floor}}$ for the same data, which isolates the approach channel of Eq.~(\ref{eq:dscaling}). The simpler transform $S_1$ gives the cleaner single-exponential decay; for $S_2$ the curves are visibly modulated, consistent with the oscillatory fine structure of Appendix~\ref{app:osc}. Note that the fitted exponents for these systems are less well converged than for the two representative systems of the main text.}
    \label{fig:main-2}
\end{figure*}

The resulting OBC spectra before and after transformation are shown in Fig.~\ref{fig:main-1}, and the corresponding error scaling in Fig.~\ref{fig:main-2}.

\newpage

\clearpage
\section{Critical NHSE Hamiltonian}



\begin{equation}
H_{3}(z)= \begin{pmatrix}
1.2z+0.4+0.2/z & 0 \\
0 & 0.2z-0.4+1.2/z
\end{pmatrix}
\label{eq:H3}
\end{equation}

\begin{figure*}[htbp]
    \centering
    \subcaptionbox{$H_3,S_2,N=10$.\label{fig:sub1-5}}
        [0.48\linewidth]{\includegraphics[width=0.52\linewidth]{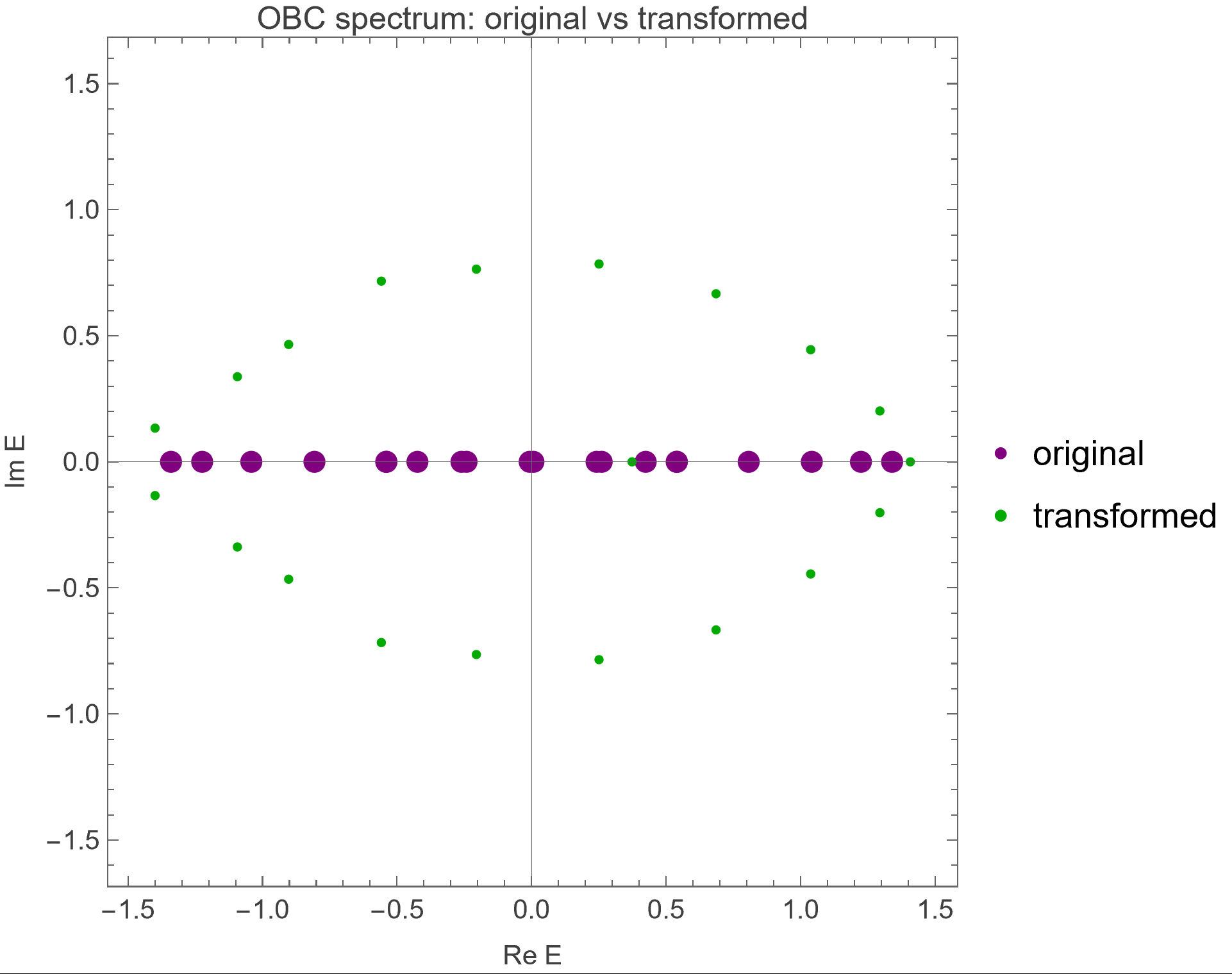}}
    \hfill
    \subcaptionbox{$H_3,S_2,N=20$.\label{fig:sub2-7}}
        [0.48\linewidth]{\includegraphics[width=0.52\linewidth]{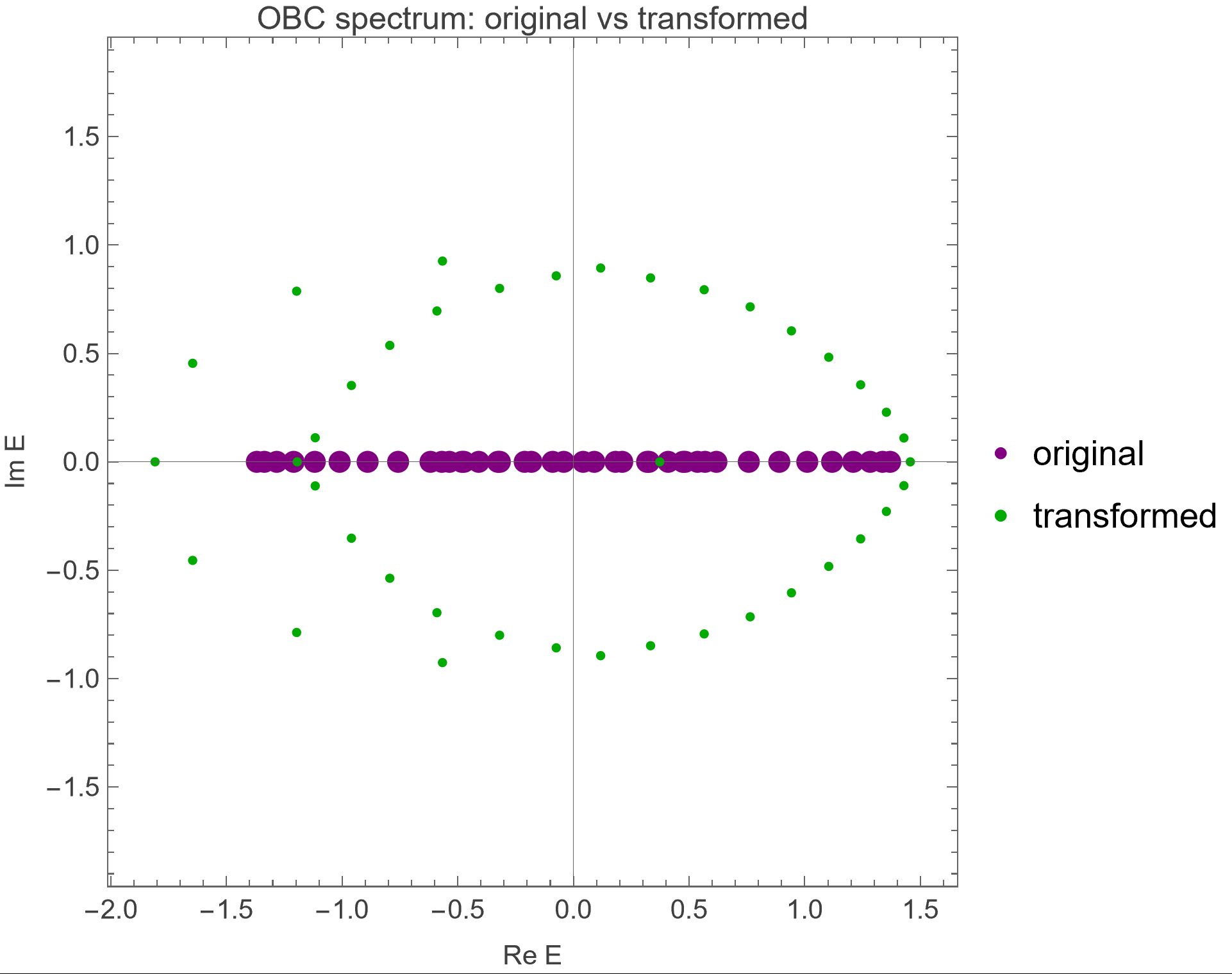}}
    \subcaptionbox{$H_3,S_2,N=30$\label{fig:sub3-4}}
        [0.48\linewidth]{\includegraphics[width=0.52\linewidth]{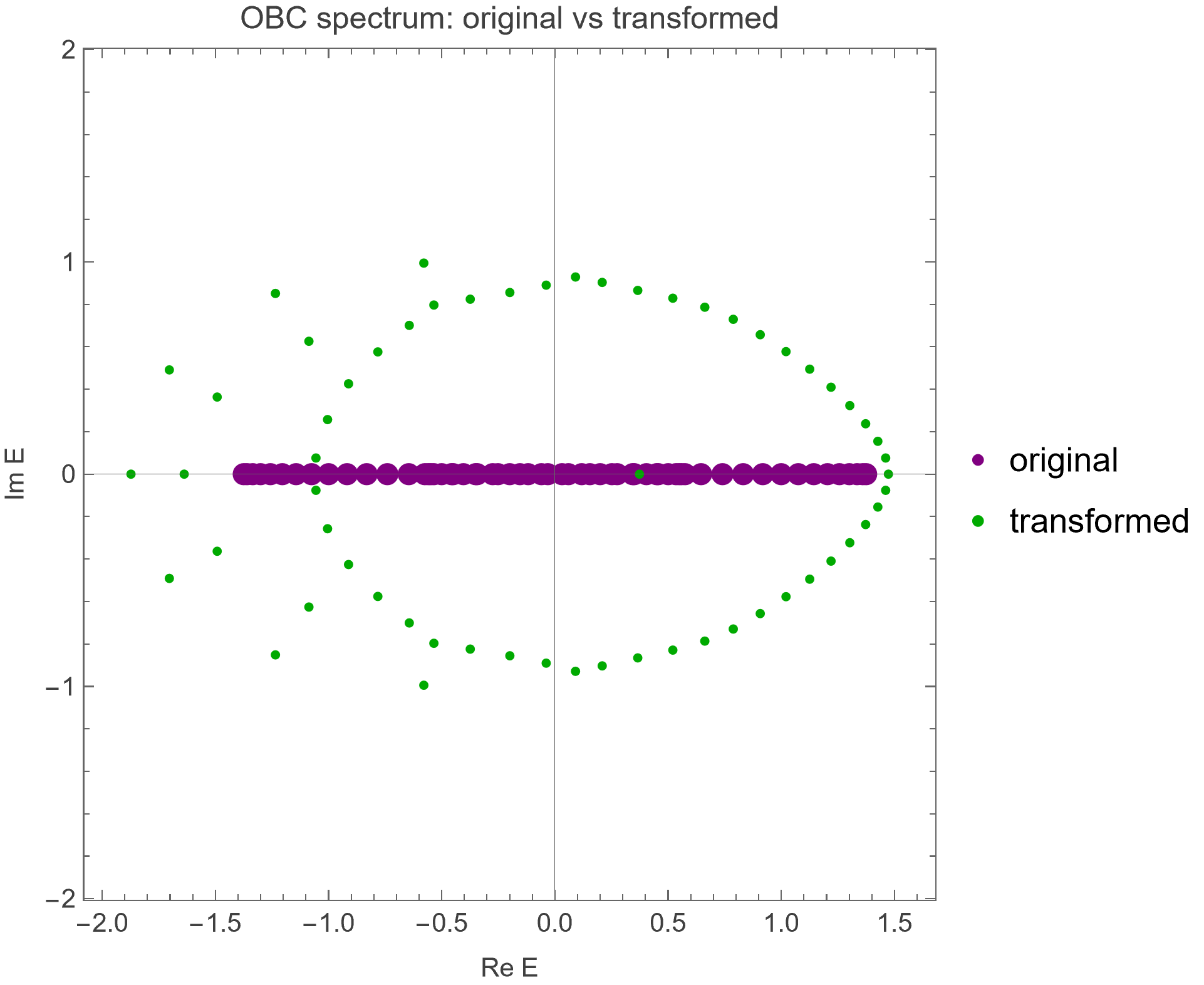}}
    \caption{The eigenspectra before and after transform with $H_3,S_2$ and finite size $N$. The transformed Hamiltonian inherits the extreme sensitivity property to certain parameters. Any infinitesimal error in this case leads to a finite difference of the OBC spectra at large $N$. }
    \label{fig:main-3}
\end{figure*}

We further investigate $H_3$ that exhibits the critical non-Hermitian skin effect, which has finite size dependent coupling properties. The corresponding OBC spectrum is unstable with respect to any slight change of the off-diagonal terms away from zero, i.e.\ the zero-coupling and finite-coupling limits do not commute. Figure~\ref{fig:main-3} shows the effect growing with system size.

\end{document}